\documentclass{article}

\usepackage{microtype}
\usepackage{graphicx}
\usepackage{subcaption}
\usepackage{booktabs} 

\usepackage{hyperref}

\usepackage[accepted]{icml2024}

\usepackage{amsmath}
\usepackage{amssymb}
\usepackage{mathtools}
\usepackage{amsthm}

\usepackage[capitalize]{cleveref}

\usepackage{multirow}
\usepackage{rotating}

\usepackage{colortbl}
\definecolor{step-2}{rgb}{1,0.949,0.875}
\definecolor{step-1_9}{rgb}{1,0.949,0.84}
\definecolor{step-1}{rgb}{1,0.949,0.8}
\definecolor{ref-0_5}{rgb}{1,0.949,0.775}
\definecolor{ref}{rgb}{1,0.949,0.729}
\definecolor{ref_5}{rgb}{1,0.935,0.729}
\definecolor{step1}{rgb}{1,0.91,0.729}
\definecolor{step1_5}{rgb}{1,0.920,0.729}
\definecolor{step1_75}{rgb}{1,0.895,0.729}
\definecolor{step1_8}{rgb}{1,0.88,0.729}
\definecolor{step2}{rgb}{1,0.865,0.729}
\definecolor{step2_5}{rgb}{1,0.865,0.729}
\definecolor{step3}{rgb}{1,0.82,0.7}
\definecolor{step3_5}{rgb}{1,0.802,0.7}
\definecolor{step4}{rgb}{1,0.789,0.7}
\definecolor{step5}{rgb}{1,0.75,0.68}

\theoremstyle{plain}

\theoremstyle{definition}

\theoremstyle{remark}

\usepackage[textsize=tiny]{todonotes}

\renewcommand{\paragraph}[1]{\textbf{#1}. }
\renewcommand{\paragraph}[1]{\textbf{#1}. }

\icmltitlerunning{Neural SPH: Improved Neural Modeling of Lagrangian Fluid Dynamics}

\begin{document}

\twocolumn[
\icmltitle{Neural SPH: Improved Neural Modeling of Lagrangian Fluid Dynamics}



\icmlsetsymbol{equal}{*}

\begin{icmlauthorlist}
\icmlauthor{Artur P. Toshev}{tum}
\icmlauthor{Jonas A. Erbesdobler}{tum}
\icmlauthor{Nikolaus A. Adams}{tum,mep}
\icmlauthor{Johannes Brandstetter}{jku,nxai}
\end{icmlauthorlist}

\icmlaffiliation{tum}{Chair of Aerodynamics and Fluid Mechanics, School of Engineering and Design, Technical University of Munich, Garching, Germany}
\icmlaffiliation{jku}{ELLIS Unit Linz, LIT AI Lab, Institute for Machine Learning, Johannes Kepler University, Linz, Austria}
\icmlaffiliation{nxai}{NXAI GmbH, Austria}
\icmlaffiliation{mep}{Munich Institute of Integrated Materials, Energy and Process Engineering, Technical University of Munich, Germany}
\icmlcorrespondingauthor{Artur P. Toshev}{artur.toshev@tum.de}

\icmlkeywords{partial differential equations, Navier-Stokes equations, learned solver, graph neural networks, physics conservation}

\vskip 0.3in
]



\printAffiliationsAndNotice{}  

\begin{abstract}
Smoothed particle hydrodynamics (SPH) is omnipresent in modern engineering and scientific disciplines. SPH is a class of Lagrangian schemes that discretize fluid dynamics via finite material points that are tracked through the evolving velocity field. Due to the particle-like nature of the simulation, graph neural networks (GNNs) have emerged as appealing and successful surrogates. However, the practical utility of such GNN-based simulators relies on their ability to faithfully model physics, providing accurate and stable predictions over long time horizons -- which is a notoriously hard problem. In this work, we identify particle clustering originating from tensile instabilities as one of the primary pitfalls. Based on these insights, we enhance both training and rollout inference of state-of-the-art GNN-based simulators with varying components from standard SPH solvers, including pressure, viscous, and external force components. All Neural SPH-enhanced simulators achieve better performance than the baseline GNNs, often by orders of magnitude in terms of rollout error, allowing for significantly longer rollouts and significantly better physics modeling. Code available under \href{https://github.com/tumaer/neuralsph}{https://github.com/tumaer/neuralsph}.
\end{abstract}

\begin{figure}[th]
    \centering
      \begin{sideways}
        \begin{minipage}{0.07\textheight}
          \centering
          GNS$_{\phantom{g}}$
        \end{minipage}
      \end{sideways}
    \includegraphics[trim={0 0cm 0 0},clip,width=0.93\linewidth]{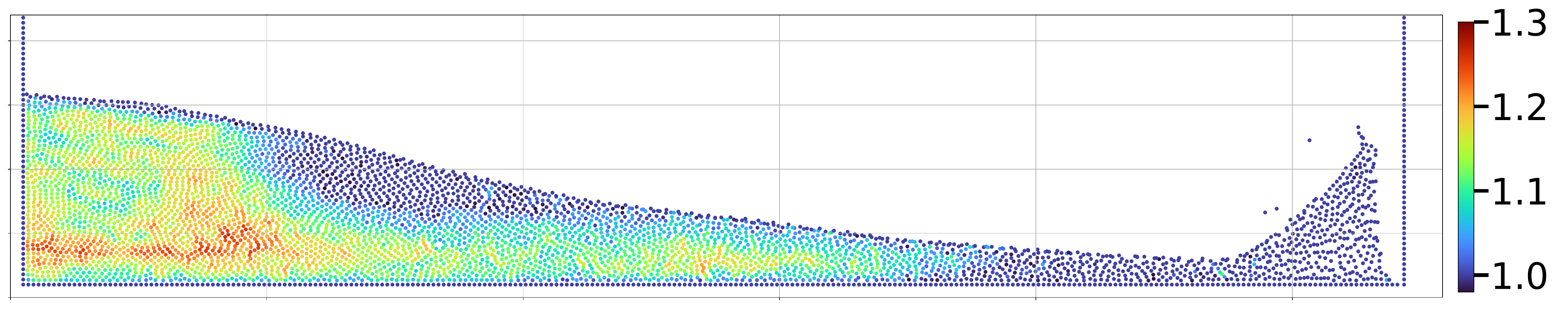}

    \centering
      \begin{sideways}
        \begin{minipage}{0.07\textheight}
          \centering 
          GNS$_g$
        \end{minipage}
      \end{sideways}
    \includegraphics[trim={0 0cm 0 0},clip,width=0.93\linewidth]{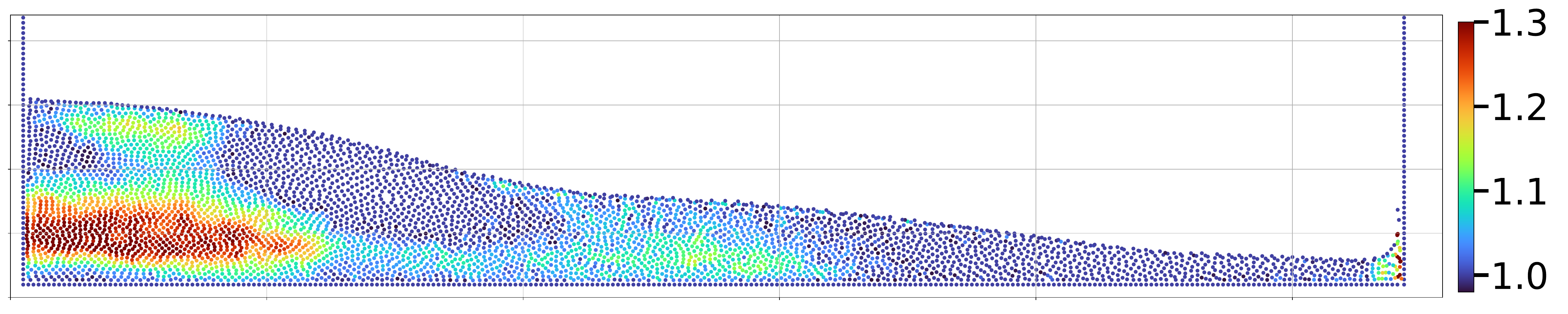}

    \centering
      \begin{sideways}
        \begin{minipage}{0.07\textheight}
          \centering
          GNS$_{g,p}$
        \end{minipage}
      \end{sideways}
    \includegraphics[trim={0 0cm 0 0},clip,width=0.93\linewidth]{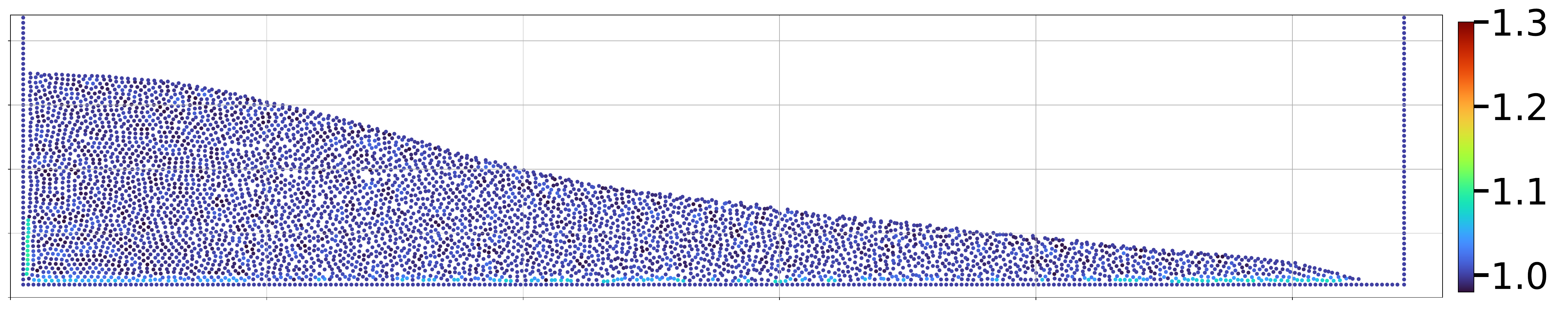}

    \centering
      \begin{sideways}
        \begin{minipage}{0.07\textheight}
          \centering
          SPH$_{\phantom{g}}$
        \end{minipage}
      \end{sideways}
    \includegraphics[trim={0 0cm 0 0},clip,width=0.93\linewidth]{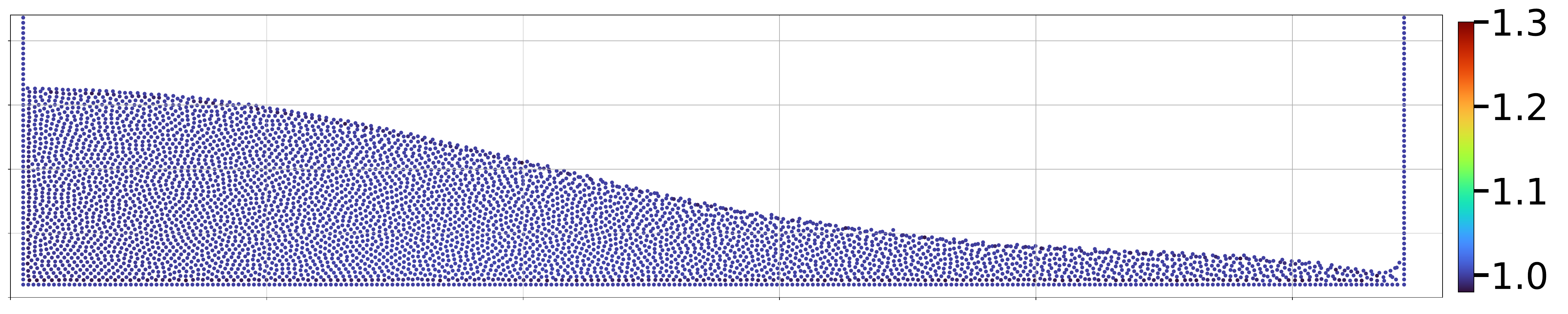}

    \caption{Neural SPH improves Lagrangian fluid dynamics, showcased by physics modeling of the 2D dam break example after 80 rollout steps. Different models exhibit different physics behaviors. From top to bottom: GNS~\citep{Sanchez:20}, GNS with corrected force only (GNS$_g$), full SPH enhanced GNS (GNS$_{g,p}$), and the ground truth SPH simulation. 
    The colors correspond to the density deviation from the reference density; the system is considered physical within 0.98-1.02.
    }
    \label{fig:dam_step80}
\end{figure}

\section{Introduction} \label{sec:intro}

In the sciences, considerable efforts have led to the development of highly complex mathematical models of our world, with many naturally formulated as partial differential equations (PDEs). Over the past years, deep neural network-based PDE surrogates have gained significant momentum as a more computationally efficient solution methodology~\citep{thuerey2021physics,brunton2023machine}, transforming amongst others computational fluid dynamics~\citep{guo2016convolutional, kochkov2021machine,li2020fourier,gupta2023towards,alkin2024universal}, weather forecasting~\citep{rasp2021data, weyn2020improving, sonderby2020metnet, pathak2022fourcastnet, lam2022graphcast, nguyen2023climax,bodnar2024aurora}, and molecular modeling \citep{gasteiger2021gemnet,batzner2022e3, batatia2022mace,Zeni:23,Merchant:23}. 

In computational fluid dynamics (CFD), we broadly categorize numerical simulation methods into two distinct families: particle-based and grid-based, better known as Lagrangian and Eulerian discretization schemes. In Eulerian schemes, space is discretized, i.e., fixed finite nodes or control volumes lead to grid-based or mesh-based models. In Lagrangian schemes, the discretization happens on finite material points, commonly known as particles, which dynamically move with the local deformation of the continuum. One of the most prominent Lagrangian discretization schemes is smoothed particle hydrodynamics (SPH), originally proposed by \citet{lucy1977numerical} and \citet{gingold1977smoothed} for applications in astrophysics.
In contrast to grid- and mesh-based approaches, SPH approximates the field properties using radial kernel interpolations over adjacent particles. The strength of the SPH method is that it does not require connectivity constraints, e.g., meshes, which is particularly useful for simulating systems with large deformations. Since its foundation, SPH has been greatly extended and is the preferred method to simulate problems with (a) free surfaces~\citep{marrone2011delta,violeau2016smoothed}, (b) complex boundaries~\citep{adami2012generalized}, (c) multi-phase flows~\citep{hu2007incompressible}, and (d) fluid-structure interactions~\citep{antoci2007numerical}.

In deep learning, graph neural networks (GNNs)~\citep{Scarselli:08, Kipf:17} are an obvious fit to model particle-based dynamics. Often, predicted accelerations at the nodes are numerically integrated to model the time evolution of the particles or the mesh, i.e., dynamics are updated in a hybrid neural-numerical fashion~\citep{Sanchez:20, pfaff2020learning, Mayr:23}. Most recent applications of GNN-based simulators involve Lagrangian fluid simulations~\citep{toshev2023learning,toshev2024lagrangebench,winchenbach2024symmetric}. One downside of these simulators is the risk of non-physical instabilities during rollout, which affects the neural and numerical components.

It is known that already standard SPH schemes exhibit tensile instability, i.e., numerical errors leading to particle clumping and void regions when negative pressure occurs within what should be an incompressible fluid~\citep{price2012smoothed}. This has led to the development of improved SPH schemes explicitly targeting regularity of particle distribution~\citep{adami2013transport, zhang2017generalized}. A review of SPH literature indicates that even methods seeking to improve other properties, like reducing artificial dissipation~\citep{zhang2017weakly} or handling violent water flows~\citep{marrone2011delta}, may also improve the particle distribution.

In this work, we present a large-scale analysis of Lagrangian physical modeling capabilities of various GNN-based simulators, i.e., a non-equivariant and an equivariant one. We identify a shared pitfall, i.e., particle clustering effects that are similar to those known from SPH schemes. Particle clustering in GNN-based simulators limits stable rollouts and accurate physics modeling.
Based on these insights, we draw inspiration from numerical SPH solvers and enhance both training and inference of state-of-the-art GNN-based simulators with varying components
from standard SPH solvers, including (i) pressure, (ii) viscous, and (iii) external force components -- all implemented in JAX~\citep{jax2018github}. Methodologically, our main contributions are two: (a) novel external force treatment during training, and (b) an additional SPH relaxation routine during inference.

We demonstrate the efficacy of Neural SPH-enhanced Lagrangian simulators by achieving better performance on seven diverse 2D and 3D Lagrangian datasets -- sometimes by orders of magnitude in terms of rollout error -- than the baseline GNN, allowing for significantly better physical modeling capabilities. We note that the introduced Neural SPH techniques may apply to a wide range of physics scenarios beyond GNNs and SPH. Our source code is available at \href{https://github.com/tumaer/neuralsph}{https://github.com/tumaer/neuralsph}.

\section{Simulating Lagrangian dynamics} \label{challenges}

\paragraph{Smoothed particle hydrodynamics}
Smoothed particle hydrodynamics (SPH) approximates the incompressible Navier-Stokes equations (NSE) by the so-called weakly compressible NSE. This is necessary because the density of the fluid is defined by radial kernel summation $\rho_i=\sum_j m_j W(r_{ij}|h) $, where $m_j$ represents the mass of the adjacent particles $j$, and $W$ the radial interpolation kernel with smoothing length $h$ that operates on the scalar distance $r_{ij}$. This summation may violate strict incompressibility. However, the weak compressibility assumption typically allows for up to $\sim1\%$ density deviation~\citep{monaghan2005smoothed}. This $\sim 1\%$ is also enforced for the weakly compressible SPH method, while evolving density and momentum:
\begin{align}
\frac{\mathrm{d}}{\mathrm{d}t}(\rho)
&=
-\rho \left( \nabla \cdot \mathbf{u} \right) ,
\label{eq:nse_mass} \\
\frac{\mathrm{d}}{\mathrm{d}t}( \mathbf{u}) &= \underbrace{- \frac{1}{\rho}\nabla p}_\text{pressure} + \underbrace{\frac{\nu}{V_{ref}L_{ref}} \nabla^2  \mathbf{u}}_\text{viscosity} + \underbrace{\mathbf{g}}_\text{ext. force}.
\label{eq:nse_momentum}
\end{align}
Herein, $\rho$ is the density, $\mathbf{u}$ the velocity vector, $p$ the pressure, $\mathbf{g}$ the external force, $\nu$ the viscosity, and $U_{ref},L_{ref}$ the reference velocity and length scale. Without loss of generality, we consider $U_{ref}=1$, $L_{ref}=1$. We note that either density summation with kernel averaging, or density evolution (\cref{eq:nse_mass}) is used to compute the density, and as we explain later, the former is the preferred and the latter the more general approach. To evolve the system in time, the above equation(s) are integrated in time by, e.g., semi-implicit Euler (see \cref{app:coarsening_appendix}).
However, solving these equations with standard SPH methods may still produce artifacts, most notably when particle clumping exceeds the 1\% density-fluctuation requirement~\citep{adami2013transport}.

\paragraph{SPH particle redistribution}
The term responsible for a homogeneous particle distribution in the SPH method is the pressure gradient term $\frac{1}{\rho}\nabla p$ in the momentum equation \cref{eq:nse_momentum}. In weakly compressible SPH, the pressure is computed from density through the equation of state
\begin{equation}
    p(\rho)= p_{ref}\left(\frac{\rho}{\rho_{ref}} - 1 \right).
    \label{eq:tait_eos}
\end{equation}
Thus, for a reliable approximation of the density $\rho$, the pressure term ensures a repulsive force of scale $p_{ref}$ whenever the density exceeds the given reference value $\rho_{ref}$, where typically $\rho_{ref}=1$. However, the pressure term is not necessarily sufficient for producing a good particle distribution, as we can see in the bottom part of Fig. 9 in~\citet{toshev2024lagrangebench}. For this reason, more advanced SPH schemes have been developed, distinguishing between the physical velocity field and the velocity by which particles are shifted~\citep{adami2013transport, zhang2017generalized}. These schemes are related to Arbitrary Lagrangian-Eulerian methods~\citep{hirt1974arbitrary} instead of being fully Lagrangian.

\paragraph{Challenges of density computation at free surfaces}
Accurately computing the density at free surfaces is a difficult task for SPH methods. In the standard SPH formulation, the density at each particle is calculated by a kernel-weighted summation of the mass of adjacent particles~\citep{gingold1977smoothed}. However, particles at free surfaces have low density when using density summation, which leads to incorrect pressure values~\citep{monaghan1994simulating}. The low-density inconsistency can be corrected for by globally and locally conservative least-squares interpolation~\citep{dilts2000mls}, adaptive kernel estimation procedure~\citep{sigalotti2006freesurf}, or by initializing the simulation by first evolving particles with a heavily damped version of the momentum conservation~\citep{becker2007weakly}.
However, most SPH methods for free surface flows resort to the continuity equation to represent the rate of change in density~\cite{monaghan1994simulating,bonet1999corrected}. In this density evolution formulation, density derivatives are integrated over time (see \cref{eq:nse_mass}). On top of the density evolution, density filters, such as periodic re-initialization, are applied~\citep{Gesteira2010stateoftheart, colagrossi2003numerical, shepard1968filter}.

\paragraph{GNN-based simulators}
The formulation of the learning problem is based on LagrangeBench~\citep{toshev2024lagrangebench}. We look at the task of autoregressive acceleration prediction of a Lagrangian particle system, which we then integrate twice using semi-implicit Euler integration to evolve the system over time (see \cref{app:coarsening_appendix}). The datasets consist of particle types per particle and particle coordinates $\textbf{P}^{t_k}$ over $k\in(0, K)$ steps, where each frame $\textbf{P}^t$ is made up of $n\in(1, N)$ particles $\textbf{p}_n^t \in \mathbb{R}^d$ in dimension $d$. The inputs to the learned surrogate are state vectors $\textbf{X}^{t_{k-H}:t_k}$, with history size $H$, each of which contains the past velocities $\mathbf{U}_k=[\mathbf{u}_{k,1},...,\mathbf{u}_{k,N}]$ inferred using the finite difference approximation of past coordinates, as well as optional features like external force vector $\mathbf{g}$, e.g., gravity. 

We use the default configuration files from LagrangeBench for training, including random walk noise~\citep{pfaff2020learning} and the pushforward trick~\citep{brandstetter2021message}. These default configurations provide the baseline models, on top of which we add our methods.

\paragraph{Pathological particle clustering during long rollouts for GNN-based simulators}
The entering point to our analysis is the realization that simulated rollouts of a learned Graph Network-based Simulators (GNS)~\citep{Sanchez:20} severely violate the $1\%$ compressibility requirement present in weakly compressible SPH methods -- see top part of \cref{fig:dam_step80}. This figure shows compression of as much as $1.4 \cdot \rho_{ref}$ in the left part, which is not only unphysical regarding the density itself but might also lead to unphysical dynamics in the sense of periodic compressions and expansions later in the rollout, see \cref{sec:experiments}. The violation -- although much worse -- resembles pressure inaccuracies in classical numerical SPH solvers.

\begin{figure}[ht]
    \centering
    \includegraphics[trim={0mm 5mm 0mm 0mm},clip,width=0.9\linewidth]{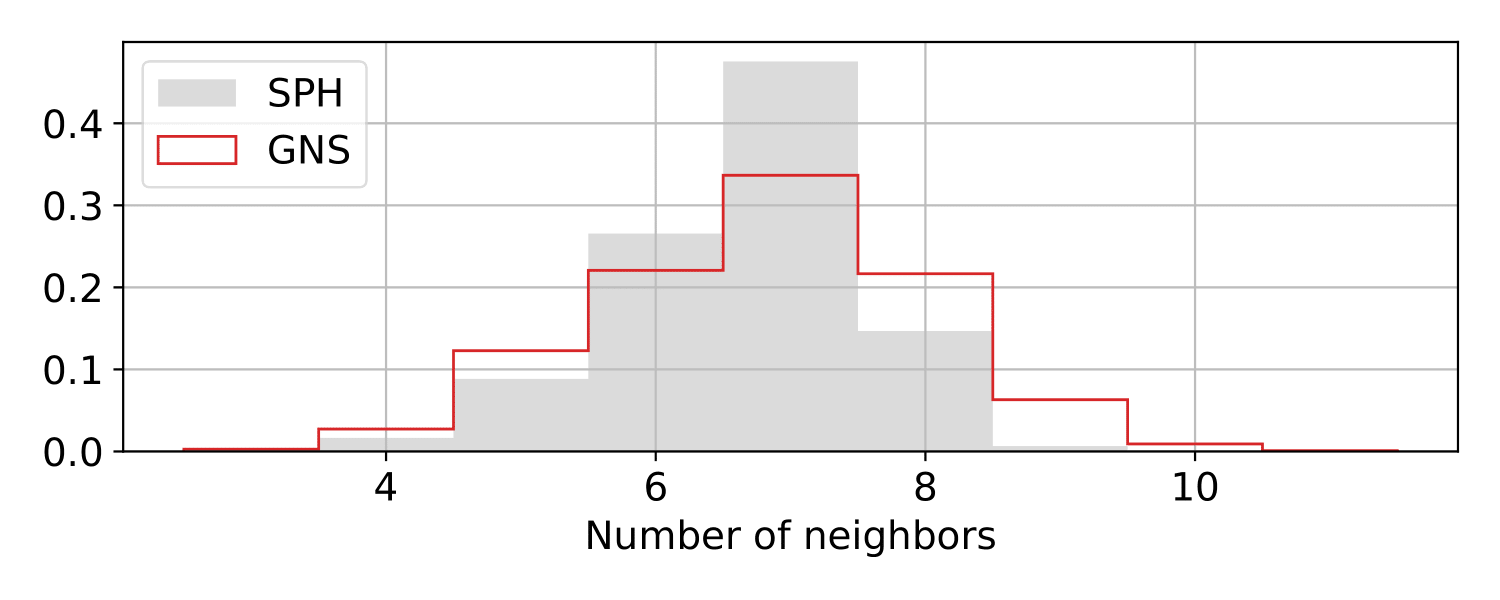}
    \caption{Number of neighbors mismatch due to particle clustering. Histogram of the number of neighbors of the 2D lid-driven cavity experiment after 400 rollout steps (average over all test rollouts).}
    \label{fig:ldc_hist_nbrs}
\end{figure}

To qualitatively understand clustering, in \cref{fig:ldc_hist_nbrs}, we plot the histogram of the per-particle number of neighbors corresponding to the left graphic of the 2D lid-driven cavity from \cref{fig:ldc_step400}, which also has regions with high particle density. In \cref{fig:ldc_hist_nbrs}, we see a pronounced increase in the number of particles with 8-10 neighbors, indicating clustering artifacts.

\paragraph{The problem of external forces}
We observe that in roughly 8 out of 25 dam break test trajectories at step 80, the front of the wave spreads out as if a virtual wall exists way in front of the actual wall -- see \cref{fig:dam_step80,app:dam_plots}. Such behavior has been discussed in literature~\citep{klimesch2022simulating}, and the current consensus is that the GNN-based simulators learn to infer the dynamics from velocity correlations. Thus, when the velocity reaches a given threshold, it has learned to model the presence of a wall. In the following, we demonstrate that by forcing the network to predict a target acceleration that excludes the external force part, the overall dynamics become more physical, and significantly fewer artifacts occur.

\section{Neural SPH} \label{sec:method}

In this section, we introduce Neural SPH, which improves both training and rollout inference of temporally coarsened GNN-based simulators. Neural SPH comprises a routine to correct for induced modeling errors due to external forces, and inference-time refinement steps of the system state based on SPH relaxation methods.

\paragraph{Correction of external forces}
In the learning problem formulation by~\citet{toshev2024lagrangebench}, the GNN-based simulators receive as node inputs a time sequence of the $H$ most recent historic velocities stacked to $\mathbf{u}_{k-H:k}=[\mathbf{u}_{k-H},...\mathbf{u}_k]$ and an optional external force vector. Consequently, the GNN-based simulators are confronted with the underlying instantaneous force and not the effective force, i.e., the force that acts on the particles upon temporal coarsening. We make two observations: 
\vspace{-4pt}
\begin{enumerate}
    \item The impact of the external force $\mathbf{g}$ is already included in the dynamics given by the past velocities $\mathbf{u}_{k-H:k}$. Thus, providing a constant force vector, i.e., gravitational force, as model input might be necessary when training equivariant models, but as~\citet{Sanchez:20} show in their appendix C2, the GNS model does not improve when external force information is added. However, in the general case of systems with spatially varying forces, having force vectors as inputs is crucial. An example is the reverse Poiseuille flow, which has a positive force in $x$ direction when $y>1$ and a negative force when $y<1$ (see \cref{app:erf_appendix}).
    \item By predicting the full acceleration $\mathbf{a}$, the GNN-based simulators are forced to model gravity implicitly. One might argue that gravity is just a bias term in the last decoder layer, and thus, a GNN-based simulator should be able to model gravitational effects quite easily. However, we observe that for a GNS model trained on dam break (see \cref{fig:dam_step80} top part), the bias term in the last layer is more than an order of magnitude smaller than the respective gravitational acceleration.
\end{enumerate}

Especially the latter observation indicates that GNN-based simulators indeed mainly learn velocity correlations as suggested by~\citet{klimesch2022simulating}. Referring to the structure of \cref{eq:nse_momentum}, and motivated by operator splitting, we suggest to bracket terms on the right-hand side of this equation as $[...] + \mathbf{g}$. If considering temporal coarsening of GNN-based simulators over $M$ SPH steps, and given that the dataset is generated by running an SPH simulation with a constant time step $\Delta t_{SPH}$, the steps over which the GNN-based simulator integrates are $M \Delta t_{SPH}$. In the case of a constant force $\mathbf{g}$, this leads to an effective external force after $M$ SPH steps of $\mathbf{g}_{M}^{FD}=(M\Delta t_{SPH})^2 \mathbf{g}$, as by double integration of acceleration to positions with a finite difference time step $\Delta t^{FD}=1$, see \cref{app:coarsening_appendix}. Thus, when removing the accumulated external force from the full acceleration, i.e., 
\begin{equation}
    \mathbf{a} = \texttt{GNN}(\textbf{X}^{t_{k-H-1}:t_k}, \mathbf{g}) + \mathbf{g}_{M}^{FD} \ ,
\end{equation}
the model is forced to disentangle the interactions between external forces and internal dynamics, i.e., the other two terms on the right-hand side of \cref{eq:nse_momentum}. We attain a powerful formulation of the learning problem since the dynamics are controlled more explicitly, as shown in \cref{fig:dam_step80} and in \cref{fig:dam_traj0,fig:dam_traj13,fig:dam_traj14,fig:dam_traj15} of \cref{app:dam_plots}.  

However, if the force $\mathbf{g}$ varies over space or time, it cannot be independently integrated over $M$ time steps. In this case, modeling the correct effective external force requires (i) precise information on the forces that act on a given particle over each of the $M$ steps we want to coarse-grain over, and (ii) taking the average over these contributions, i.e., $\mathbf{g}_{M}^{FD}=(M\Delta t_{SPH})^2 \frac{1}{M}\sum_{m=1}^M\mathbf{g}_m$. Since we typically do not have access to such information, we propose a convolution-based solution. In the case of a spatially varying but constant in time force field, we use the standard deviation of velocities over the dataset $\sigma_u$ as a proxy of how much a particle moves perpendicularly to the force field, as this perpendicular motion is what we want to smoothen for. We then convolve the force function with a Gaussian distribution $\mathcal{N}(0,\sigma_u^2)$ with the standard deviation $\sigma_u$ and thus smoothen the force function to account for the effective force exerted on a particle that moves across regions with variable forcing. 

This convolution can be implemented in two ways: (i) If the function is simple enough, i.e., an analytical solution exists, we can use it directly. (ii) Alternatively, we may evaluate the instantaneous external force at the current particle coordinates and then apply an SPH kernel convolution, which is very similar to a convolution with a Gaussian, except that it has compact support. Applying a kernel $W(r|h)$ with $h=\sigma_u$ enables us to effectively smoothen any given force function. As a side remark, applying a convolution with an SPH kernel $W(\cdot|h)$ of a particular $h$ over the mass of each adjacent particle is exactly what density summation does.

\paragraph{Correction of particle distribution via SPH relaxation}
In order to correct the pathological particle clustering of learned GNN-based simulators, we add an intermediate step during the rollout of a learned Lagrangian solver, namely an \emph{SPH relaxation step}. The idea is that if the learned solver pushes the system to an unphysical particle configuration, we can reduce density fluctuations by running an SPH relaxation simulation of up to 5 steps. By SPH relaxation, we refer to the process of taking the point cloud right after the temporal update of the learned model, and then -- solely based on the particle coordinates -- applying an SPH update with the assumption of zero initial velocities~\citep{litvinov2015towards,fan2024analysis}.  We can apply SPH relaxation using  the \textbf{pressure term} in \cref{eq:nse_momentum} or the \textbf{viscous term} in \cref{eq:nse_momentum}. One update step of relaxation corresponds to
\begin{align}
\mathbf{a} &= \alpha \frac{-1}{\rho}\nabla p + \alpha \beta \nabla^2  \mathbf{u} \ , \label{eq:nse_momentum_learned} \\ 
\mathbf{p} &= \mathbf{p} + \mathbf{a} \label{eq:nse_integration_learned} \ , 
\end{align}
where we hide the time step and the pre-factors in the hyperparameters $\alpha$ and $\beta$. Adding and fine-tuning these hyperparameters is essential for various reasons: (a) in SPH, it proves challenging to identify a reference velocity, which is needed for determining the time step size; (b) adhering to the Courant-Friedrichs-Lewy (CFL) condition \citep{courant1928partiellen} would most certainly result in smaller time steps, and most importantly, (c) the step size is implicitly determined by how much the GNN-based simulator distorts the system. This largest distortion depends on many factors, such as temporal coarsening steps $M$ and the choice of the GNN-based simulator. We propose fine-tuning these hyperparameters as shown later in this section.

\paragraph{Correction of density at walls and free surfaces}
Recall that also existing SPH methods encounter challenges when predicting the density at free surfaces. On the one hand, density summation, which is the preferred method for density computation due to implicit mass conservation, is not directly applicable to free surfaces since it encounters density inconsistencies. On the other hand, density-transport equations abandon exact mass conservation.

For GNN-based simulators, we propose a novel way of estimating the density of a system at free surfaces.  Our approach combines the SPH requirement that density fluctuations should not exceed $\sim 1\%$ -- which we round up to $2\%$ -- with density summation. We extend density summation by (a) setting all values $<0.98 \rho_{ref}$ to $\rho_{ref}$, and (b) clipping all values $>1.02 \rho_{ref}$, i.e. setting them to $1.02 \rho_{ref}$. Modification (a) guarantees that particles at free surfaces are set to the reference condition, preventing surface instabilities. Modification (b) truncates large outliers akin to gradient clipping when training a neural network, stabilizing the relaxation dynamics. Our approach is closely related to cavitation modeling, where it is common to use tensile instability control (TIC) \citep{sun2018multi} to avoid negative pressure values that increase the particle disorder and eventually lead to the occurrence of particle clustering and clumping~\citep{lyu2023towards}. The main idea of TIC is to change the pressure gradient formulation according to the particle location, e.g., at a free surface, and the sign of its pressure value~\citep{sun2018multi}. With this novel density computation routine, we can easily work with wall discretizations consisting of one wall layer, whereas standard SPH typically requires three or more wall layers~\citep{adami2012generalized}. To complete the discussion on wall boundaries, we use the generalized wall boundary condition approach by~\citet{adami2012generalized} to enforce the impermeability of the walls.

\paragraph{SPH Relaxation parameter tuning}
We propose a three-step parameter-tuning process for the SPH relaxation parameters (see \cref{app:ablations_ldc} for examples):
\begin{enumerate}
    \item Tune $\alpha$ while number of relaxation steps $l=1$ and $\beta=0$. Typically, $\alpha\in(0.005,0.05)$.
    \item Tune $l$ with optimal $\alpha$ and $\beta=0$. Typically, $l\in(1,5)$.
    \item Tune $\beta$ with optimal $\alpha$ and $l$. Typically, $\beta\in(0.1, 1)$.
\end{enumerate}
The measures we use while tuning are the position MSE, Sinkhorn divergence, kinetic energy MSE, MAE of density deviation from the reference $\rho_{ref}$, Dirichlet energy~\citep{zhou2005regularization} of the density field, and Chamfer distance, see \cref{app:ablations} for more details.

\textbf{Related work.}
We want to stress that except for the proposed treatment of external forces, our method does not require retraining the GNN-based simulator. This differentiates our work from an orthogonal line of research, which has experienced a surge in recent years, namely using differentiable solvers as part of the machine learning model \citep{um2020sol}. On the spectrum of classical numerical solvers to black-box end-to-end ML models, one also finds the class of hybrid models, which are ML models utilizing algorithmic ideas from classical solvers \citep{toshev2023relationships,lienen_fen2022,karlbauer2022composing,kochkov2021machine,li2022graph,brandstetter2021message}. Yet, all of these approaches construct a neural network that needs to be trained, whereas our SPH relaxation happens only during inference. 

Conceptually closest to our work is the recent PDE-Refiner model class~\citep{lippe2024pderefiner}. PDE-Refiner draws inspiration from diffusion models to apply a small number of refinement steps on learned Eulerian solvers. The refinement steps substantially improve the modeling of high frequency components, which yields more stable long-term predictions and better physics modeling, at the cost of increased inference time and a dedicated training routine. We point out that because PDE-Refiner is designed for Eulerian systems, it does not have the notion of dynamic particle coordinates underlying Lagrangian methods. Thus, extending PDE-Refiner to the Lagrangian description is not trivial, as one could choose to refine the accelerations or velocities or directly the particle coordinates, and such investigations are beyond the scope of this work. Furthermore, for particle systems, we do not have efficient ways to accurately evaluate high spatial frequencies over point clouds akin to the FFT on grids, and additionally, the physical setup of our problems does not involve high spatial frequencies.

\section{Experiments} \label{sec:experiments}

Our analyses are based on the datasets of \citet{toshev_2023_10021926}, accompanying the LagrangeBench paper~\citep{toshev2024lagrangebench}. These datasets represent challenging coarse-grained temporal dynamics and contain long trajectories, i.e., up to thousands of steps. 
We test the performance difference of two popular GNN-based simulators when: (i) external forces are removed from the model target ($\square_g$), (ii) an SPH relaxation with pressure term is applied ($\square_p$), and (iii) an SPH relaxation with viscous term is applied ($\square_{\nu}$).  

\paragraph{GNN-based simulators}
The Graph Network-based Simulator (GNS) model~\citep{Sanchez:20} is a popular learned surrogate for physical particle-based simulations and our main model. The architecture is kept simple, based on the encoder-processor-decoder principle,
where the processor consists of multiple graph network blocks~\citep{battaglia2018relational}. Our second model, the Steerable E(3)-equivariant Graph Neural Network (SEGNN)~\citep{brandstetter2021geometric} is a general implementation of an E($3$) equivariant GNN, where layers are directly conditioned on steerable attributes for both nodes and edges. The main building block is the steerable MLP, i.e., a stack of learnable linear Clebsch-Gordan tensor products interleaved with gated non-linearities~\citep{weiler20183d}.
SEGNN layers are message-passing layers~\citep{gilmer2017neural} where steerable MLPs replace the traditional non-equivariant MLPs for both message and node update functions. These two models were chosen as they present the current state-of-the-art surrogates for Lagrangian fluid dynamics \citep{toshev2024lagrangebench}, and also because they are representative of two fundamentally different classes of GNNs: non-equivariant (GNS) and equivariant (SEGNN).

\paragraph{Implementation of SPH relaxation}
In our experience, it suffices to perform the relaxation operation for 1-5 iterations ($l$), depending on the problem. We summarize the used hyperparameters in \cref{tab:hyperparams,app:hyperparams}. Given that the learned surrogate is trained on every 100th SPH step, these additional SPH relaxation steps only marginally increase the rollout time -- by a factor of 1.05-1.15 per relaxation step for a 10-layer 128-dimensional GNS model simulating the 2D RPF case, see \cref{tab:timing,app:inference_speed}. In the same table, we observe an increase in runtime for 3D RPF and GNS-10-128 of roughly 1.4x per relaxation step, but we believe that this comes from the much more compute-intense neighbor search, which is reevaluated at every relaxation step. However, as the relaxation does not need to be implemented in a differentiable framework (we currently adopt JAX-SPH~\citep{toshev2024jax}), more efficient implementations, e.g. in C++, can significantly reduce these runtimes. For more compute-intense models like SEGNN the slowdown factor reduces, as the relaxation has a fixed computational cost independent of the particular GNN model. 

Most of the computational overhead of the relaxation is due to its neighbor list, which has significantly more edges than the default neighbor list of the GNN-based simulators. The GNN graph generation uses the default radial cutoff distance from LagrangeBench, which corresponds to roughly $1.5$ average particle distances. In contrast, the SPH relaxation uses the Quintic spline kernel with a cutoff of $3$ average particle distances, i.e., the SPH relaxation operates on $2^{d}$ more edges, with dimension $d\in \{2, 3\}$. Therefore, our approach can be regarded as a multiscale approach, similar to the learned multi-scale interatomic potential presented by~\citep{fu2023learning}. The difference is that in our approach, only the part using the smaller cutoff is a neural network, and the longer-range interactions simply stabilize the system in terms of better density distributions.

\paragraph{Training with SPH relaxation}
An appealing idea is to use the SPH relaxation as a regularization during training, in the hope that we can omit running relaxations at inference time. We tried various ways of implementing this idea, but none of them improved rollout performance, see \cref{app:relaxed_training}.

\paragraph{Overview of results}
Our results on 400-step rollouts using the GNS model are summarized in \cref{tab:results} and are averaged over all test trajectories and over the trajectory length. See \cref{tab:segnn_results} for the SEGNN results. As error measures, we use (a) the mean-squared error of positions (MSE$_{400}$), (b) the Sinkhorn divergence, which quantifies the conservation of the particle distribution, and (c) the kinetic energy error (MSE$_{Ekin}$) as a global measure of the physical behavior. The viscous term is shown only for reverse Poiseuille flow because it did not improve the performance on the other datasets. We note that by splitting the test sets into sequences of length 400, we obtain only 12-25 test trajectories, leading to noisy performance estimates. We discuss the necessity for larger datasets later in this section.
For various parameter ablations, the evolution of error metrics with error bounds, and three more error metrics (density MAE, Dirichlet energy, and Chamfer distance), see \cref{app:ablations}.
Overall, all Neural SPH-enhanced simulators achieve better performance than the baseline GNNs, often by orders of magnitude, allowing for significantly longer rollouts and significantly better physics modeling.

\paragraph{Note on error thresholds}
We note that upon tuning the parameters of our method, it either improves performance or converges to the baseline, with the latter being what mainly happens to RPF 3D according to \cref{app:hyperparams}. We hypothesize that the baseline already produces very good particle distributions, and there is little potential for improvement. 
It thus seems necessary to define a threshold of when a learned simulator performs \textit{well enough} in the sense of the requirements of the downstream task of interest. We refer to physical thresholds like the \textit{chemical accuracy} in computational chemistry or the \textit{energy and forces within threshold} measure used in the Open Catalyst project~\citep{chanussot2021open}, both of which are designed to quantify whether a computational model is useful for practical applications. We stress the importance and leave the derivations of such thresholds for Lagrangian fluid simulations to future work.

\begin{table}[ht]
\centering
\footnotesize
\begin{tabular}{cllll}
                        & Model      & $\text{MSE}_{400}$ & Sinkhorn & $\text{MSE}_{\text{Ekin}}$  \\
\hline 
\multirow{2}{*}{\parbox{8mm}{\centering 2D TGV}} & GNS     & \cellcolor{step3}$5.3e-4$ & \cellcolor{step3}$5.4e-7$ & \cellcolor{step3}$5.6e-7$ \\
                                                 & GNS$_p$ & \cellcolor{step1}$4.8e-4$ & \cellcolor{ref}$1.7e-8$ & \cellcolor{step1}$4.8e-7$ \\ 
\hline
\multirow{4}{*}{\parbox{8mm}{\centering 2D RPF}} & GNS     & \cellcolor{step3}$2.7e-2$ & \cellcolor{step3}$3.6e-7$  & \cellcolor{step3}$4.3e-3$  \\ 
                                                 & GNS$_g$ & \cellcolor{step3}$2.7e-2$ & \cellcolor{step2}$2.7e-7$ & \cellcolor{step1}$3.7e-4$  \\
                                                 & GNS$_{g,p}$ & \cellcolor{step3}$2.7e-2$ & \cellcolor{ref}$2.9e-8$ & \cellcolor{step1_5}$4.1e-4$  \\ 
                                                 & GNS$_{g,p,\nu}$ & \cellcolor{step3}$2.7e-2$ & \cellcolor{ref_5}$3.0e-8$ & \cellcolor{ref}$1.4e-4$ \\ 
\hline
\multirow{2}{*}{\parbox{8mm}{\centering 2D LDC}} & GNS     & \cellcolor{step3}$3.3e-2$ & \cellcolor{step3}$3.1e-4$ & \cellcolor{step3}$1.1e-4$ \\
                                                 & GNS$_p$ & \cellcolor{step1}$1.6e-2$ & \cellcolor{step-2}$2.8e-7$ & \cellcolor{step-1}$1.2e-6$ \\ 
\hline
\multirow{4}{*}{\parbox{8mm}{\centering 2D DAM}} & GNS     & \cellcolor{step3}$1.9e-1$ & \cellcolor{step3}$3.8e-2$ & \cellcolor{step3}$4.6e-2$ \\
                                                 & GNS$_g$ & \cellcolor{ref}$8.0e-2$ & \cellcolor{step1_8}$1.3e-2$ & \cellcolor{step2}$9.4e-3$ \\
                                                 & GNS$_p$ & \cellcolor{step1_5}$9.7e-2$ & \cellcolor{ref}$7.1e-3$ & \cellcolor{step1_75}$5.8e-3$ \\
                                                 & GNS$_{g,p}$ & \cellcolor{ref_5}$8.4e-2$ & \cellcolor{ref_5}$7.5e-3$ & \cellcolor{ref}$2.1e-3$ \\ 
\hline
\multirow{2}{*}{\parbox{8mm}{\centering 3D TGV}} & GNS     & \cellcolor{step3}$4.8e-2$ & \cellcolor{step3}$4.1e-6$ & \cellcolor{step3}$3.6e-2$ \\
                                                 & GNS$_p$ & \cellcolor{step1_8}$4.6e-2$ & \cellcolor{ref}$9.0e-7$ & \cellcolor{step1}$4.2e-2$ \\ 
\hline
\multirow{3}{*}{\parbox{8mm}{\centering 3D RPF}} & GNS     & \cellcolor{step3}$2.3e-2$ & \cellcolor{step3}$4.4e-7$ & \cellcolor{step3}$1.7e-5$ \\
                                                 & GNS$_g$ & \cellcolor{step3}$2.3e-2$ & \cellcolor{step3}$4.4e-7$ & \cellcolor{step4}$4.1e-5$ \\
                                                 & GNS$_p$ & \cellcolor{step3}$2.3e-2$ & \cellcolor{step1}$1.0e-7$ & \cellcolor{step2}$1.5e-5$ \\ 
                                                 & GNS$_{g,p}$ & \cellcolor{step3}$2.3e-2$ & \cellcolor{step1_75}$1.3e-7$ & \cellcolor{step4}$4.1e-5$ \\
\hline
\multirow{2}{*}{\parbox{8mm}{\centering 3D LDC}} & GNS     & \cellcolor{step3}$3.2e-2$ & \cellcolor{step3}$2.0e-5$ & \cellcolor{step3}$1.3e-7$ \\
                                                 & GNS$_p$ & \cellcolor{step3}$3.2e-2$ & \cellcolor{ref}$1.1e-6$ & \cellcolor{ref}$2.9e-8$ \\ 
\end{tabular}
\caption{Performance measures averaged over a rollout of 400-steps. An additional subscript $g$ indicates that external forces are removed from the model outputs, subscript $p$ indicates that the SPH relaxation has a pressure term, and subscript ${\nu}$ that the viscosity term is added to the SPH relaxation. The numbers in the table are averaged over all test trajectories. $\text{MSE}_{400}$ corresponds to: $\text{MSE}_{120}$ for 2D TGV, $\text{MSE}_{55}$ for 3D TGV, and $\text{MSE}_{395}$ for 2D DAM, as these are the full trajectory lengths excluding initial history size $H=5$. }
\label{tab:results}
\end{table}

\subsection{External Force Treatment}
In this section, we study the influence of the proposed external force treatment without combining it with the SPH relaxation. As only the dam break and reverse Poiseuille flow datasets have external force features, we focus on them. 

\paragraph{Dam break (DAM)}
We saw a major performance boost on dam break when removing external forces from the target ($\text{GNS}_g$), see \cref{tab:results,app:ablations_dam}. This simple modification of the training objective improves all considered measures by at least a factor of 2 and by as much as a factor of 5 on a rollout of the full dam break trajectory, i.e., 400 steps. 
For up to 20-step rollouts, $\text{GNS}_g$ training does not improve the position error, which is in accordance with~\citet{Sanchez:20} and their Fig. C1. However, as the simulation length goes beyond 50 steps, numerical errors quickly accumulate and lead to artifacts like the one visible in the top part of \cref{fig:dam_step80}. This particular failure mode in the front part of the dam break wave develops by first compressing the fluid to as much as $1.5\rho_{ref}$, and then the smallest instability in the tip causes particles to detach from the free surface. From there on, GNS starts acting as if the right wall has already been reached and fails to model the double wave structure from the reference solution, see \cref{app:dam_plots}.

\paragraph{Force smoothing in reverse Poiseuille Flow}
The external force of the reverse Poiseuille flow datasets is provided as a function corresponding to the instantaneous force, but when we train towards the effective dynamics over multiple original solver steps, we need to adjust this force. In particular, when predicting the dynamics over $M=100$ temporal coarse-graining steps provided by LagrangeBench, an RPF particle might jump back and forth across the boundary separating the left- and right-ward forcing. Thus, it is not possible to infer the aggregated external force directly only knowing the particle coordinates at step $M$. We, therefore, apply a convolution of a Gaussian function with the force function. Since the forcing in RPF is a step function, this specific convolution has an analytical solution, i.e., the error function $\text{erf}( \cdot )$. We use $\text{erf}( \cdot )$ as a replacement for the original force function. See \cref{app:erf_appendix} for more details and visualization of the force before and after the convolution.

\begin{figure}[ht]
    \centering
    \includegraphics[trim={12mm 113mm -2mm 7mm},clip,width=1.03\linewidth]{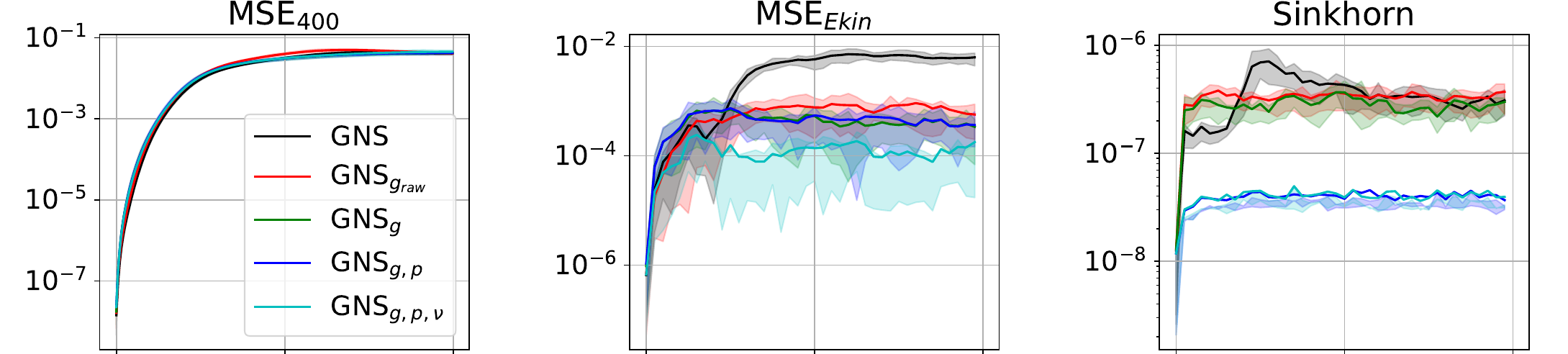} \vspace{-6mm} \\
    \includegraphics[trim={12mm 14mm -2mm 182mm},clip,width=1.03\linewidth]{figs/rpf2d_gns_ext.pdf} \vspace{-6mm} \\
    {\tiny \hspace{5mm} Step \hspace{24mm} Step \hspace{24mm} Step}
    \caption{Ablations on RPF 2D with GNS-10-128 over the simulation length. Adapted from \cref{fig:rpf2d_gns_ext} in \cref{app:ablations_rpf}. \label{fig:rpf2d_gns_ext_upperhalf}}
\end{figure}

\paragraph{Reverse Poiseuille flow (RPF)} 
See \cref{fig:rpf2d_gns_ext_upperhalf} for a subset of our ablation results on RPF 2D with GNS-10-128, or the full results on RPF 2D/3D and GNS/SEGNN in \cref{app:ablations_rpf}. When removing external forces from the target of the GNS model ($\text{GNS}_g$), we observed that using the original, i.e., not smoothed, force leads to highly unstable dynamics in the shearing region, which causes the failure of the dynamics after less than 50 steps, see GNS$_{g_{raw}}$ in \cref{fig:rpf2d_segnn_ext,fig:rpf3d_gns_ext}. When switching to the smoothed force function, the system becomes much more stable to perturbations and significantly improves the kinetic energy error. It is important to note that the kinetic energy is paramount to RPF, as this physical system is characterized by constant kinetic energy up to small fluctuations. \\
Looking at the 20-step position MSE reported in LagrangeBench, the $\text{GNS}_g$ training leads to worse performance, roughly by a factor of 1.5 (see the beginning of the evolution in \cref{fig:rpf2d_gns_ext_upperhalf}). This is important to note because we trade off worse short-term behavior in favor of better long-rollout performance, with the latter being the practical use-case we target. In this context, the LagrangeBench datasets pre-define a split of 50/25/25, which is far from sufficient if we want stable error estimates on rollouts of 400-step length,
as also discussed, e.g., in~\citet{fu2023forces}.

\subsection{SPH Relaxation}
This section presents the results of our SPH relaxation on its own, and also in combination with the proposed external force treatment. We divide the discussion based on common characteristics of the datasets into periodic boundary cases, cases with wall boundaries, and free surface problems.

\subsubsection{Periodic boundaries}

\paragraph{Taylor-Green vortex (TGV)}
We did not expect the SPH relaxation to be very beneficial to the Taylor-Green vortex cases because (a) the trajectories are rather short with 125 and 60 steps in the 2D and 3D cases, respectively, and also (b) TGV represents a decaying problem, making it less prone to clustering in later stages of the trajectory. But according to \cref{tab:results}, we get a consistent improvement of the position error MSE$_{400}$ of $\sim5\%$ and significant Sinkhorn divergence improvements on the 2D and 3D datasets.

\paragraph{Viscous term}
In addition to external force subtraction, we found it beneficial to use the pressure ($p$) and viscous (${\nu}$) terms during relaxation, termed $\text{GNS}_{p,\nu}$.
Viscosity, which manifests itself in shearing forces, in general, refers to the idea that if two fluid elements are close to each other but move in opposite directions, then they should both decelerate. Thus, to apply viscosity, we need to again approximate velocities by finite differences between consecutive positions of particles.

\begin{figure}[ht]
    \centering
    \includegraphics[trim={0cm 5mm 0mm 0mm},clip,width=0.9\linewidth]{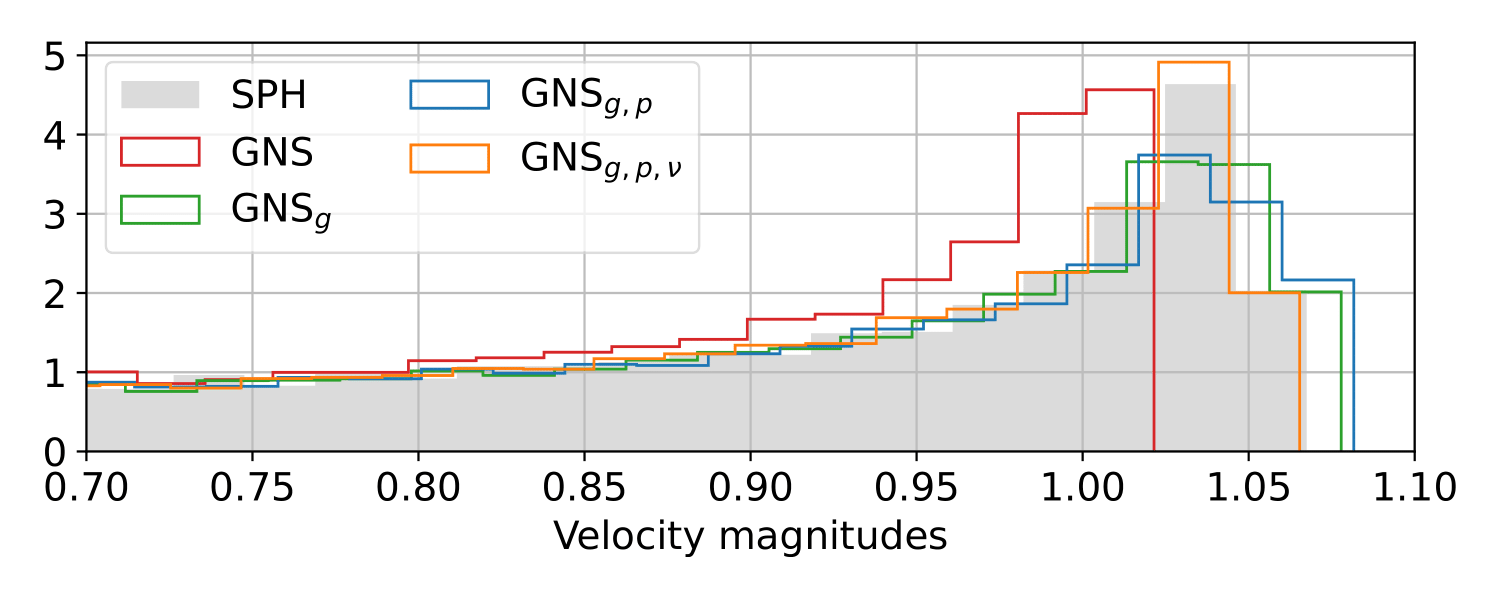}
    \caption{Velocity magnitudes histogram of 2D reverse Poiseuille flow after 400 rollout steps (averaged over all rollouts). Our $\text{GNS}_{g,p,\nu}$ matches the ground truth distribution of SPH.}  
    \label{fig:rpf_hist_min}
\end{figure}

\paragraph{Reverse Poiseuille flow (RPF)}
In \cref{fig:rpf_hist_min,fig:rpf_hist}, we show histograms over velocity magnitudes to quantify how the different RPF correction terms impact the dynamics. Firstly, the original GNS model loses its high-velocity components over time, resembling a diffusion process, which makes it more stable with respect to perturbations, but, at the same time, leads to wrong kinetic energy. Secondly, simply changing the training objective by removing the external force (see GNS$_g$) already mitigates the problem of missing high velocities. And by adding the viscous term, which is especially relevant in the shearing region, to the pressure gradient term, we almost perfectly recover the target velocity distribution. See \cref{fig:rpf2d_gns_ext_upperhalf,app:ablations_rpf} for further details.

\subsubsection{Wall boundaries}

\begin{figure}[ht]
    \centering
      \begin{sideways}
        \begin{minipage}{0.12\textheight}
          \centering
          Density
        \end{minipage}
      \end{sideways}
    \includegraphics[trim={0 0cm 0 0},clip,width=0.31\linewidth]{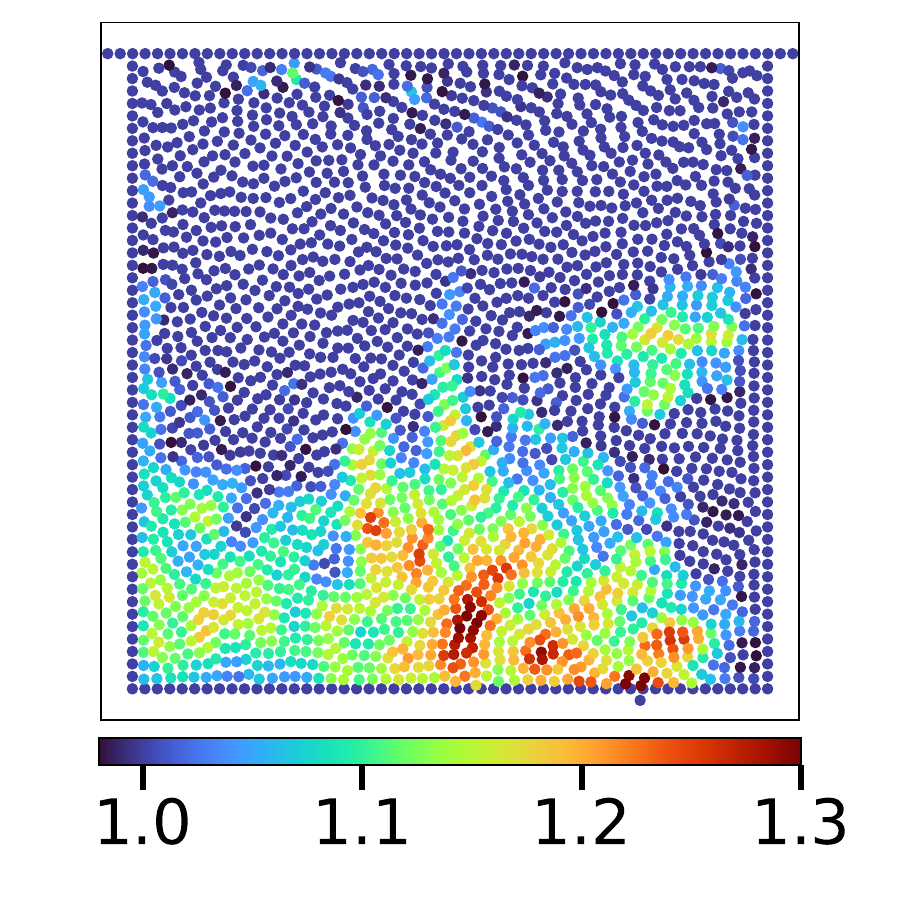}
    \includegraphics[trim={0 0cm 0 0},clip,width=0.31\linewidth]{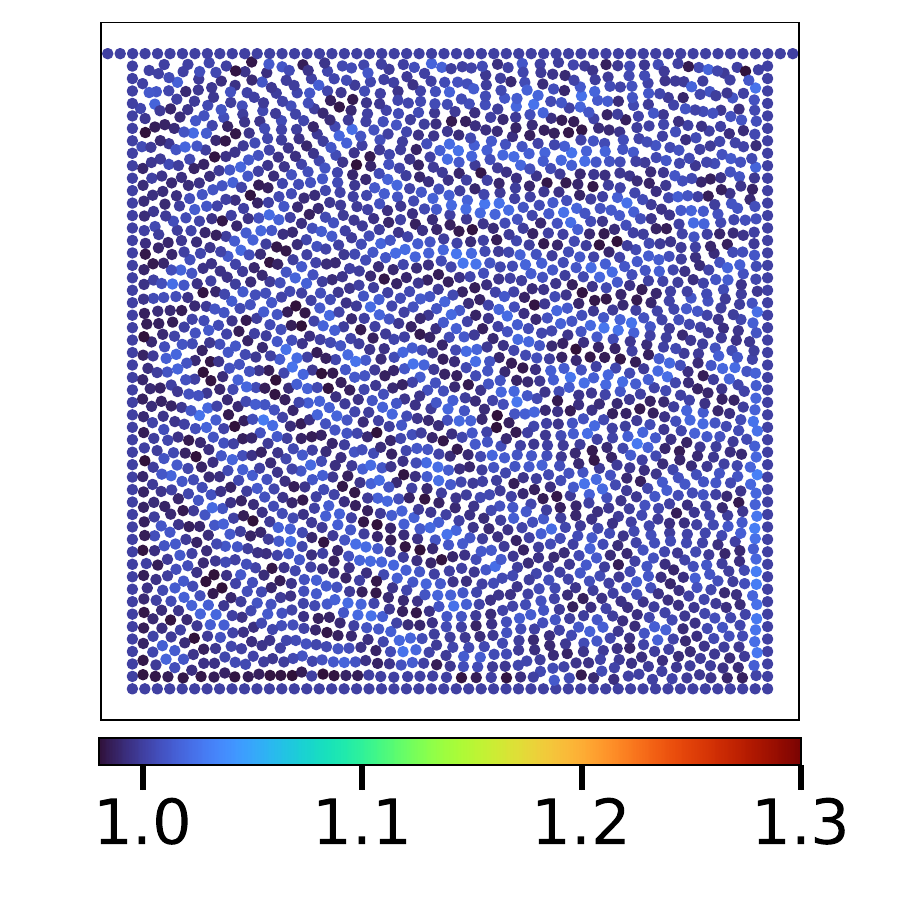}
    \includegraphics[trim={0 0cm 0 0},clip,width=0.31\linewidth]{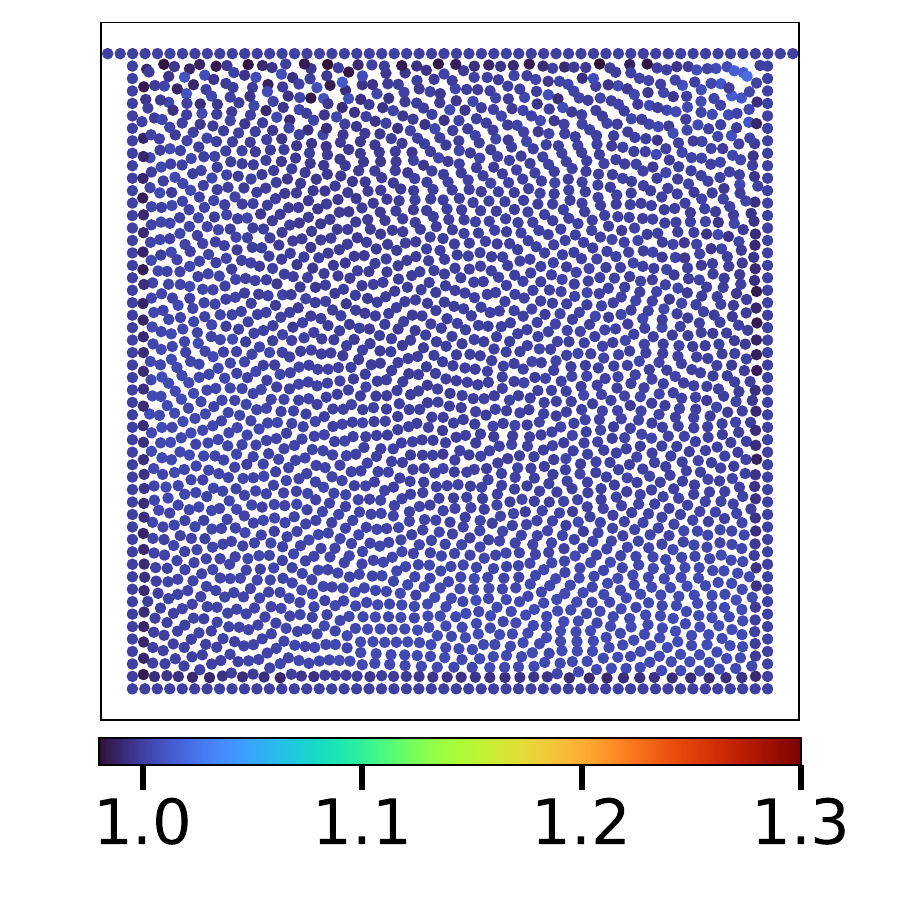}

    \centering
      \begin{sideways}
        \begin{minipage}{0.11\textheight}
          \centering
          Vel. magnitude
        \end{minipage}
      \end{sideways}
    \includegraphics[trim={0 5mm 0 0},clip,width=0.31\linewidth]{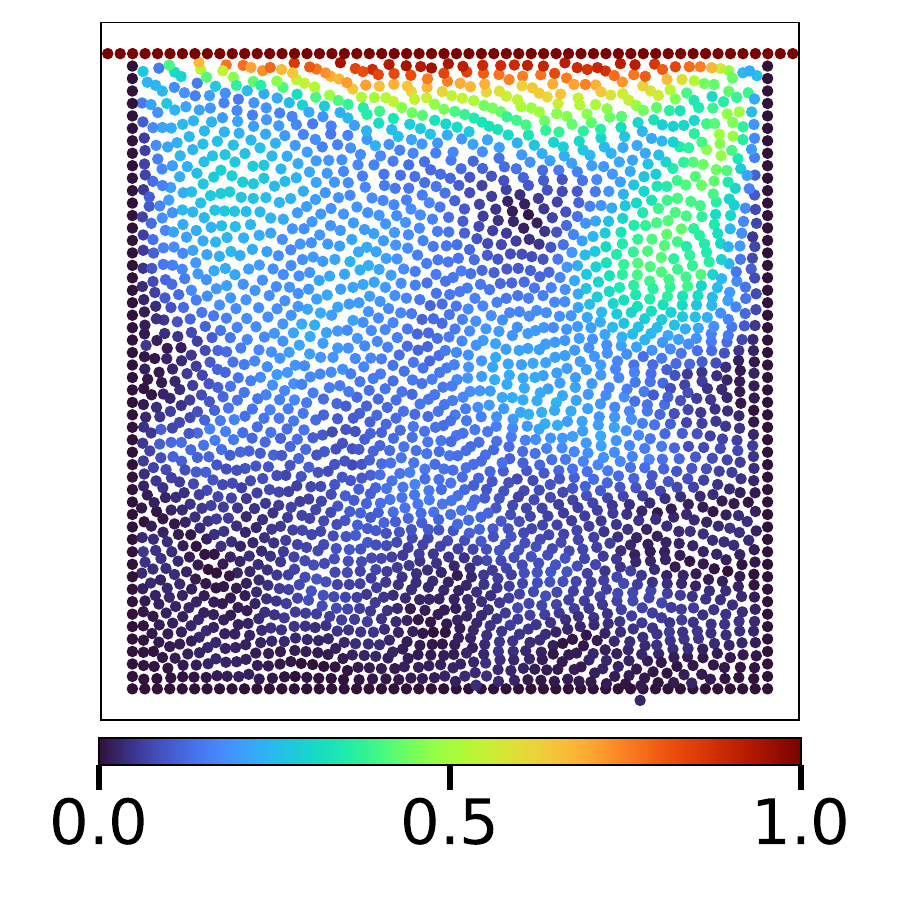}
    \includegraphics[trim={0 5mm 0 0},clip,width=0.31\linewidth]{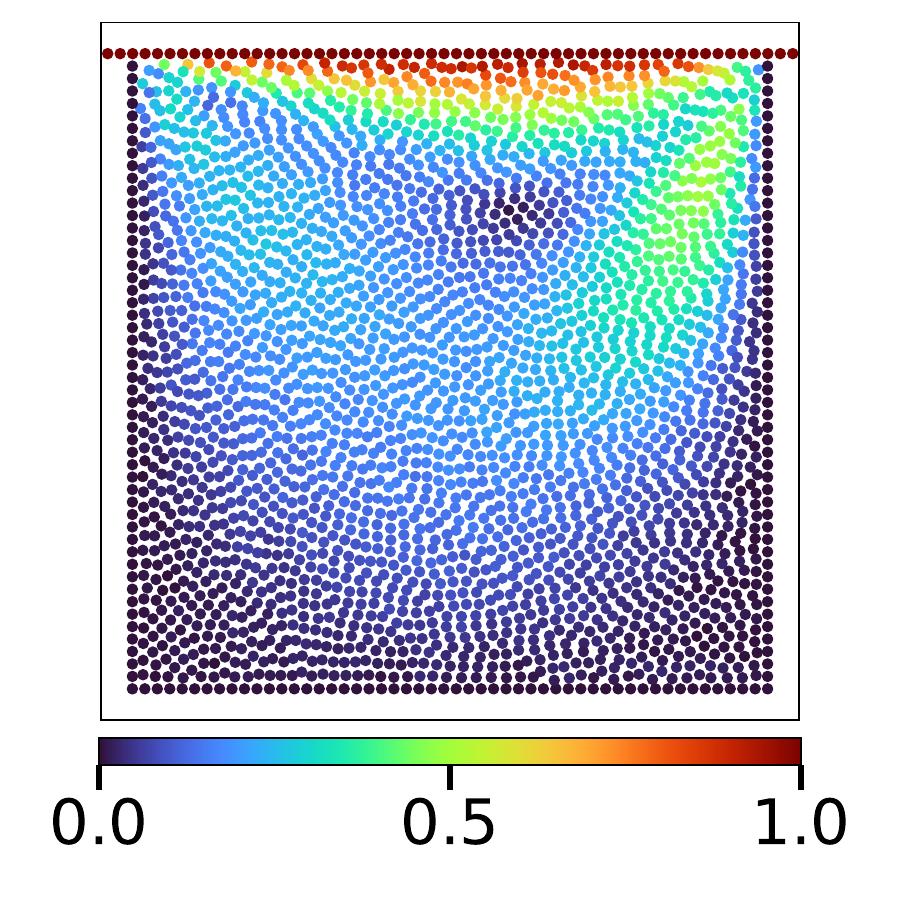}
    \includegraphics[trim={0 5mm 0 0},clip,width=0.31\linewidth]{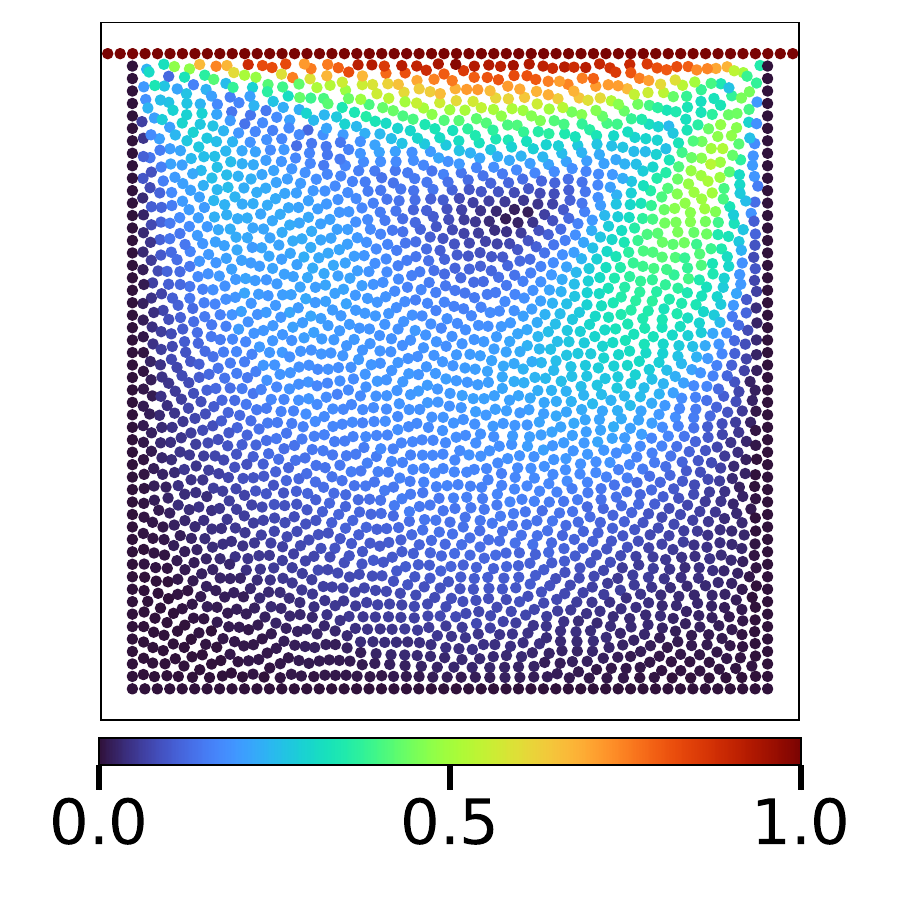}

    \hspace{1.2cm} GNS \hspace{1.7cm} GNS$_p$ \hspace{1.7cm} SPH \hspace{0.7cm}
    \caption{Density and velocity magnitude of 2D lid-driven cavity after 400 rollout steps (left to right): GNS, GNS$_p$, SPH. The colors in the first row correspond to the density deviation from the reference density; the system is considered physical within 0.98-1.02.
    \label{fig:ldc_step400}}
\end{figure}

A typical failure mode of learned solvers is that one or multiple particles penetrate what should be a solid wall, see top left part of \cref{fig:ldc_step400} for LDC 2D and top part of \cref{fig:dam_traj14} in \cref{app:dam_plots} for DAM 2D. We solve this problem nearly completely with our SPH relaxations.

\paragraph{Relaxation at wall boundaries}
The only part we have not discussed yet is how to ensure that particles do not escape the computational domain by passing through the walls. We use the simple and effective approach laid out in the generalized wall boundary condition paper by~\citet{adami2012generalized}. The idea of this approach is to enforce the impermeability of the walls by setting the pressure of the dummy wall particles to the average pressure of their adjacent fluid neighbors, see Eq.~(27) in~\citet{adami2012generalized}, and, thus, constructing a setting of zero pressure gradients normal to the walls. With this boundary condition implementation, we obtain the following one-step relaxation algorithm: 1. density computation for fluid particles, 2. pressure computation for fluid particles through the equation of state, 3. computation of pressure of wall particles via weighted summation over the pressure of adjacent fluid particles, and 4. evaluation of the pressure gradient term, which gives the forces used to integrate the momentum equation \cref{eq:nse_momentum_learned} through \cref{eq:nse_integration_learned}.

\paragraph{Lid-driven cavity (LDC)}
In the lid-driven cavity example, we observe that the learned model pushes particles away from the fast-moving lid into the lower half of the domain, which has profound consequences. On the one hand, the pressure at the bottom increases to an extent such that one or more particles gradually pass through the bottom wall. On the other hand, since too few particles reside close to the lid, the shearing forces are underrepresented, yielding a loss of kinetic energy, i.e., dynamics are lost. We fix both these issues with an SPH relaxation, forcing particles to be homogeneously distributed within the domain, see \cref{fig:ldc_step400,fig:dam_traj14}. See \cref{app:ablations_ldc} for various hyperparameter sensitivity ablations on LDC 2D/3D and GNS/SEGNN. While tuning the parameters is crucial, once tuned, they seem to work fairly reliably. 

\subsubsection{Free surfaces}
A major difference between dam break and the other datasets we benchmark is that in dam break we not only care about the particle distribution within the fluid, but also about the volume filled with fluid. The latter is the focus of this section, and it is reflected in the MSE$_{400}$ and Sinkhorn divergence measures, but not in MSE$_{Ekin}$. 

\paragraph{Dam break (DAM)}
Interestingly, by either our external force treatment or the SPH relaxation, we seem to fix the problem of the fan-like spreading of the wavefront. We interpret this as a confirmation that the reason for this failure mode is the high compression at the tip.
However, fixing the high compression levels in the bulk fluid requires our SPH relaxation, which we run with as few as three steps. The $\text{GNS}_{g,p}$ setup then recovers the correct dynamics with a significantly higher precision as measured by the Sinkhorn divergence, but also the kinetic energy MSE, indicating that the fluid also evolves more physically. Regarding the fluid surface, if we carefully look at the height of the fluid in \cref{fig:dam_traj0,fig:dam_traj13,fig:dam_traj14,fig:dam_traj15}, we see that the $\text{GNS}_{g,p}$ case very closely resembles the ground truth. See \cref{app:ablations_dam} for ablations. 

\subsection{SEGNN Results}
We applied the same external force treatment and SPH relaxations to the SEGNN model~\citep{brandstetter2021geometric} without further tuning of the Neural SPH hyperparameters (see \cref{app:hyperparams}) and summarize the results in \cref{tab:segnn_results}. This comparison is useful not only for better comparability but also to show that proper SPH relaxation often depends more on the dataset than on the model -- for example, moving the external force out of the 2D RPF case results in a 40 times lower kinetic energy error.
However, in some cases, the GNS and SEGNN models behave quite differently. In most cases, SEGNN performs on par with GNS on long trajectories, with the notable SEGNN blowups on LDC 2D, DAM 2D, and RPF 3D. In particular, when we change the treatment of the external force in dam break without applying additional wall boundary conditions, we observe many particles falling through the bottom wall around step 200. Adding the relaxation with wall boundary conditions solves this problem, but investigating the qualitative differences between GNS and SEGNN would be an interesting future work. See \cref{app:ablations} for our hyperparameter ablations.

\begin{table}[ht]
\centering
\footnotesize
\begin{tabular}{cllll}
                        & Model      & $\text{MSE}_{400}$ & Sinkhorn & $\text{MSE}_{\text{Ekin}}$  \\
\hline 
\multirow{2}{*}{\parbox{8mm}{\centering 2D TGV}} & SEGNN     & \cellcolor{step3}$4.0e-4$ & \cellcolor{step3}$4.4e-7$ & \cellcolor{step3}$3.9e-7$ \\
                                                 & SEGNN$_p$ & \cellcolor{step2}$3.8e-4$ & \cellcolor{ref}$1.5e-8$ & \cellcolor{step1}$2.8e-7$ \\ 
\hline
\multirow{4}{*}{\parbox{8mm}{\centering 2D RPF}} & SEGNN     & \cellcolor{step3}$2.7e-2$ & \cellcolor{step3}$3.3e-7$  & \cellcolor{step3}$4.3e-3$  \\ 
                                                 & SEGNN$_g$ & \cellcolor{step3_5}$2.8e-2$ & \cellcolor{step3}$3.3e-7$ & \cellcolor{ref}$1.2e-4$  \\
                                                 & SEGNN$_{g,p}$ & \cellcolor{step3_5}$2.8e-2$ & \cellcolor{ref}$3.5e-8$ & \cellcolor{ref_5}$1.6e-4$  \\ 
                                                 & SEGNN$_{g,p,\nu}$ & \cellcolor{step3_5}$2.8e-2$ & \cellcolor{ref_5}$3.8e-8$ & \cellcolor{step1_8}$7.3e-4$ \\ 
\hline
\multirow{2}{*}{\parbox{8mm}{\centering 2D LDC}} & SEGNN     & \cellcolor{step3}$7.6e-2$ & \cellcolor{step3}$2.3e-3$ & \cellcolor{step3}$9.1e+0$ \\
                                                 & SEGNN$_p$ & \cellcolor{step1_8}$1.8e-2$ & \cellcolor{step-2}$5.8e-7$ & \cellcolor{step-2}$1.6e-5$ \\ 
\hline
\multirow{4}{*}{\parbox{8mm}{\centering 2D DAM}} & SEGNN     & \cellcolor{step3}$1.5e-1$ & \cellcolor{step3}$3.4e-2$ & \cellcolor{step3}$1.9e-2$ \\
                                                 & SEGNN$_g$ & \cellcolor{step3_5}$1.6e-1$ & \cellcolor{step2}$2.1e-2$ & \cellcolor{step5}$1.9e+1$ \\
                                                 & SEGNN$_p$ & \cellcolor{step2_5}$1.2e-1$ & \cellcolor{step1_75}$9.4e-3$ & \cellcolor{step2_5}$1.2e-2$ \\
                                                 & SEGNN$_{g,p}$ & \cellcolor{step1}$8.6e-2$ & \cellcolor{ref}$4.9e-3$ & \cellcolor{ref_5}$2.6e-3$ \\ 
\hline
\multirow{2}{*}{\parbox{8mm}{\centering 3D TGV}} & SEGNN     & \cellcolor{step3}$4.2e-2$ & \cellcolor{step3}$6.1e-6$ & \cellcolor{step3}$2.4e-2$ \\
                                                 & SEGNN$_p$ & \cellcolor{step2_5}$4.1e-2$ & \cellcolor{step1}$6.0e-7$ & \cellcolor{step3_5}$2.7e-2$ \\ 
\hline
\multirow{4}{*}{\parbox{8mm}{\centering 3D RPF}} & SEGNN     & \cellcolor{step3}$1.2e-1$ & \cellcolor{step3}$1.0e-4$ & \cellcolor{step3}$1.5e+3$ \\
                                                 & SEGNN$_p$ & \cellcolor{ref_5}$2.6e-2$ & \cellcolor{ref}$1.3e-5$ & \cellcolor{ref-0_5}$1.8e-2$ \\ 
                                                 & SEGNN$_g$ & \cellcolor{step1_5}$2.7e-2$ & \cellcolor{step-1}$2.6e-6$ & \cellcolor{step-1_9}$9.5e-3$ \\
                                                 & SEGNN$_{g,p}$ & \cellcolor{ref_5}$2.6e-2$ & \cellcolor{step-2}$7.9e-7$ & \cellcolor{step-2}$5.7e-3$ \\ 
\hline
\multirow{2}{*}{\parbox{8mm}{\centering 3D LDC}} & SEGNN     & \cellcolor{step3}$3.3e-2$ & \cellcolor{step3}$2.3e-5$ & \cellcolor{step3}$1.7e-7$ \\
                                                 & SEGNN$_p$ & \cellcolor{step3}$3.3e-2$ & \cellcolor{ref}$2.0e-6$ & \cellcolor{step3_5}$1.8e-7$ \\ 
\end{tabular}
\caption{SEGNN-10-64 results. Same structure as \cref{tab:results}.}
\label{tab:segnn_results}
\end{table}

\section{Concluding Remarks}
We introduce Neural SPH, a framework for improved training and inference of GNN-based simulators for Lagrangian fluid dynamics simulations. We demonstrate the utility of our toolkit on seven diverse 2D and 3D datasets and on two state-of-the-art GNN-based simulators, GNS and SEGNN. 
We identify particle clustering originating from tensile instabilities as one of the primary pitfalls of GNN-based simulators.
Through the proposed external force treatment and SPH relaxation step, distribution-induced errors are minimized, leading to more robust and physically consistent dynamics.
Compared to other methods, Neural SPH does not require a differentiable solver and increases the inference time only by a fixed and rather small amount. 

\paragraph{Limitations and future work}
We observe that tuning the hyperparameters of the particle relaxation is crucial since redistributing the particles inherently translates to modified velocity histories, which directly enter the next autoregressive update step. Thus, the learned solver may become unstable by bringing the past velocities out-of-distribution. Although using the proposed hyperparameter tuning recipe leads to a fairly stable inference routine of the learned solvers, further improving this recipe might be beneficial. 
Another potential limitation concerns the handling of external forces, namely, that information on the timestep and coarsening level of the dataset is required.
Finally, and related to the parameter tuning, we point out the necessity of defining physical thresholds akin to the \textit{energy and force within threshold} by~\citep{chanussot2021open}, to identify whether our Neural SPH improvements are needed in the first place. Our work shows what is possible by integrating machine learning models with established simulation routines like enforcing boundary conditions or improving particle spreading, but one can extend this idea by adding arbitrarily many terms from the enormous body of literature on classical numerics. We point out that the proposed alternation of learned and classical solver terms is a framework, applicable to any combination of compatible methods, extending beyond GNNs and Lagrangian systems.

\clearpage
\section*{Impact Statement}
Smoothed particle hydrodynamics plays a crucial role in computational fluid dynamics. Examples can be found in aerodynamics, astrophysics, or plasma physics. Given the widespread application of computational fluid dynamics, obtaining shortcuts or alternatives for computationally expensive simulations is essential for advancing scientific research, and has direct or indirect implications for reducing our carbon footprint.
However, it is important to note that relying on simulations always necessitates thorough cross-checks and monitoring, especially when employing a "learning to simulate" methodology.

\section*{Acknowledgements}
The authors thank Fabian Thiery, Christopher Zöller, and Steffen Schmidt for helpful discussions on SPH at free surfaces.

\section*{Author Contributions}
A.T. conceived the ideas of SPH relaxation and the proposed external force treatment, implemented them, ran the experiments, and wrote the first version of the manuscript. J.E. contributed the Dirichlet energy metric and wrote the literature review on density summation at free surfaces. N.A. and J.B. supervised the project from conception to design of experiments and analysis of the results. All authors contributed to the manuscript.

\bibliography{literature}
\bibliographystyle{icml2024}


\newpage
\appendix
\onecolumn

\section{Dam Break Plots} \label{app:dam_plots}

\begin{figure}[ht]
    \centering
      \begin{sideways}
        \begin{minipage}{0.07\textheight}
          \centering
          GNS$_{\phantom{g}}$
        \end{minipage}
      \end{sideways}
    \includegraphics[trim={0 0cm 0 0},clip,width=0.48\linewidth]{figs/dam2d_80_rho_gns.pdf}
    \includegraphics[trim={0 0cm 0 0},clip,width=0.48\linewidth]{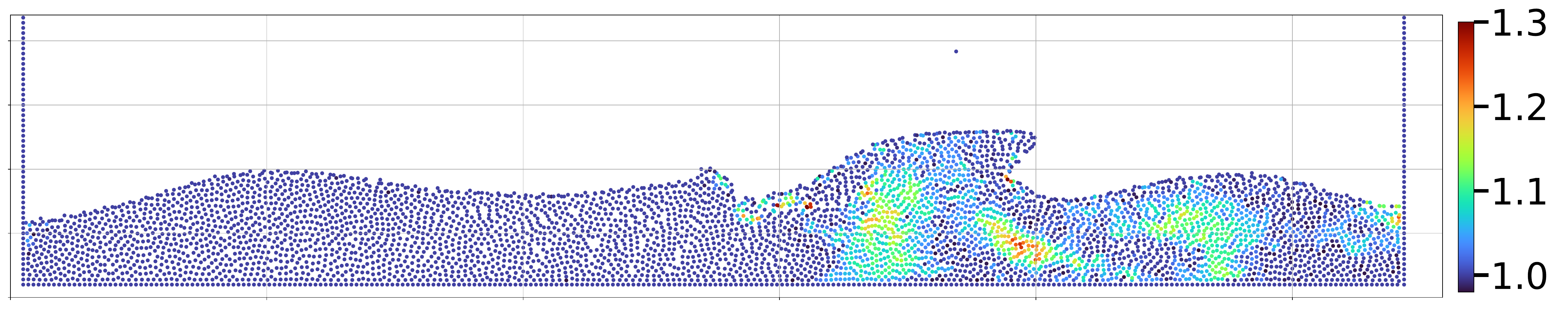}
    
    \centering
      \begin{sideways}
        \begin{minipage}{0.07\textheight}
          \centering 
          GNS$_g$
        \end{minipage}
      \end{sideways}
    \includegraphics[trim={0 0cm 0 0},clip,width=0.48\linewidth]{figs/dam2d_80_rho_sphgnn-g.pdf}
    \includegraphics[trim={0 0cm 0 0},clip,width=0.48\linewidth]{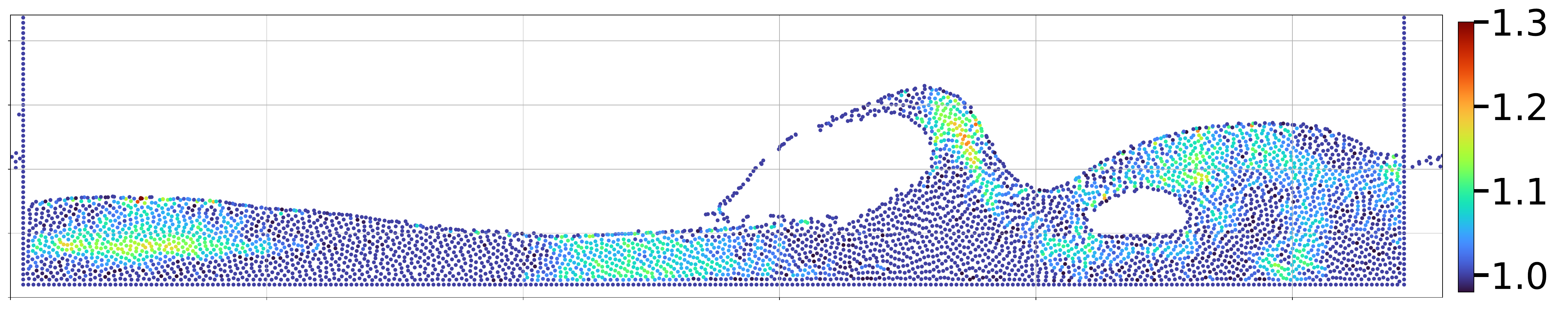}

    \centering
      \begin{sideways}
        \begin{minipage}{0.07\textheight}
          \centering
          GNS$_{g,p}$
        \end{minipage}
      \end{sideways}
    \includegraphics[trim={0 0cm 0 0},clip,width=0.48\linewidth]{figs/dam2d_80_rho_sphgnn-3.pdf}
    \includegraphics[trim={0 0cm 0 0},clip,width=0.48\linewidth]{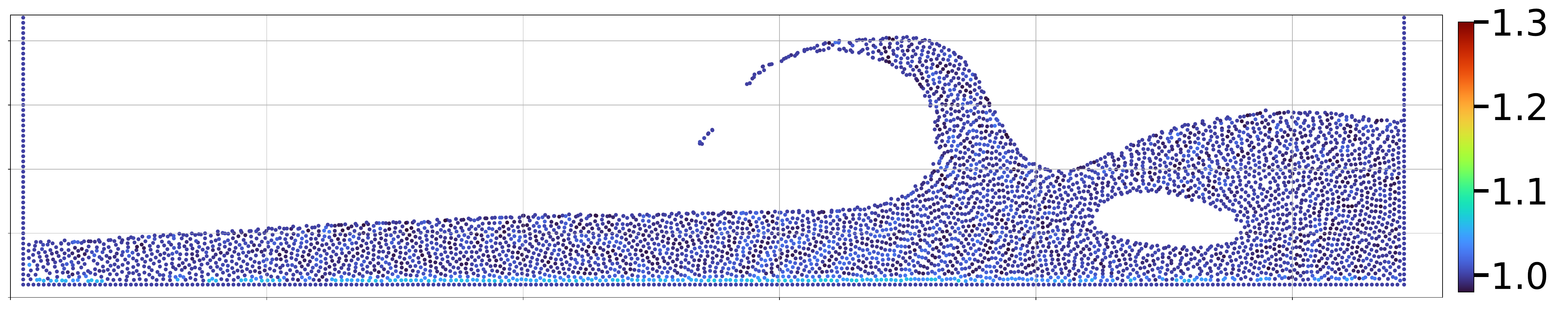}

    \centering
      \begin{sideways}
        \begin{minipage}{0.07\textheight}
          \centering
          SPH$_{\phantom{g}}$
        \end{minipage}
      \end{sideways}
    \includegraphics[trim={0 0cm 0 0},clip,width=0.48\linewidth]{figs/dam2d_80_rho_ref.pdf}
    \includegraphics[trim={0 0cm 0 0},clip,width=0.48\linewidth]{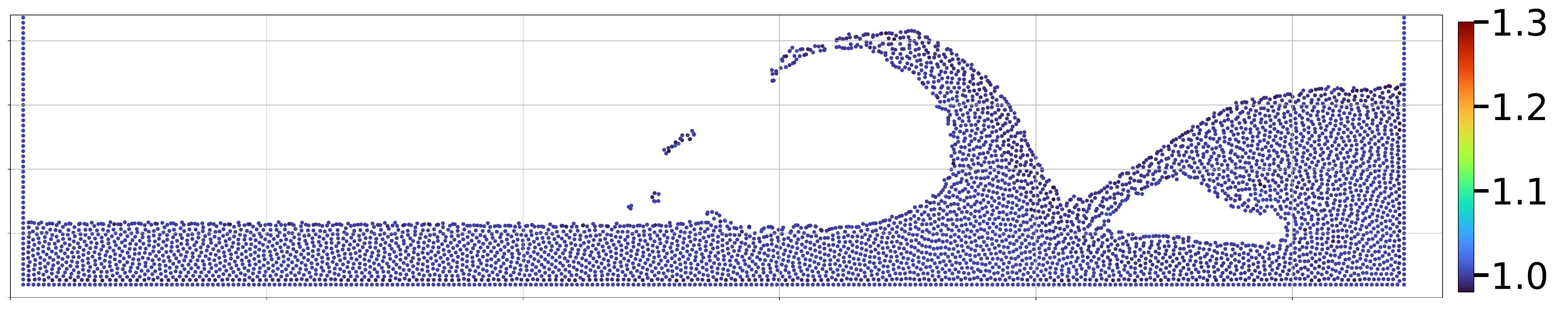}

    Step 80 \hspace{7cm} Step 240

    \caption{Dam break steps 80 and 240 of test rollout 0. Extends \cref{fig:dam_step80}.}  
    \label{fig:dam_traj0}
\end{figure}

\begin{figure}[ht]
    \centering
      \begin{sideways}
        \begin{minipage}{0.07\textheight}
          \centering
          GNS$_{\phantom{g}}$
        \end{minipage}
      \end{sideways}
    \includegraphics[trim={0 0cm 0 0},clip,width=0.48\linewidth]{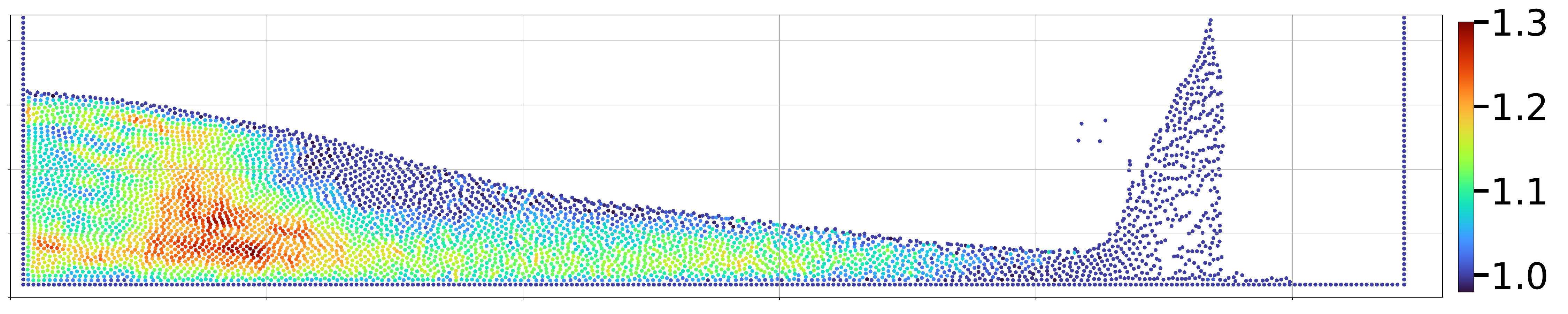}
    \includegraphics[trim={0 0cm 0 0},clip,width=0.48\linewidth]{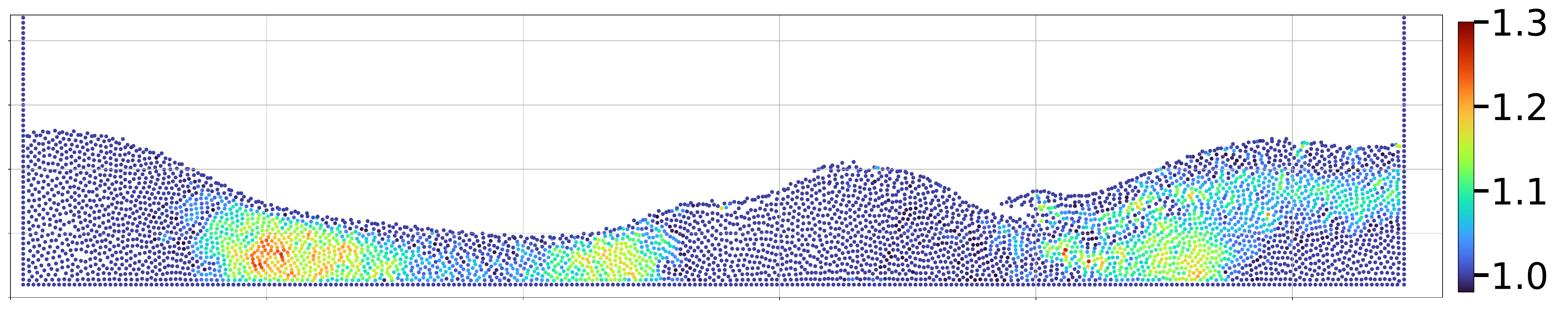}
    
    \centering
      \begin{sideways}
        \begin{minipage}{0.07\textheight}
          \centering 
          GNS$_g$
        \end{minipage}
      \end{sideways}
    \includegraphics[trim={0 0cm 0 0},clip,width=0.48\linewidth]{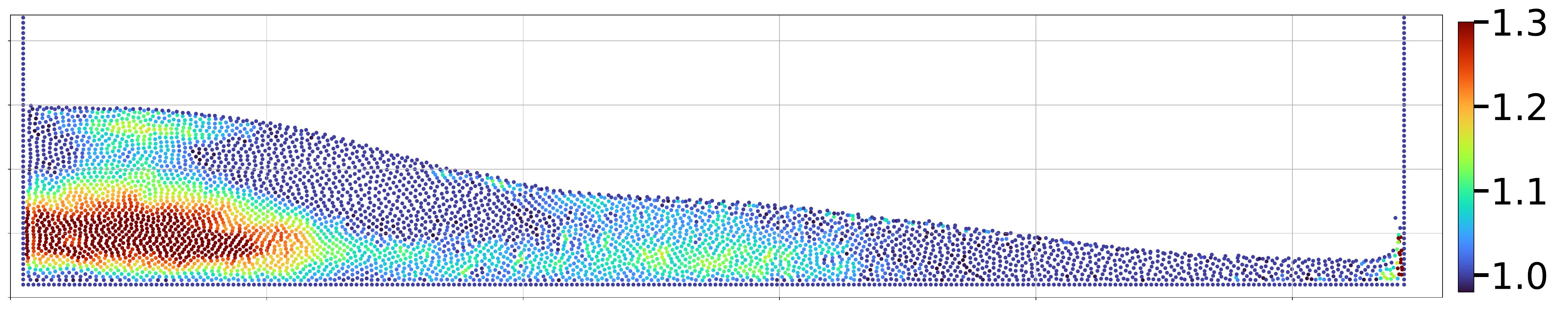}
    \includegraphics[trim={0 0cm 0 0},clip,width=0.48\linewidth]{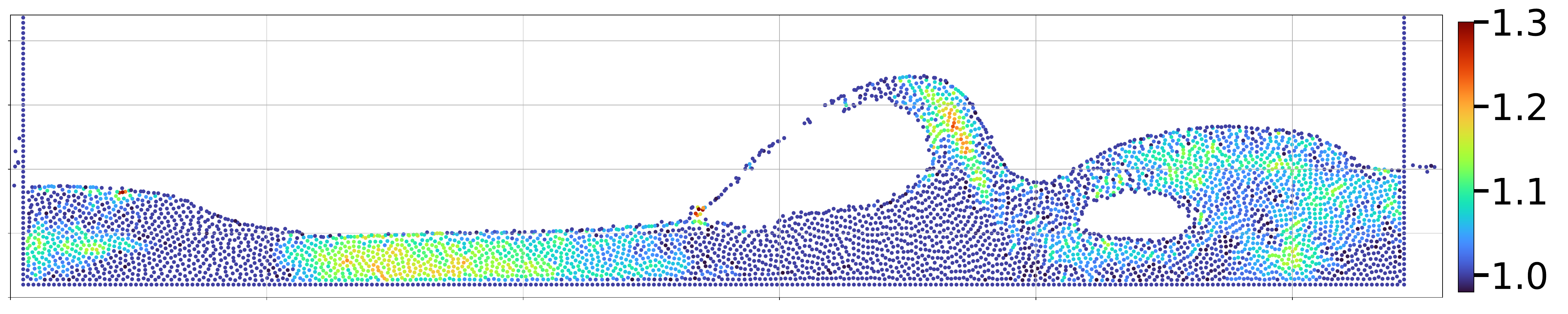}

    \centering
      \begin{sideways}
        \begin{minipage}{0.07\textheight}
          \centering
          GNS$_{g,p}$
        \end{minipage}
      \end{sideways}
    \includegraphics[trim={0 0cm 0 0},clip,width=0.48\linewidth]{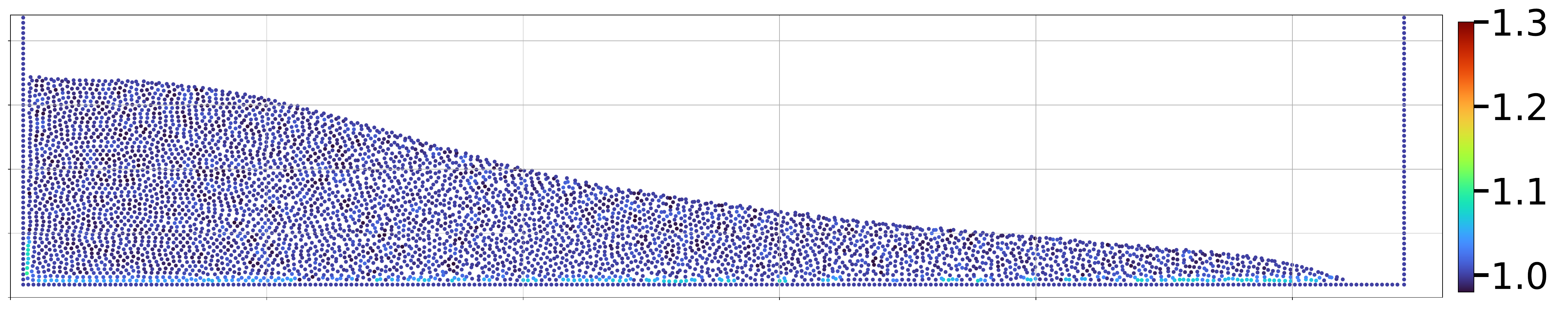}
    \includegraphics[trim={0 0cm 0 0},clip,width=0.48\linewidth]{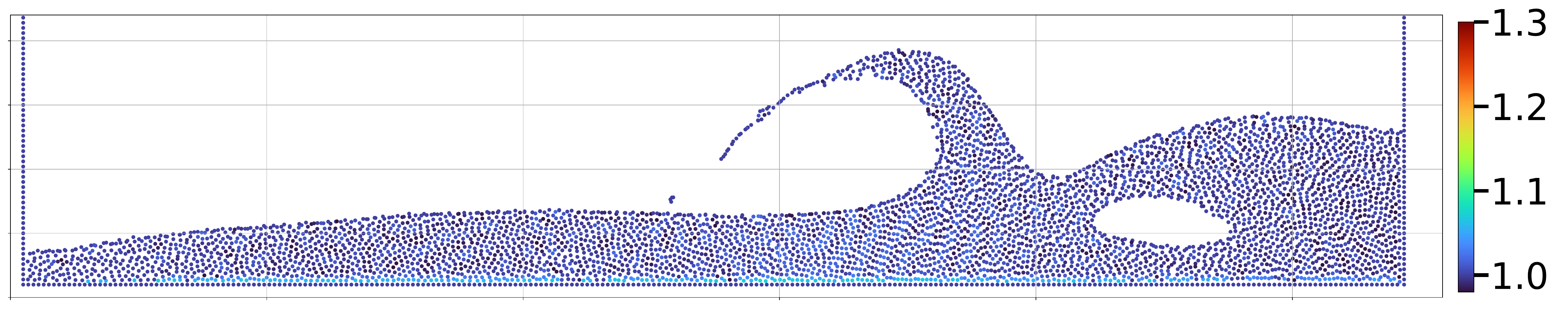}

    \centering
      \begin{sideways}
        \begin{minipage}{0.07\textheight}
          \centering
          SPH$_{\phantom{g}}$
        \end{minipage}
      \end{sideways}
    \includegraphics[trim={0 0cm 0 0},clip,width=0.48\linewidth]{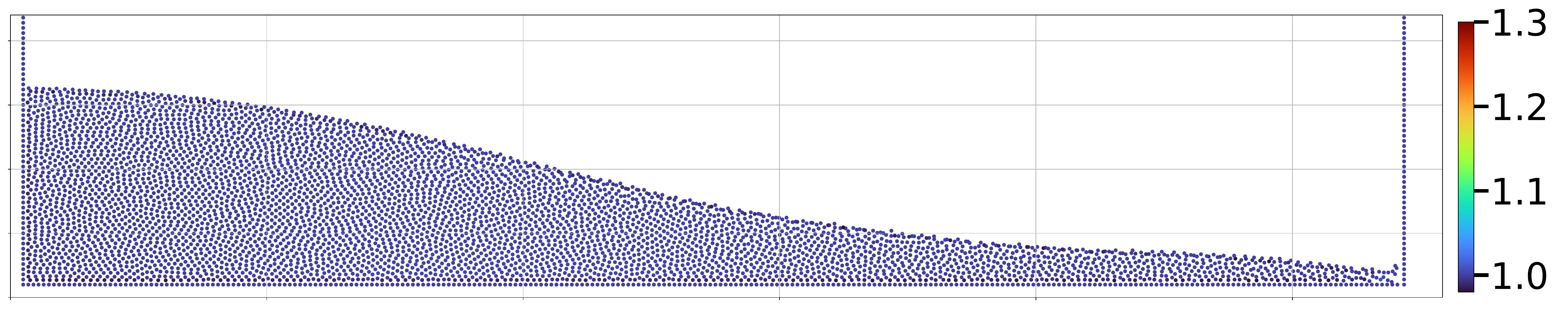}
    \includegraphics[trim={0 0cm 0 0},clip,width=0.48\linewidth]{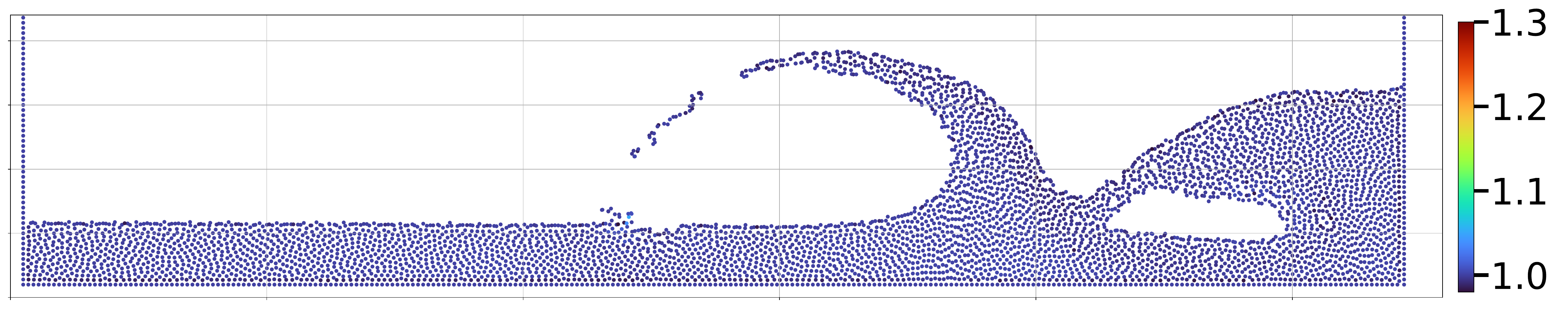}

    Step 80 \hspace{7cm} Step 240

    \caption{Dam break steps 80 and 240 of test rollout 13.}  
    \label{fig:dam_traj13}
\end{figure}

\begin{figure}[ht]
    \centering
      \begin{sideways}
        \begin{minipage}{0.07\textheight}
          \centering
          GNS$_{\phantom{g}}$
        \end{minipage}
      \end{sideways}
    \includegraphics[trim={0 0cm 0 0},clip,width=0.48\linewidth]{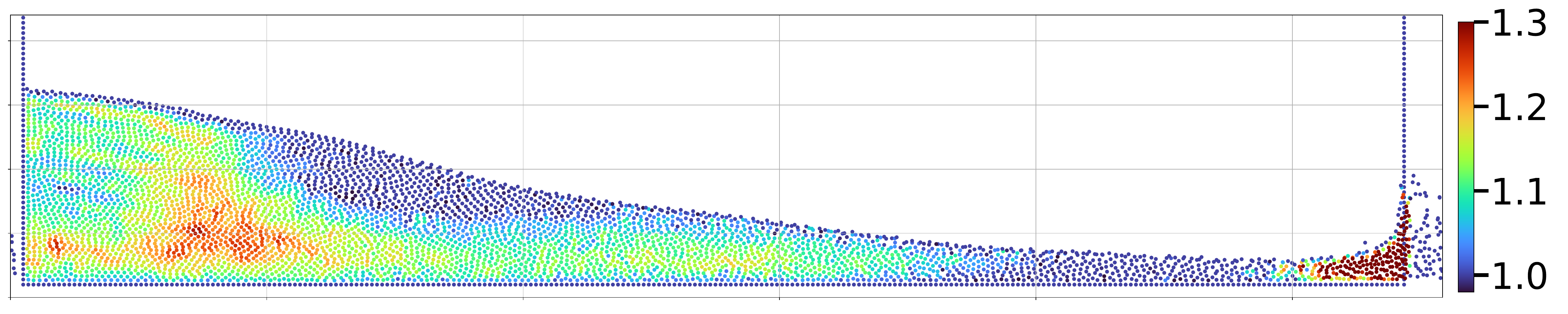}
    \includegraphics[trim={0 0cm 0 0},clip,width=0.48\linewidth]{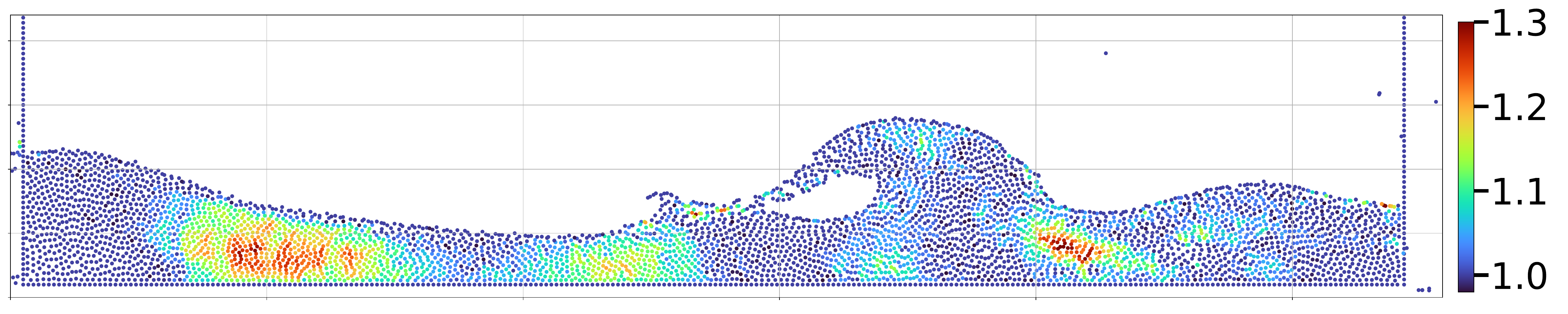}
    
    \centering
      \begin{sideways}
        \begin{minipage}{0.07\textheight}
          \centering 
          GNS$_g$
        \end{minipage}
      \end{sideways}
    \includegraphics[trim={0 0cm 0 0},clip,width=0.48\linewidth]{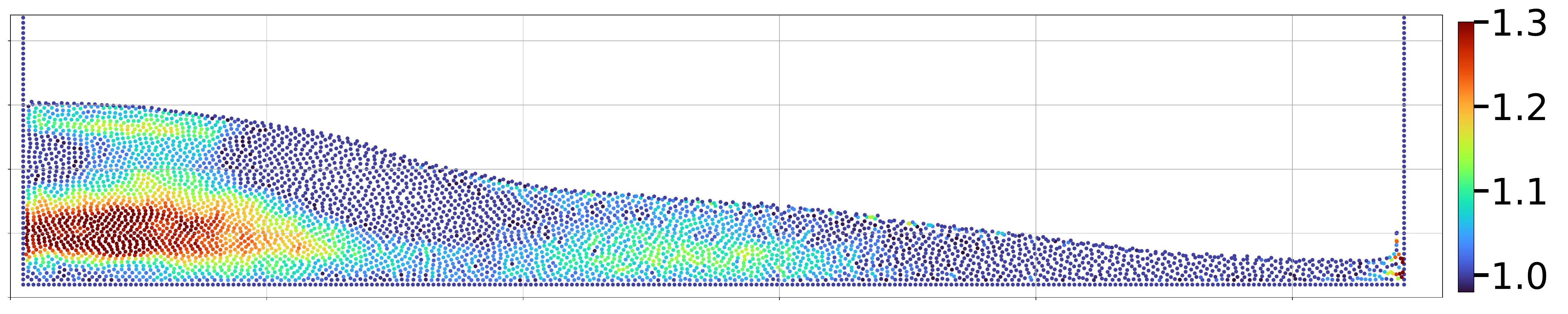}
    \includegraphics[trim={0 0cm 0 0},clip,width=0.48\linewidth]{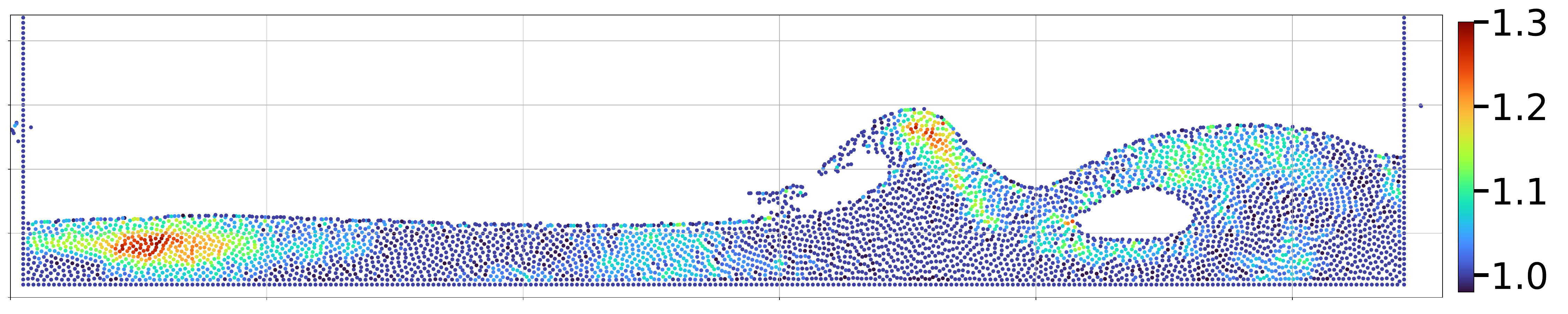}

    \centering
      \begin{sideways}
        \begin{minipage}{0.07\textheight}
          \centering
          GNS$_{g,p}$
        \end{minipage}
      \end{sideways}
    \includegraphics[trim={0 0cm 0 0},clip,width=0.48\linewidth]{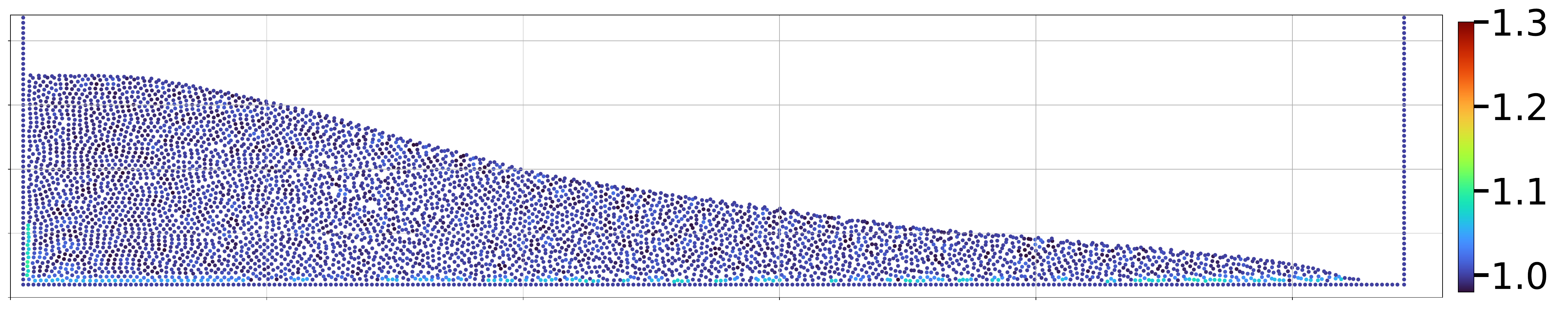}
    \includegraphics[trim={0 0cm 0 0},clip,width=0.48\linewidth]{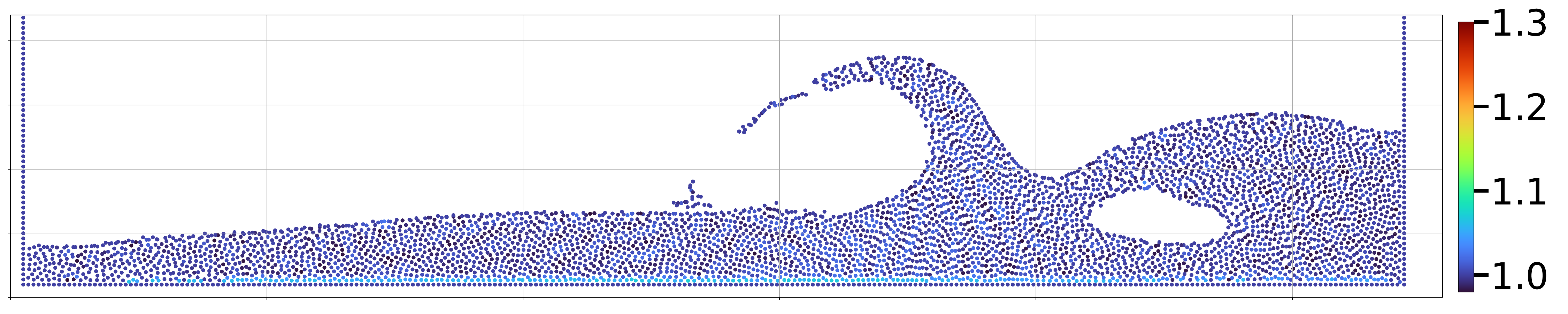}

    \centering
      \begin{sideways}
        \begin{minipage}{0.07\textheight}
          \centering
          SPH$_{\phantom{g}}$
        \end{minipage}
      \end{sideways}
    \includegraphics[trim={0 0cm 0 0},clip,width=0.48\linewidth]{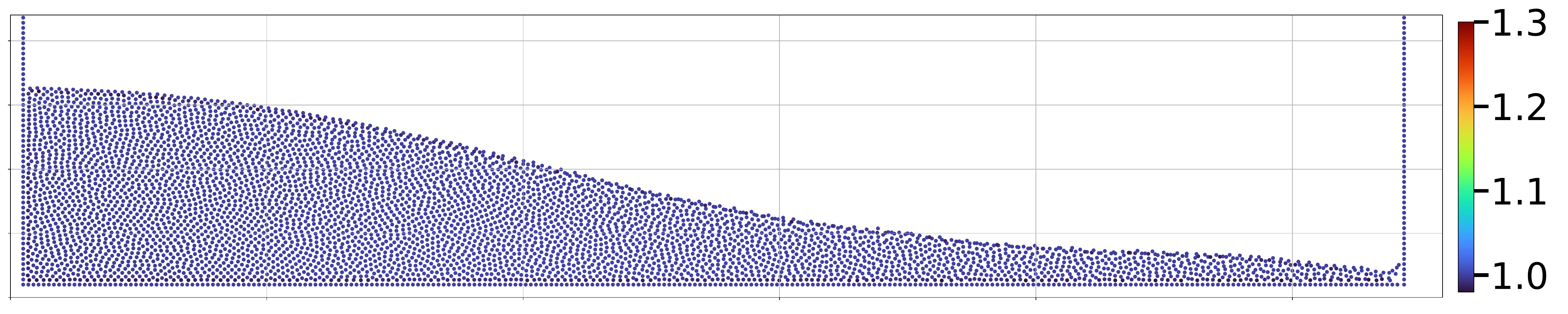}
    \includegraphics[trim={0 0cm 0 0},clip,width=0.48\linewidth]{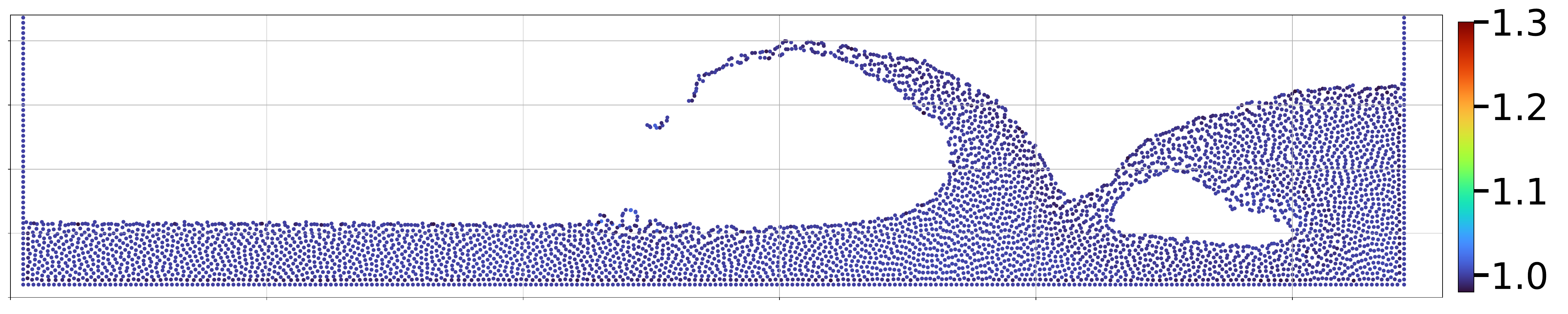}

    Step 80 \hspace{7cm} Step 240

    \caption{Dam break steps 80 and 240 of test rollout 14.}  
    \label{fig:dam_traj14}
\end{figure}

\begin{figure}[H]
    \centering
      \begin{sideways}
        \begin{minipage}{0.07\textheight}
          \centering
          GNS$_{\phantom{g}}$
        \end{minipage}
      \end{sideways}
    \includegraphics[trim={0 0cm 0 0},clip,width=0.48\linewidth]{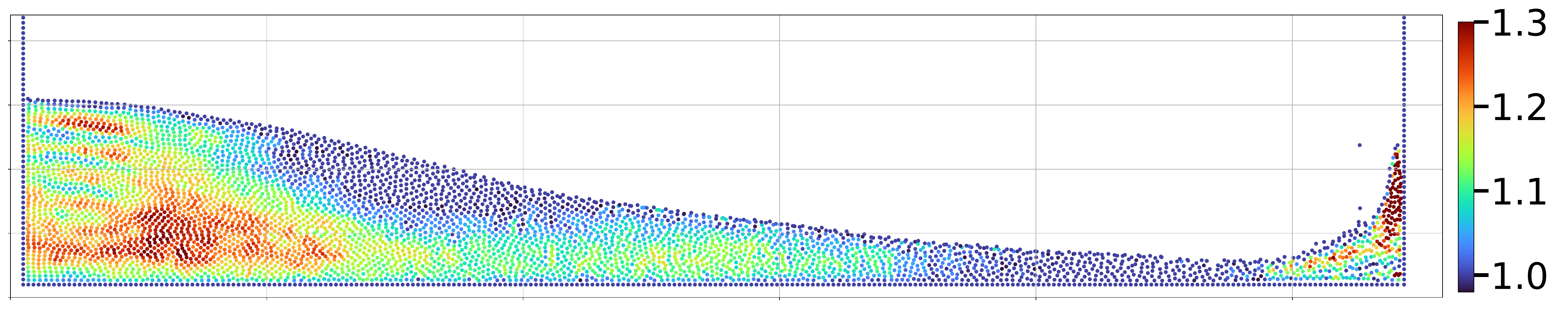}
    \includegraphics[trim={0 0cm 0 0},clip,width=0.48\linewidth]{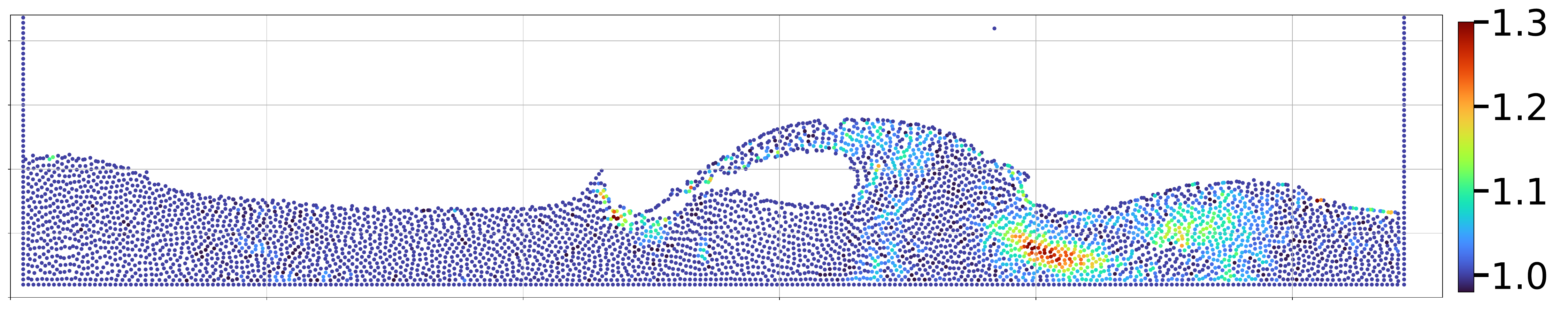}
    
    \centering
      \begin{sideways}
        \begin{minipage}{0.07\textheight}
          \centering 
          GNS$_g$
        \end{minipage}
      \end{sideways}
    \includegraphics[trim={0 0cm 0 0},clip,width=0.48\linewidth]{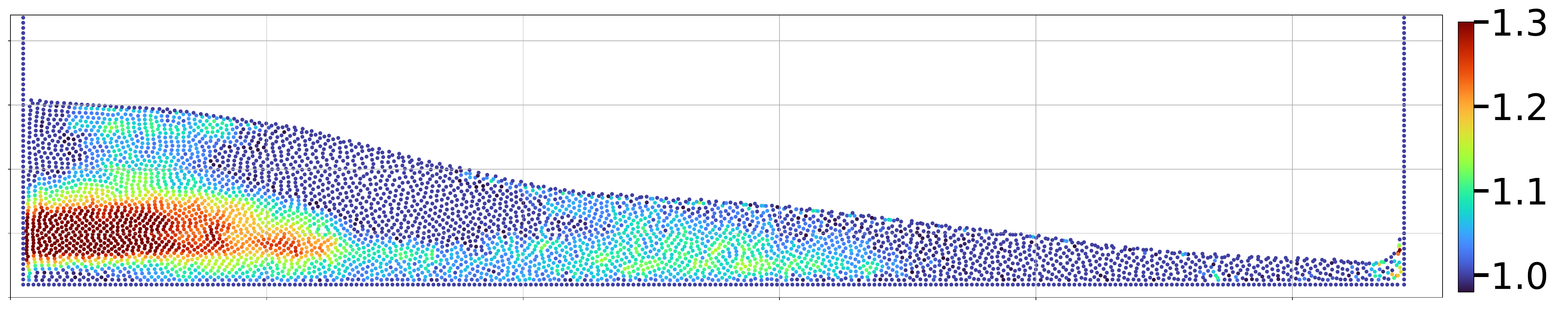}
    \includegraphics[trim={0 0cm 0 0},clip,width=0.48\linewidth]{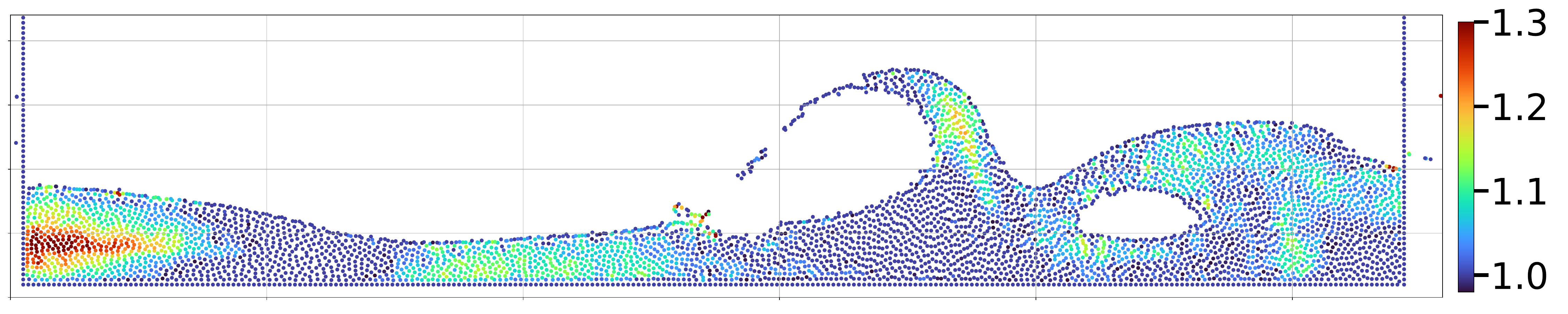}

    \centering
      \begin{sideways}
        \begin{minipage}{0.07\textheight}
          \centering
          GNS$_{g,p}$
        \end{minipage}
      \end{sideways}
    \includegraphics[trim={0 0cm 0 0},clip,width=0.48\linewidth]{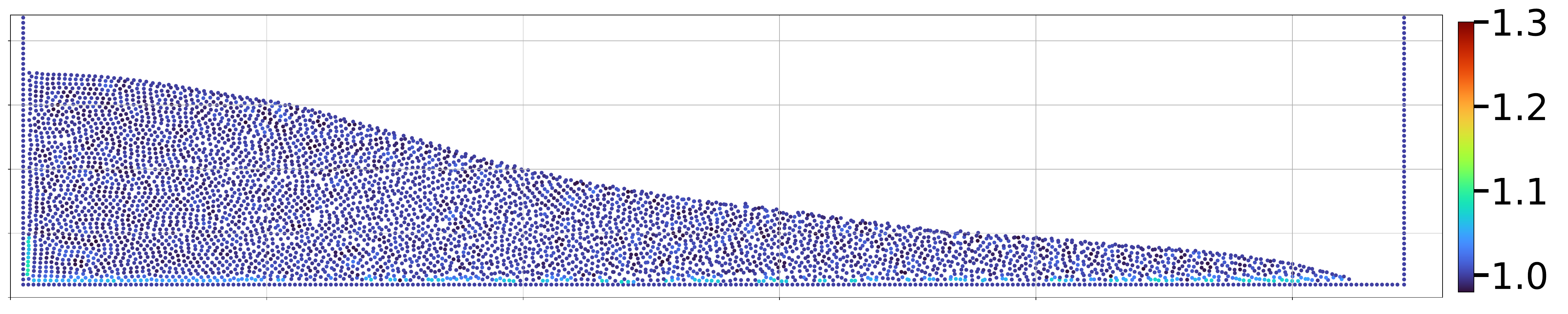}
    \includegraphics[trim={0 0cm 0 0},clip,width=0.48\linewidth]{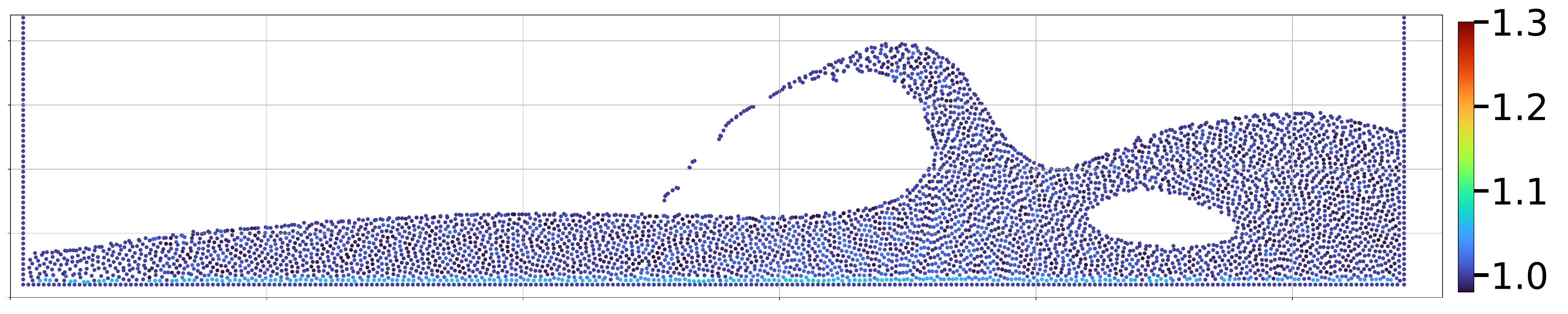}

    \centering
      \begin{sideways}
        \begin{minipage}{0.07\textheight}
          \centering
          SPH$_{\phantom{g}}$
        \end{minipage}
      \end{sideways}
    \includegraphics[trim={0 0cm 0 0},clip,width=0.48\linewidth]{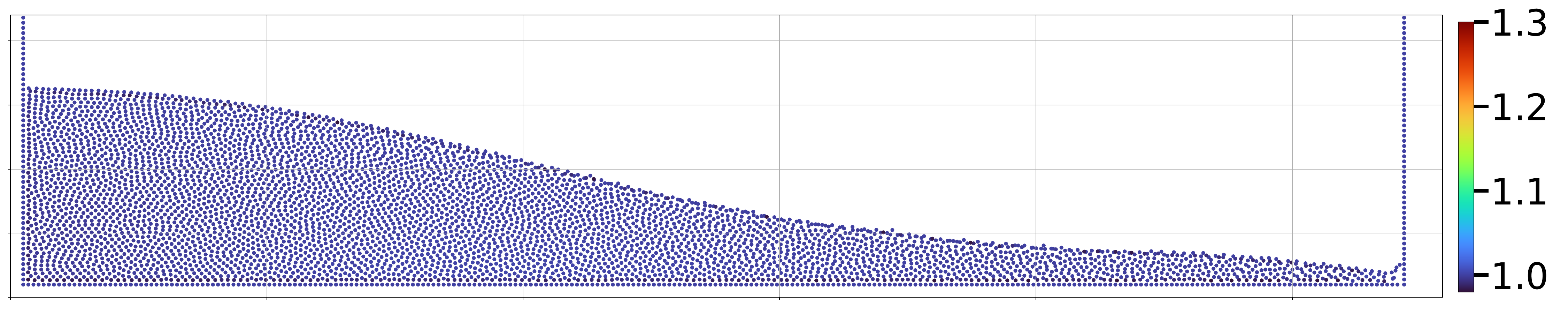}
    \includegraphics[trim={0 0cm 0 0},clip,width=0.48\linewidth]{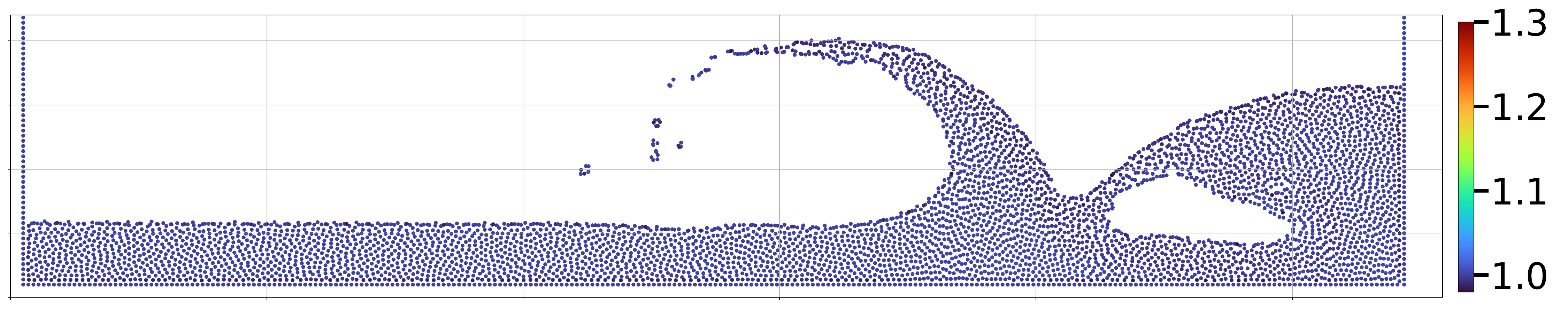}

    Step 80 \hspace{7cm} Step 240

    \caption{Dam break steps 80 and 240 of test rollout 15.}  
    \label{fig:dam_traj15}
\end{figure}

In this section, we show some more examples of dam break trajectories. Roughly one-third of GNS trajectories have the same artifacts at step 80 as test trajectory 0 (see \cref{fig:dam_traj0,fig:dam_traj13}). Roughly half of the GNS trajectories show large amounts of particles leaving the box on the right at step 80 (see \cref{fig:dam_traj14}). Only a few GNS simulations behave better at step 80 (see \cref{fig:dam_traj15}).

\section{Hyperparameters of GNS model} \label{app:hyperparams}

\begin{table}[ht]
    \centering
    \begin{tabular}{cccc}
    \hline
    Dataset & loops & $\alpha$ &  $\beta$ \\
    \hline
    2D TGV  & 5 & 0.02 & -- \\
    2D RPF  & 3 & 0.02 & 0.2 \\
    2D LDC  & 5 & 0.03 & -- \\
    2D DAM  & 3 & 0.03 & -- \\
    3D TGV  & 1 & 0.01 & -- \\
    3D RPF  & 1 & 0.005 & -- \\
    3D LDC  & 1 & 0.02 & -- \\
    \hline
    \end{tabular}
    \caption{SPH relaxation hyperparameters used in our experiments. These hyperparameters were tuned on the GNS-10-128 model.}
    \label{tab:hyperparams}
\end{table}

\pagebreak
\section{RPF 2D Plots} \label{app:rpf_plots}

\begin{figure}[H]
    \centering
    \includegraphics[trim={0 0 0cm 0},clip,width=0.98\linewidth]{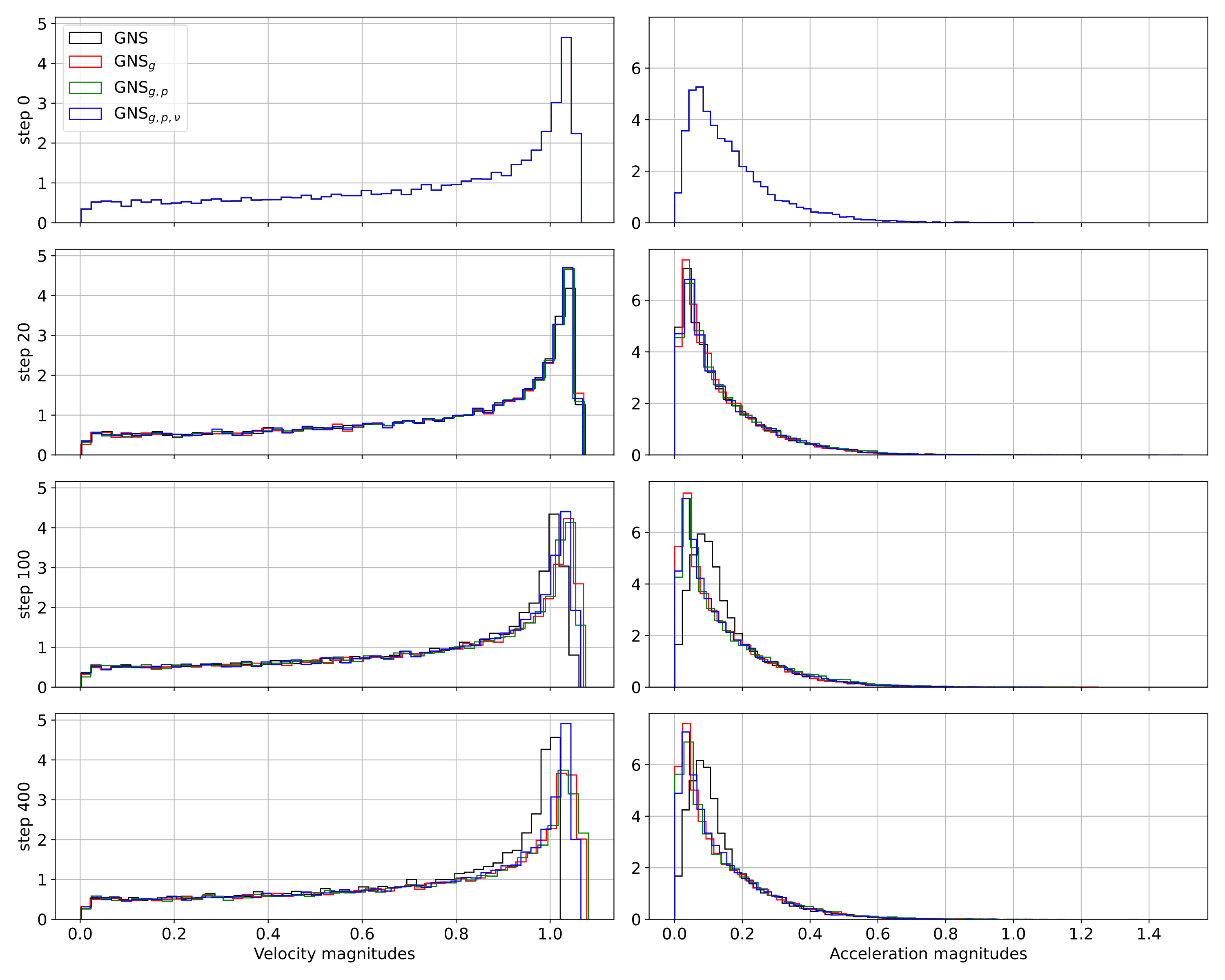}
    \caption{Velocity and acceleration magnitude histogram of 2D reverse Poiseuille flow after 400 rollout steps (average over all rollouts). Extends \cref{fig:rpf_hist_min}.}
    \label{fig:rpf_hist}
\end{figure}

\section{Forcing of Reverse Poiseuille Flow} \label{app:erf_appendix}

The forcing step function of the reverse Poiseuille flow (RPF) is given by:
\begin{align}
    f(x,y,z) = \begin{cases}
    [-1, 0, 0] \ ,     & \text{if } y > 1\\
    [1, 0, 0] \,      & \text{otherwise} \ .
    \end{cases}
\label{eq:rpf_step}
\end{align}

For the two-dimensional case, the $z$ value can be ignored.
We use the analytical solution of the convolution of the forcing step function with a Gaussian kernel of width that corresponds to the standard deviation of the velocities over the dataset. In this special case, the convolution has an analytical solution given by the error function $\text{erf}$. For the jump in the middle, we obtain the solution
\begin{align}
    f_{smooth}(x,y,z) = [- \text{erf}\left(\frac{y - 1}{\sqrt{2} \sigma} \right), 0, 0] \ .
\label{eq:rpf_step_smooth}
\end{align}

\begin{figure}[h]
    \centering
    \includegraphics[trim={0 4mm 0cm 0},clip,width=0.3\linewidth]{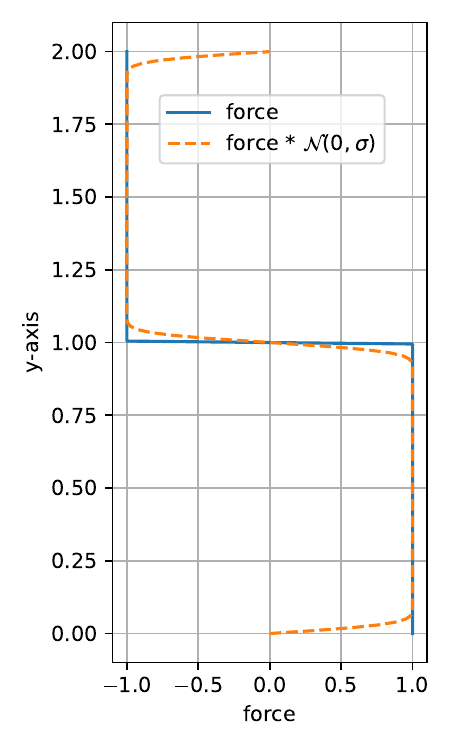}
    \caption{Forcing step function of the 2D reverse Poiseuille flow before (blue) and after convolution with normal distribution $\mathcal{N}(0,0.025^2)$ (orange).}
    \label{fig:rpf_force_smooth}
\end{figure}

We use the finite difference approximation between consecutive coordinate frames to approximate the standard deviation of the velocity. For 2D RPF, the velocity standard deviation is $[0.036, 0.00069]$, and for 3D RPF $[0.074, 0.0014, 0.0011]$. We first convert these two standard deviation vectors to their isotropic versions, assuming that the velocity components are independent Gaussian random variables, i.e., using the quadratic mean. This leads to $\sigma_{2D}=0.025$ and $\sigma_{3D}=0.043$. We round the numbers and use the values $\sigma_{2D}=0.025$ and $\sigma_{3D}=0.05$ in our experiments. The result of this smoothing procedure can be seen in \cref{fig:rpf_force_smooth}.
        
\section{Inference Speed} \label{app:inference_speed}
We measured the inference speed of GNS-10-128 and SEGNN-10-64 on the 2D and 3D reverse Poiseuille flow datasets with 0, 1, 3, or 5 relaxation steps $l$ and summarize the results in \cref{tab:timing}. This table provides more quantitative results to the discussion on inference speed in \cref{sec:experiments}.

\begin{table}[!ht]
    \centering
    \begin{tabular}{ccccccc}
    \hline
        Dataset & $t_{gt}$ [ms] & Model & $t_{l=0}$ [ms] & $t_{l=1}$ [ms] & $t_{l=3}$ [ms] & $t_{l=5}$ [ms] \\ \hline
        \multirow{2}{*}{\centering 2D RPF} & \multirow{2}{*}{\centering 43.0} & GNS & 10.7 & 11.0 & 13.3 & 14.4 \\ \cline{3-7}
           & & SEGNN  & 24.9 & 25.9 & 28.4 & 30.4 \\ \hline
        \multirow{2}{*}{\centering 3D RPF}  & \multirow{2}{*}{\centering 424} & GNS & 23.8 & 32.5 & 50.4 & 68.0 \\ \cline{3-7}
         & & SEGNN & 97.9 & 106 & 124 & 141 \\ \hline
    \end{tabular}
    \caption{Timing experiments on RPF datasets with GNS-10-128 model. With $t_{gt}$, we denote the time the ground truth SPH solver takes to simulate 100 steps, as the LagrangeBench datasets consist of every 100th solver state. We took the values $t_{gt}=43.0$ and $t_{gt}=424$ from Table 4 in~\citet{toshev2024lagrangebench}. Timing runs are averaged over 10k forward calls to the model and consecutive position relaxations.}
    \label{tab:timing}
\end{table}

\clearpage
\section{Temporal Coarsening} \label{app:coarsening_appendix}

Semi-implicit Euler:
\begin{align}
\mathbf{u}_1 &= \mathbf{u}_0 + \Delta t \mathbf{a}_0 \\
\mathbf{p}_1 &= \mathbf{p}_0 + \Delta t \mathbf{u}_1 \\
    &= \mathbf{p}_0 + \Delta t \mathbf{u}_0 + \Delta t^2 \mathbf{a}_0 \\
\mathbf{u}_2 &= \mathbf{u}_1 + \Delta t \mathbf{a}_1 \\
    &= \mathbf{u}_0 + \Delta t (\mathbf{a}_0 + \mathbf{a}_1) \\
\mathbf{p}_2 &= \mathbf{p}_1 + \Delta t \mathbf{u}_2 \\
    &= (\mathbf{p}_0 + \Delta t \mathbf{u}_0 + \Delta t^2 \mathbf{a}_0) + \Delta t (\mathbf{u}_0 + \Delta t (\mathbf{a}_0 + \mathbf{a}_1))\\
    &= \mathbf{p}_0 + \Delta t 2 \mathbf{u}_0 + \Delta t^2 (2\mathbf{a}_0 + \mathbf{a}_1) \\
    & \hspace{0.5em} \vdots \nonumber \\
\mathbf{u}_M &= \mathbf{u}_0 + \Delta t \sum_{m=0}^{M-1} \mathbf{a}_m \\
\mathbf{p}_M &= \mathbf{p}_0 + M \Delta t \mathbf{u}_0 + \Delta t^2 \sum_{m=0}^{M-1} (M-m) \mathbf{a}_m \ .
\end{align}

If $\mathbf{a}_m$ is a constant number, we can simplify the last part to:
\begin{align}
\mathbf{u}_M &= \mathbf{u}_0 + M \Delta t \mathbf{a} \\
\mathbf{p}_M &= \mathbf{p}_0 + M \Delta t \mathbf{u}_0 + 0.5 M (M+1) \Delta t^2 \mathbf{a} \ .
\end{align}

If we now compute the target effective acceleration by finite differences of positions, we end up with 
\begin{align}
\mathbf{u}_0^{FD} &= (\mathbf{p}_0 - \mathbf{p}_{-M})/ \Delta t^{FD} \\
\mathbf{u}_M^{FD} &= (\mathbf{p}_M - \mathbf{p}_{0})/ \Delta t^{FD} \\
\mathbf{a}_0^{FD} &= (\mathbf{u}_M^{FD} - \mathbf{u}_{0}^{FD})/\Delta t^{FD} = (\mathbf{p}_M - 2\mathbf{p}_{0} + \mathbf{p}_{-M})/ {\Delta t^{FD}}^2 .
\end{align}

By substituting the semi-implicit Euler rule after $M$ steps into this finite differences approximation and setting $\Delta t^{FD}=1$ for simplicity, we get an effective acceleration of

\begin{align}
    \mathbf{a}_{iM}^{FD} &= \mathbf{p}_{(i+1)M} - 2\mathbf{p}_{iM} + \mathbf{p}_{(i-1)M}  \\
    \begin{split}
        &=  M ( \Delta t \mathbf{u}_0 ((i+1) - 2i + (i-1)) \\
        &+ 0.5 \Delta t^2 \mathbf{a} (((i+1)^2M+(i+1)) - 2(i^2M + i) + ((i-1)^2M+(i-1))) ) \\
    \end{split}\\
    &=  M \left( 0 + 0.5 \Delta t^2 \mathbf{a} ( 2M ) \right) \\
    &= (M \Delta t)^2 \mathbf{a} \ .
\end{align}

\section{Ablations} \label{app:ablations}
We extend the results from the main paper by running multiple ablation studies mainly focusing on (a) the individual and combined impact of SPH relaxation and force treatment on the example of \textbf{dam break}, (b) the sensitivity of the parameters governing the proposed SPH relaxation on the example of \textbf{lid-driven cavity}, and (c) the impact of smoothing the external force function on the example of the \textbf{reverse Poiseuille flow} datasets. We believe that this exhaustive analysis of the hyperparameters is essential for practitioners who would consider using our proposed methods. To increase the value of the analysis we add (A) the evolution of the metrics over the simulation length, (B) error bars representing the 0.25 and 0.75 quantiles over the test trajectories, and (C) three more metrics compared to the main paper. The six metrics we use are:
\begin{enumerate}
    \item MSE$_{400}$ -- position MSE over 400 steps.
    \item MSE$_{Ekin}$ -- kinetic energy MSE between the predicted and ground truth frames.
    \item Sinkhorn -- Sinkhorn divergence between the particle distribution of predicted and ground truth frames. Measures how much effort it would take to move the particle mass between the two states. Scales as $\mathcal{O}(N^2)$ with the number of particles $N$ and is more compute intense than the model inference on all our datasets. 
    \item MAE$_{\rho}$ -- density MAE error measuring the deviation of the density from the reference density $\rho_{ref}$. In all our experiments $\rho_{ref}=1.0$.
    \item Dirichlet -- Dirichlet energy~\citep{zhou2005regularization} of density field $E_D \left( \rho \right) = \frac{1}{2} \int \left\lVert \nabla \rho \right\rVert_{2}^2 dx$, based on \citet{taheri2009dirichlet,diening2011soloblev}. It measures both high-frequency (e.g. clustering) and low-frequency (e.g. instabilities) density fluctuations. Lower is better and means less steep gradients~\citep{cai2020note,giovanni2023energy}.
    \item Chamfer -- symmetric Chamfer distance $d_{CD}(X,Y)=\sum_{x \in X}\min_{y \in Y} ||x-y||_2^2 + \sum_{y\in Y}\min_{x \in X} ||x-y||_2^2$ between predicted and ground truth frames. Similar to Sinkhorn, but only considers nearest neighbors, and thus much more compute efficient.
\end{enumerate}

For all these measures applies: lower is better, and $0.0$ is best.

\subsection{Dam Break} \label{app:ablations_dam}
\begin{figure}[ht]
    \centering
    \includegraphics[width=0.48\textwidth]{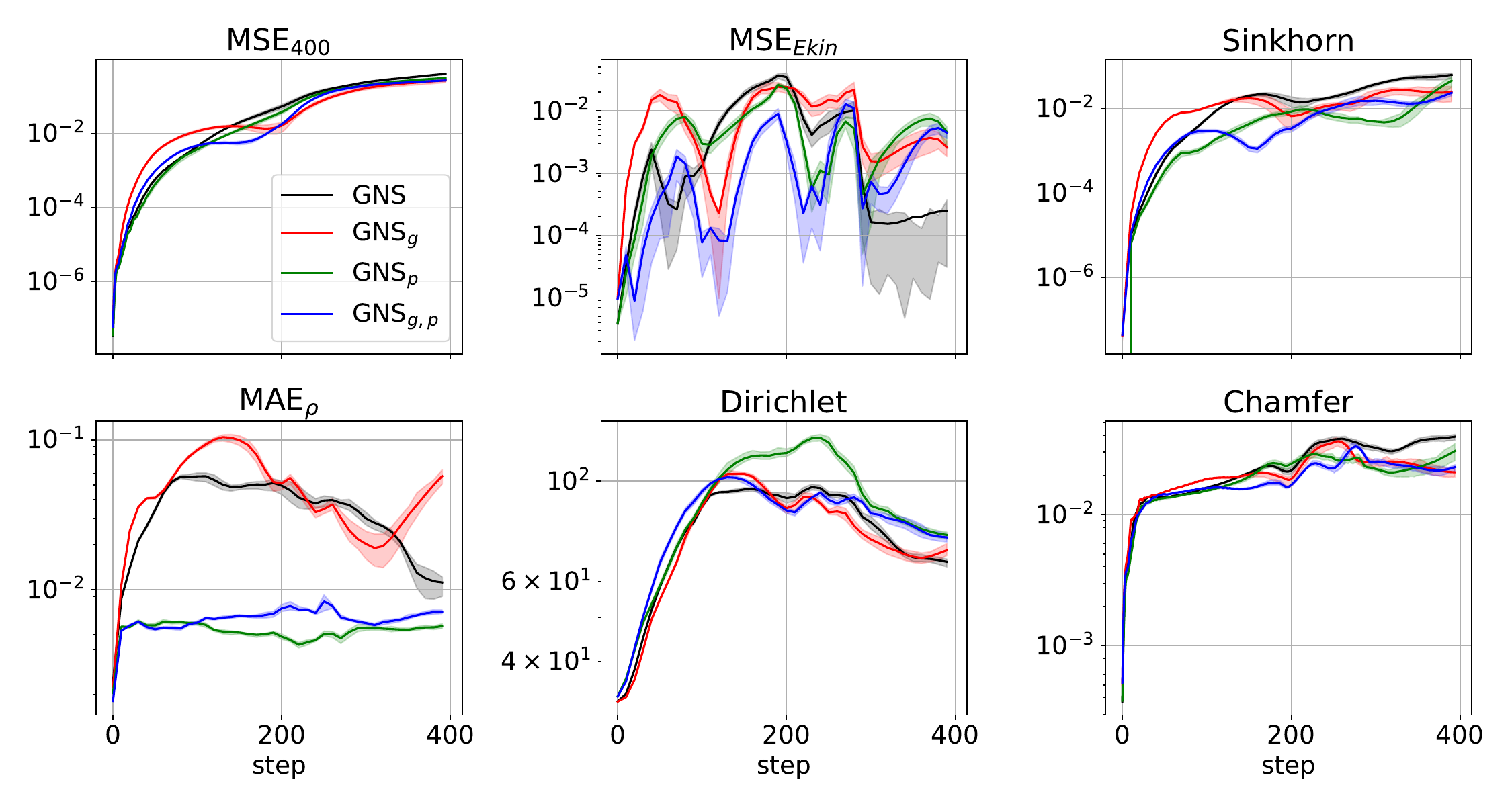}
    \hspace{10px}
    \includegraphics[width=0.48\textwidth]{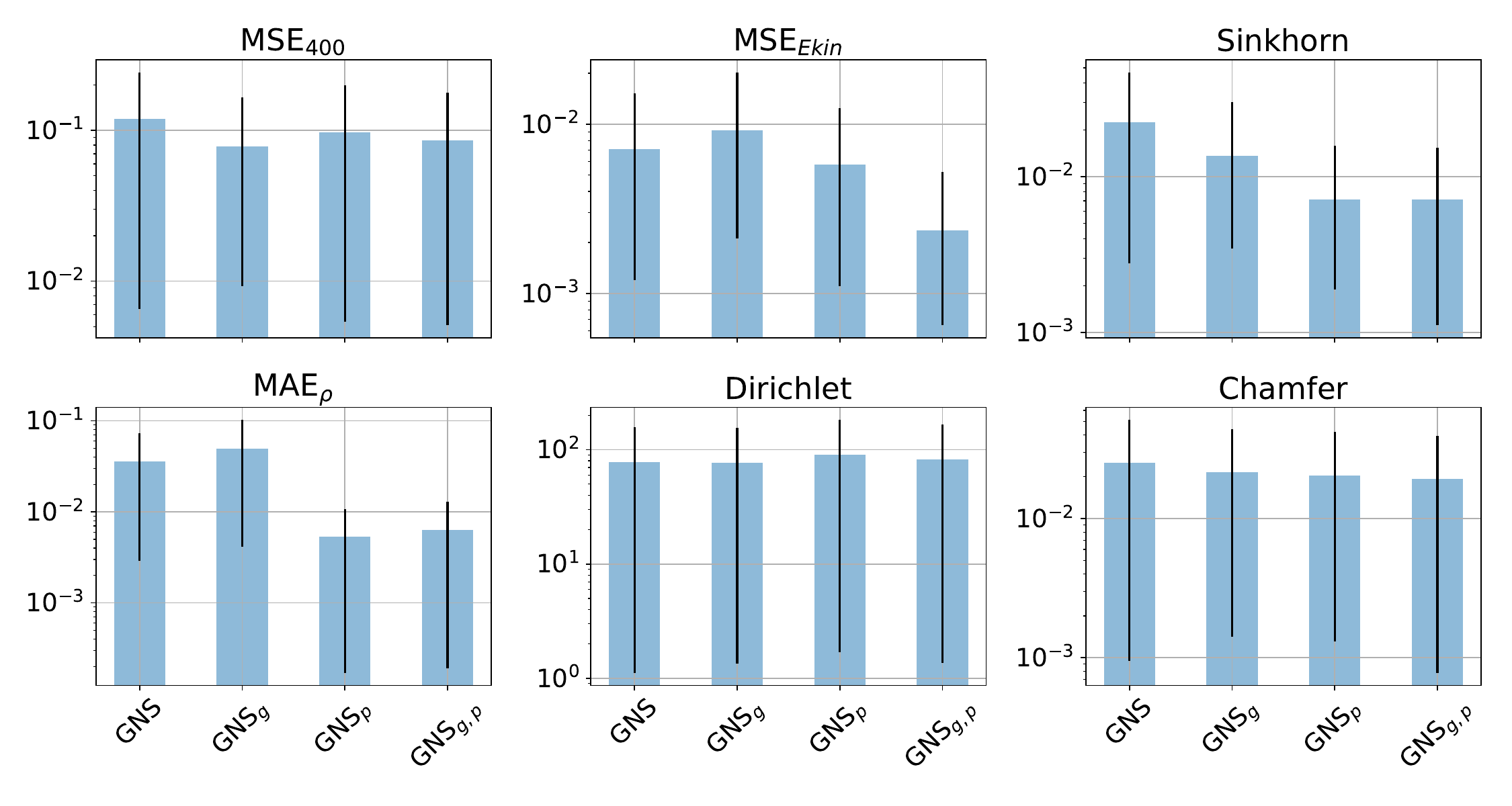}
    \caption{Ablations on DAM 2D with GNS-10-128 over the simulation length (left) and the average thereof (right). \label{fig:dam2d_gns_ext}}
\end{figure}

\begin{figure}[ht]
    \centering  
    \includegraphics[width=0.48\textwidth]{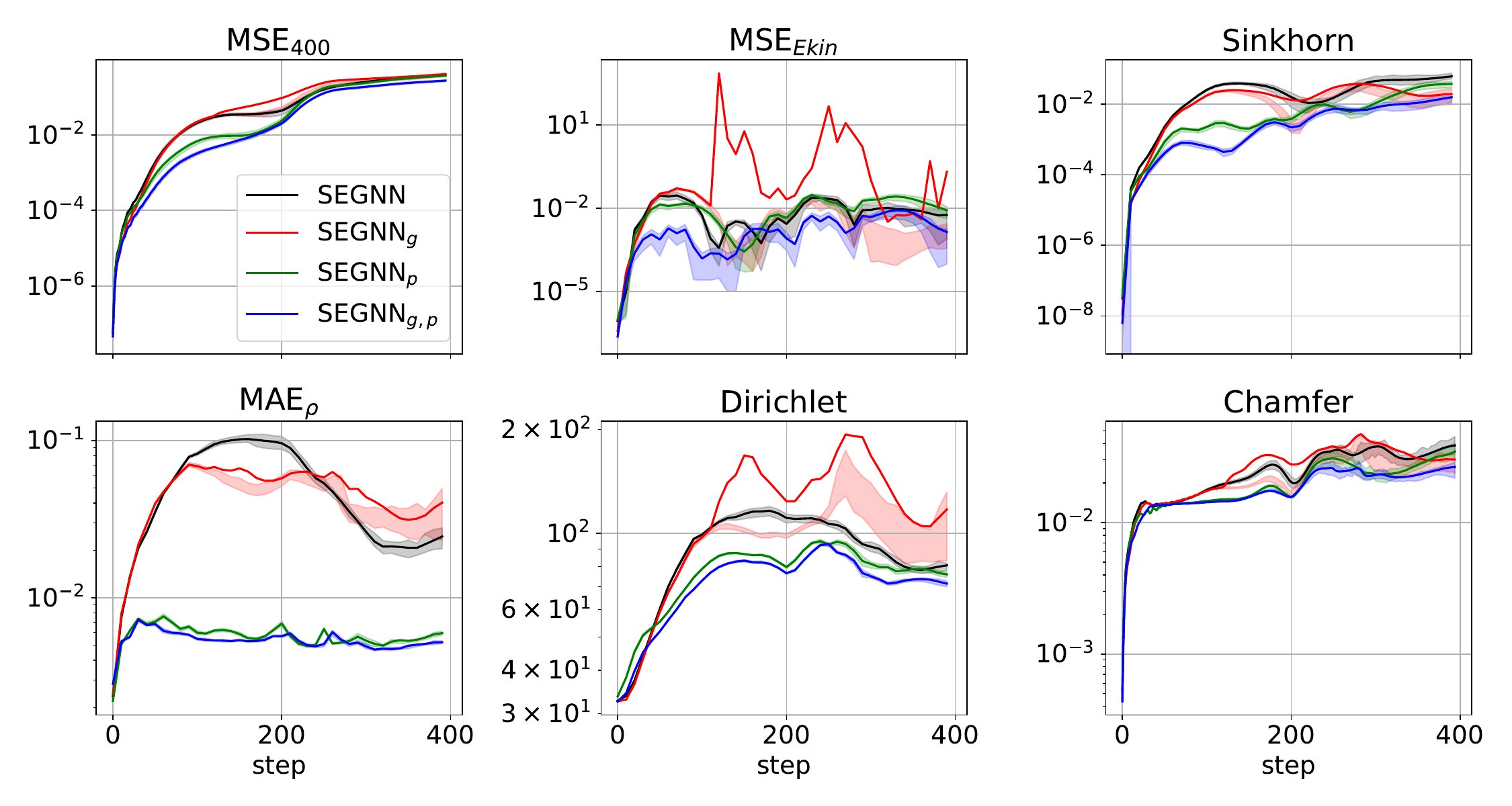}
    \hspace{10px}
    \includegraphics[width=0.48\textwidth]{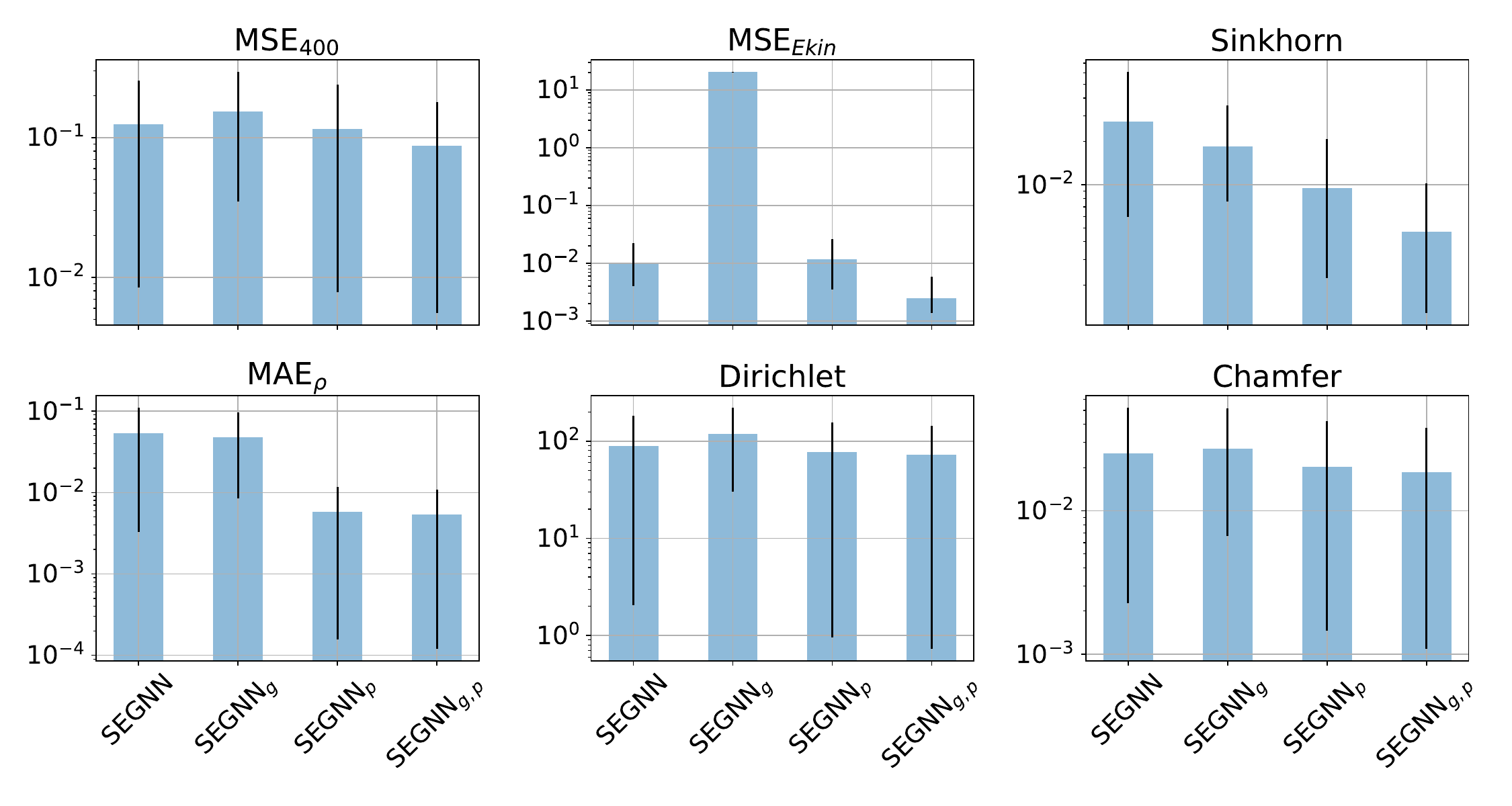}
    \caption{Ablations on DAM 2D with SEGNN-10-64 over the simulation length (left) and the average thereof (right). \label{fig:dam2d_segnn_ext}}
\end{figure}

We compare the impact of our external force treatment ($\square_{g}$), our SPH relaxation with parameters from \cref{tab:hyperparams} ($\square_{p}$), and combination of both ($\square_{g,p}$) on the dam break dataset using the GNS (\cref{fig:dam2d_gns_ext}) and SEGNN (\cref{fig:dam2d_segnn_ext}). On the MSE$_{Ekin}$, we see that only through the combination of our force treatment and SPH relaxation we achieve significant performance boosts with both the GNS and SEGNN models.

\subsection{Lid-Driven Cavity} \label{app:ablations_ldc}
We investigate the influence of the relaxation hyperparameters $\alpha$ and $\beta$ from \cref{eq:nse_momentum_learned} and the number of relaxation steps/loops.
The evolution of the six error measures over the 400 steps is shown on the left, and the average for each hyperparameter configuration is shown on the right. Intervals indicate the 0.25 and 0.75 quantiles over the 12 test trajectories (left) and the average of those values over the 400 steps (right).

\subsubsection{LDC 2D with GNS}
\begin{figure}[ht]
    \centering
    \includegraphics[width=0.48\textwidth]{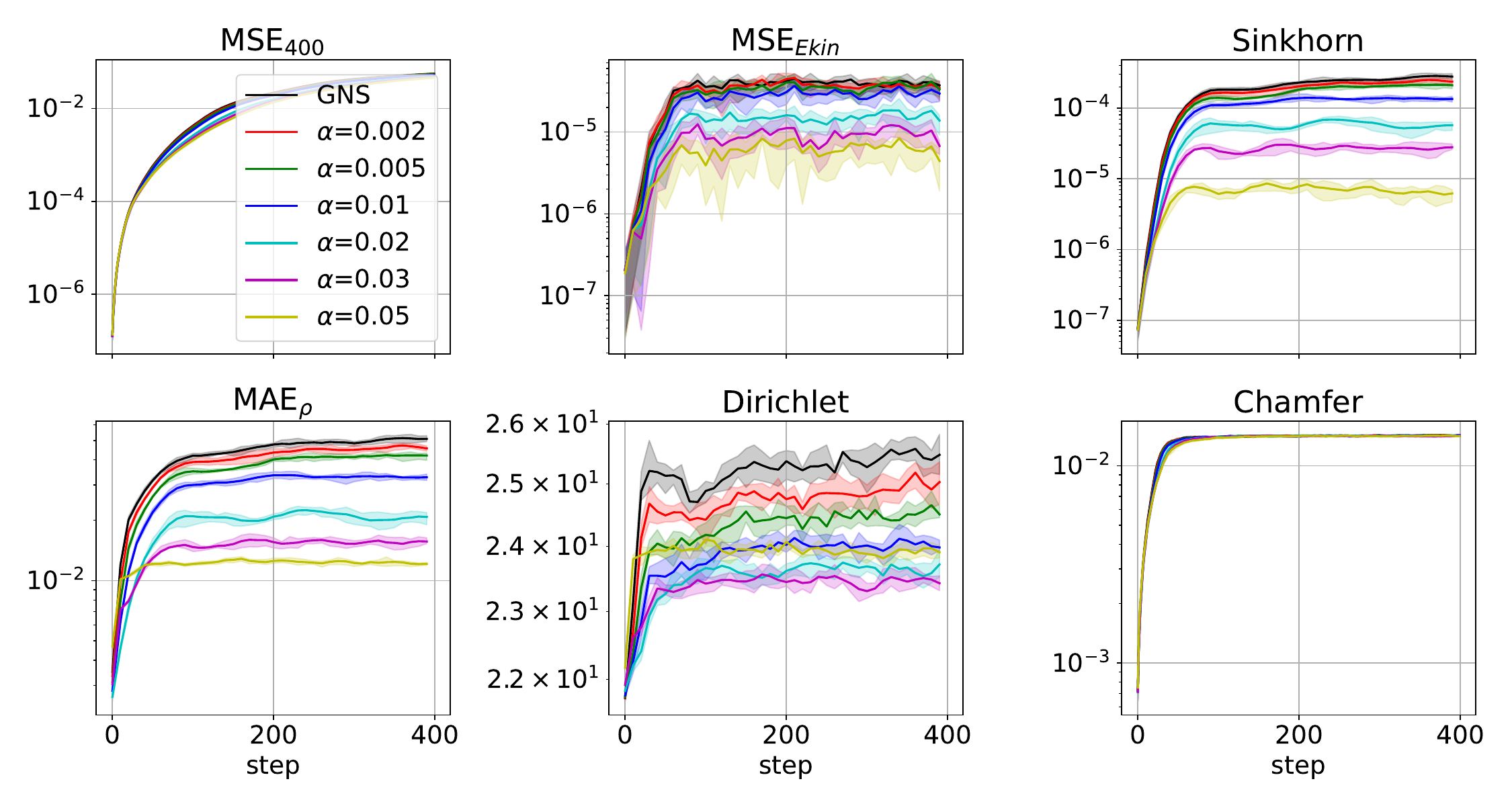}
    \hspace{10px}
    \includegraphics[width=0.48\textwidth]{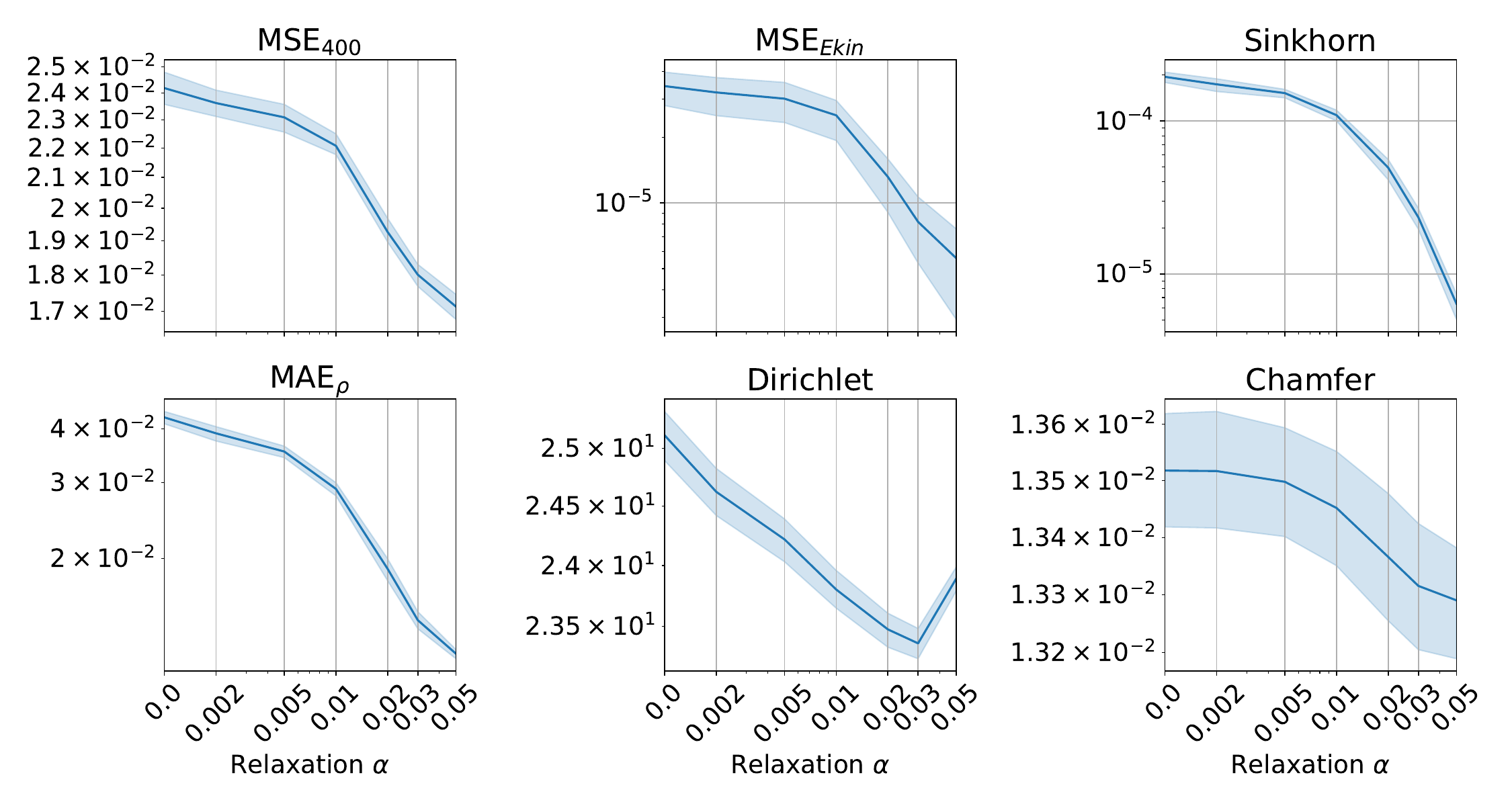}
    \caption{Ablations on LDC 2D with GNS-10-128 ($l=1$) regarding relaxation parameter $\alpha$. \label{fig:ldc2d_gns_alphas}}
\end{figure}

\begin{figure}[ht]
    \centering  
    \includegraphics[width=0.48\textwidth]{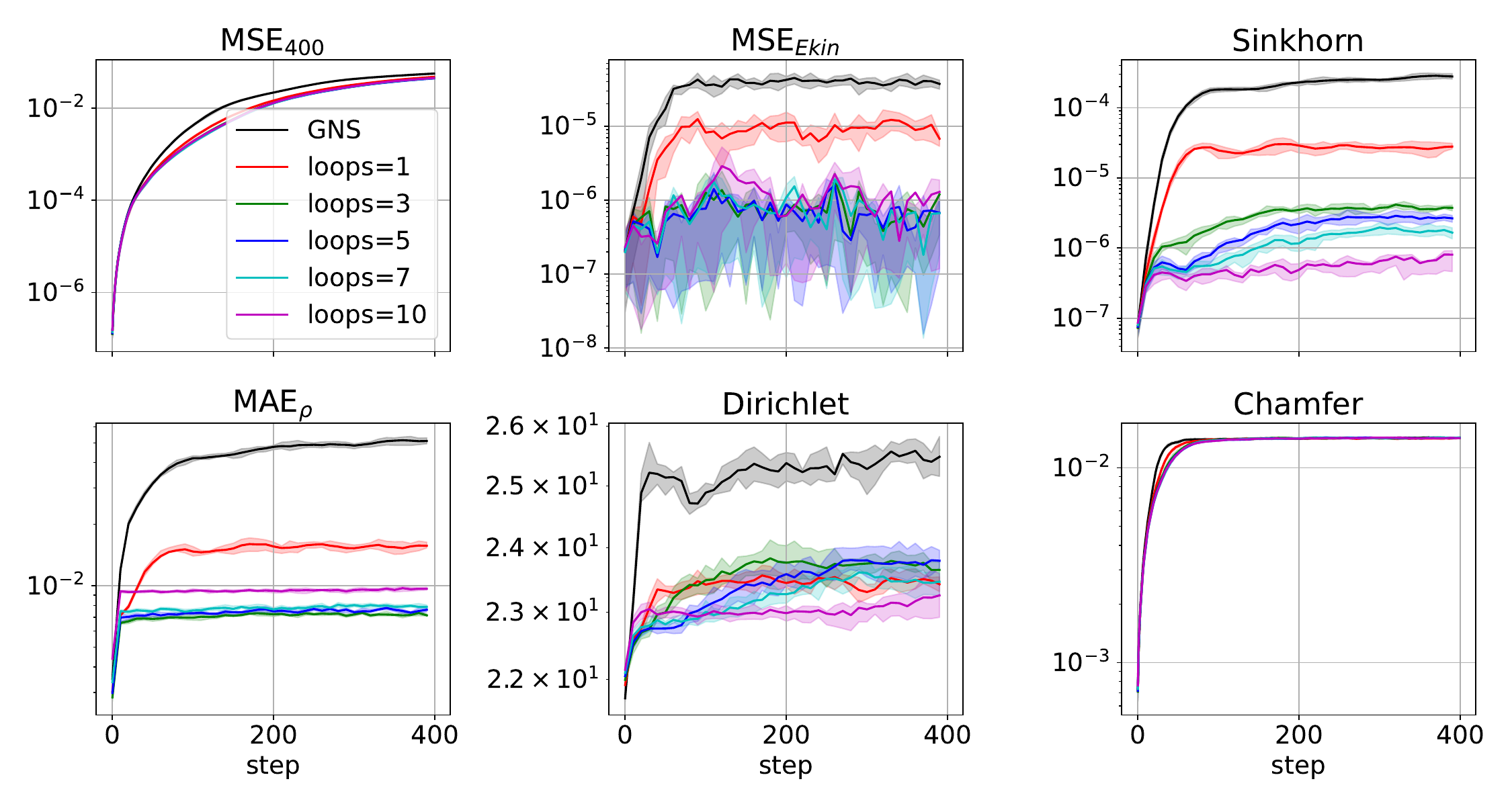}
    \hspace{10px}
    \includegraphics[width=0.48\textwidth]{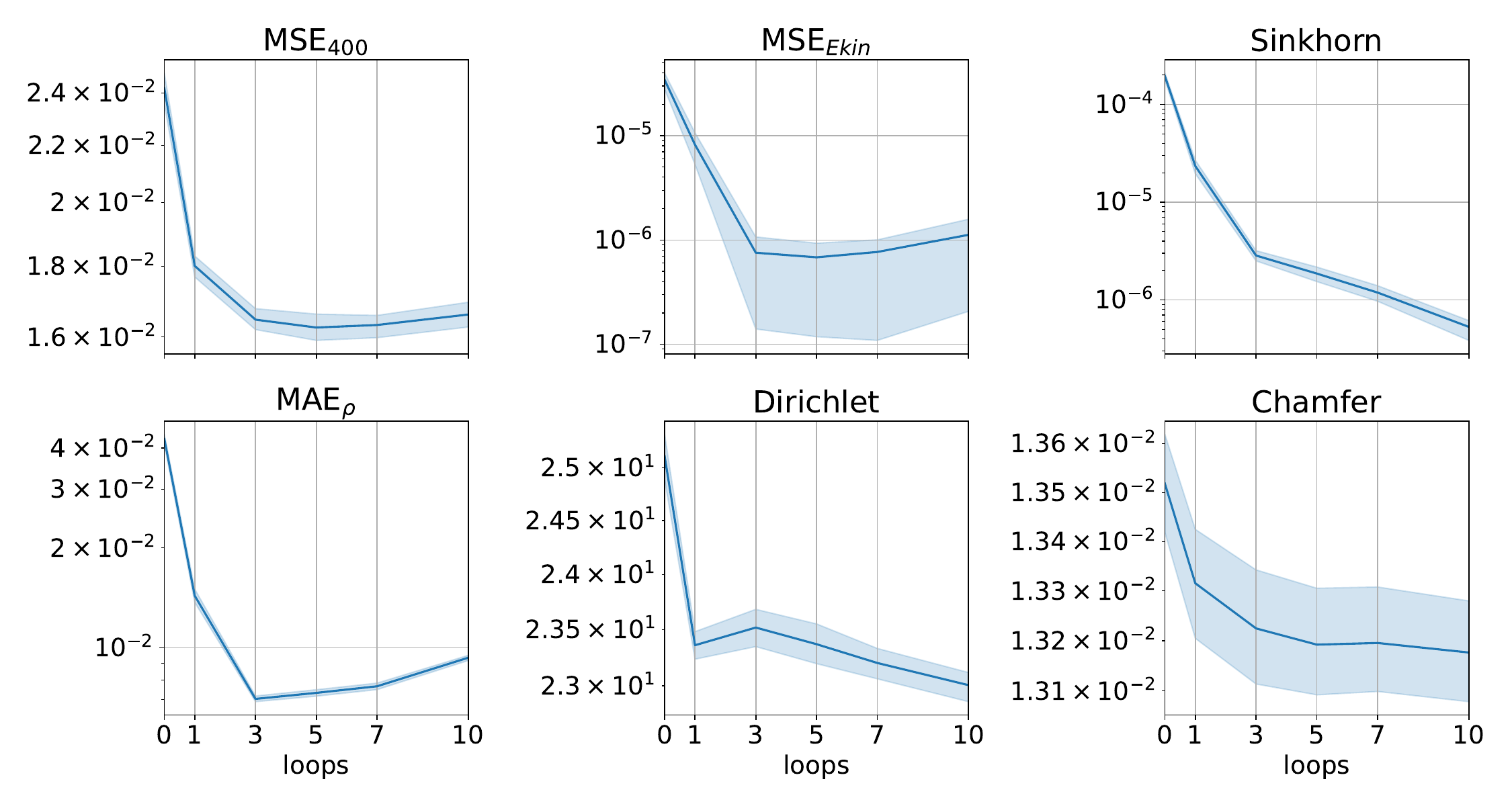}
    \caption{Ablations on LDC 2D with GNS-10-128 ($\alpha=0.03$) regarding the number of relaxation steps/loops. \label{fig:ldc2d_gns_loops}}
\end{figure}

\begin{figure}[ht]
    \centering
    \includegraphics[width=0.48\textwidth]{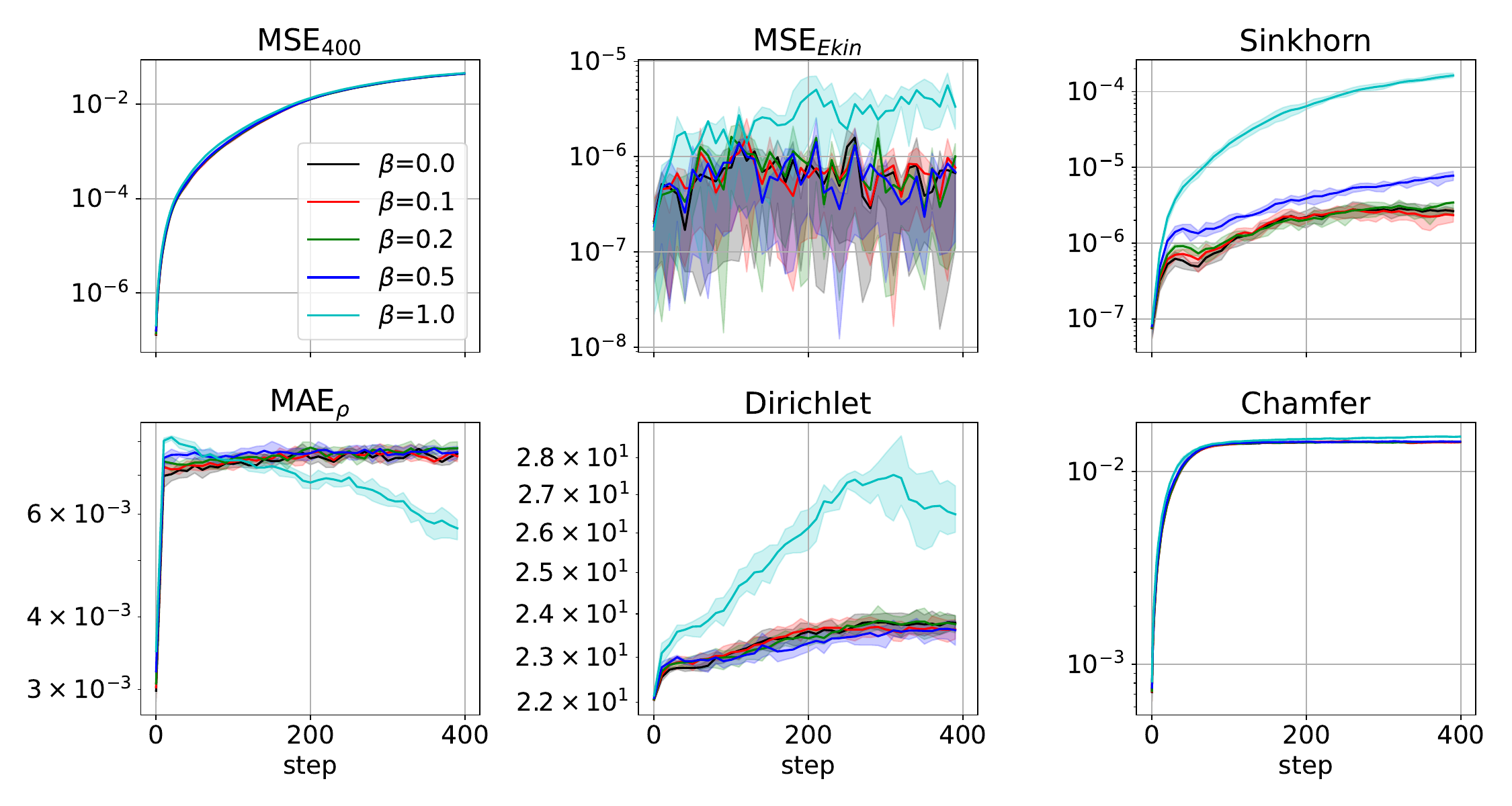}
    \hspace{10px}
    \includegraphics[width=0.48\textwidth]{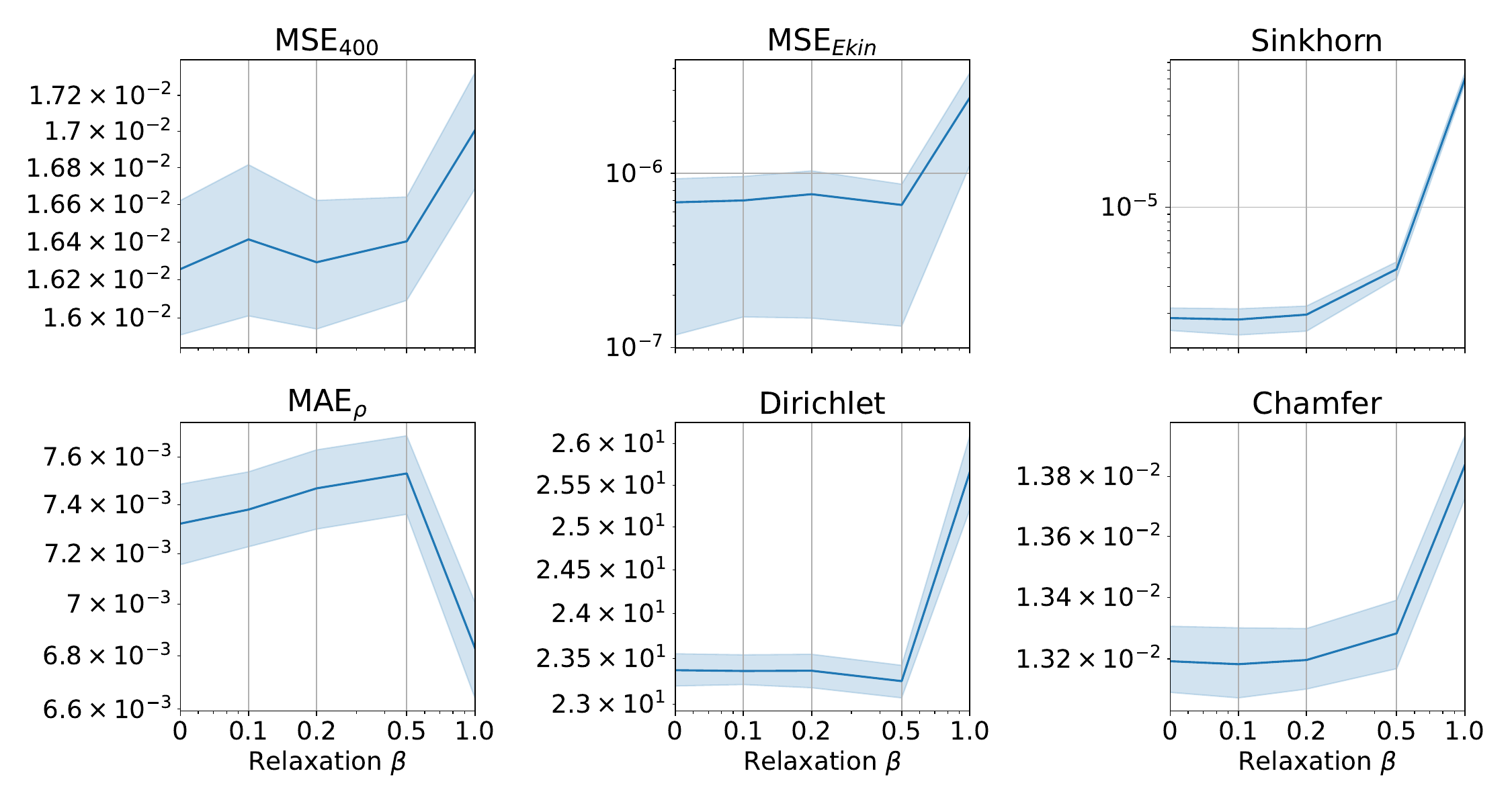}
    \caption{Ablations on LDC 2D with GNS-10-128 ($\alpha=0.03$, $l=5$) regarding relaxation parameter $\beta$.
    \label{fig:ldc2d_gns_betas}}
\end{figure}

Based on \cref{fig:ldc2d_gns_alphas}, we choose $\alpha=0.03$ as beyond this value, the Dirichlet energy starts increasing, indicating instabilities. In \cref{fig:ldc2d_gns_loops}, we see on $MSE_{400}$ and $MSE_{Ekin}$ that beyond 5 iterations the accuracy drops, so we choose $l=5$ loops. In \cref{fig:ldc2d_gns_betas}, we do not see performance gains using the viscous term, so we decide not to use it.

\subsubsection{LDC 2D with SEGNN}
We again stress that the relaxation hyperparameters were optimized on GNS and we only ablate their influence on the performance of SEGNN. But we indeed observe similar behavior between GNS and SEGNN. We do stress the dramatic improvement in performance upon 5 and more relaxation steps visible in \cref{fig:ldc2d_segnn_loops}. In contrast to GNS, we do see positive impact of the viscous term on SEGNN, and would recommend using $\beta=0.5$, see \cref{fig:ldc2d_segnn_betas}.
\begin{figure}[ht]
    \centering
    \includegraphics[width=0.48\textwidth]{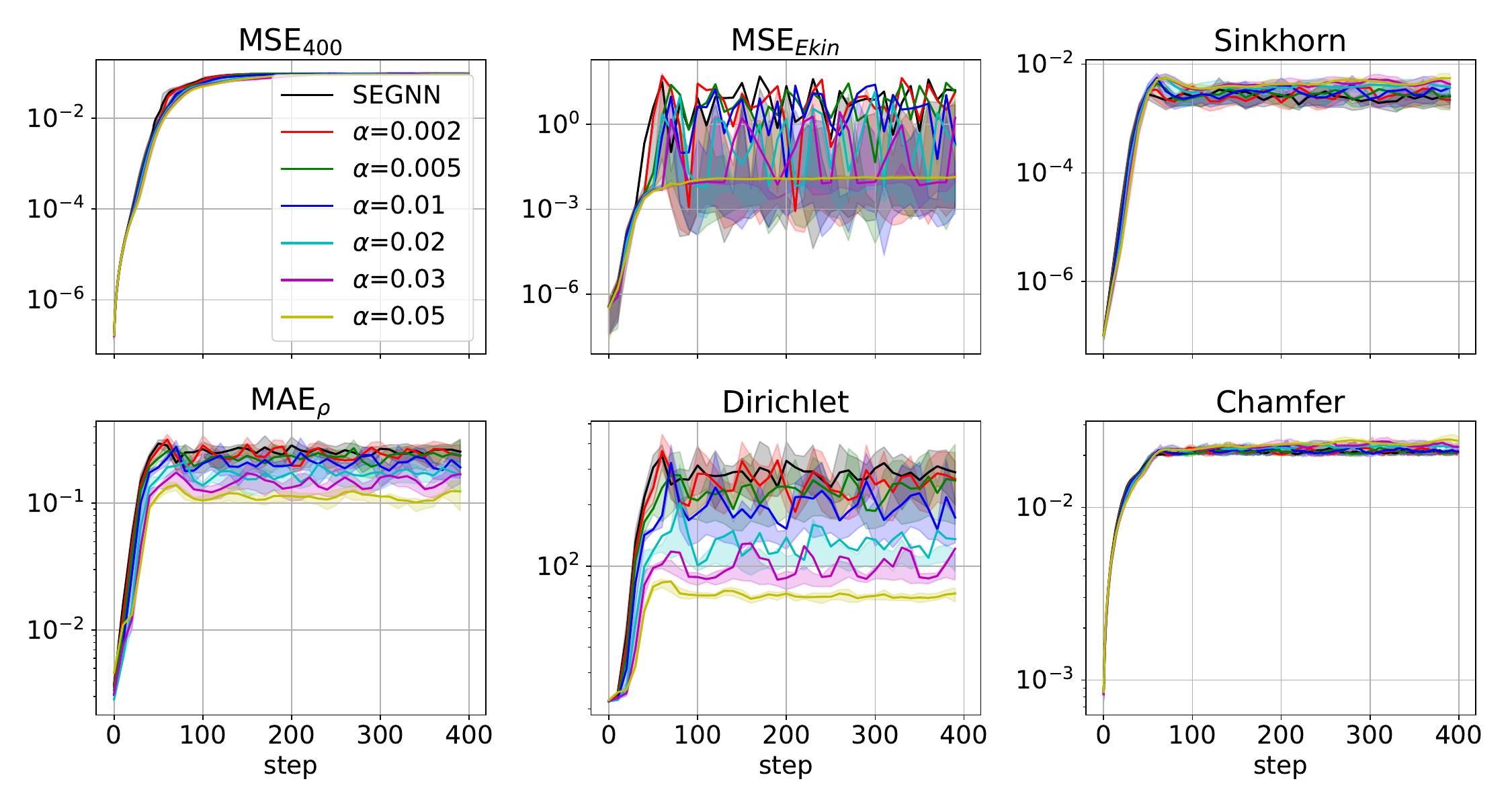}
    \hspace{10px}
    \includegraphics[width=0.48\textwidth]{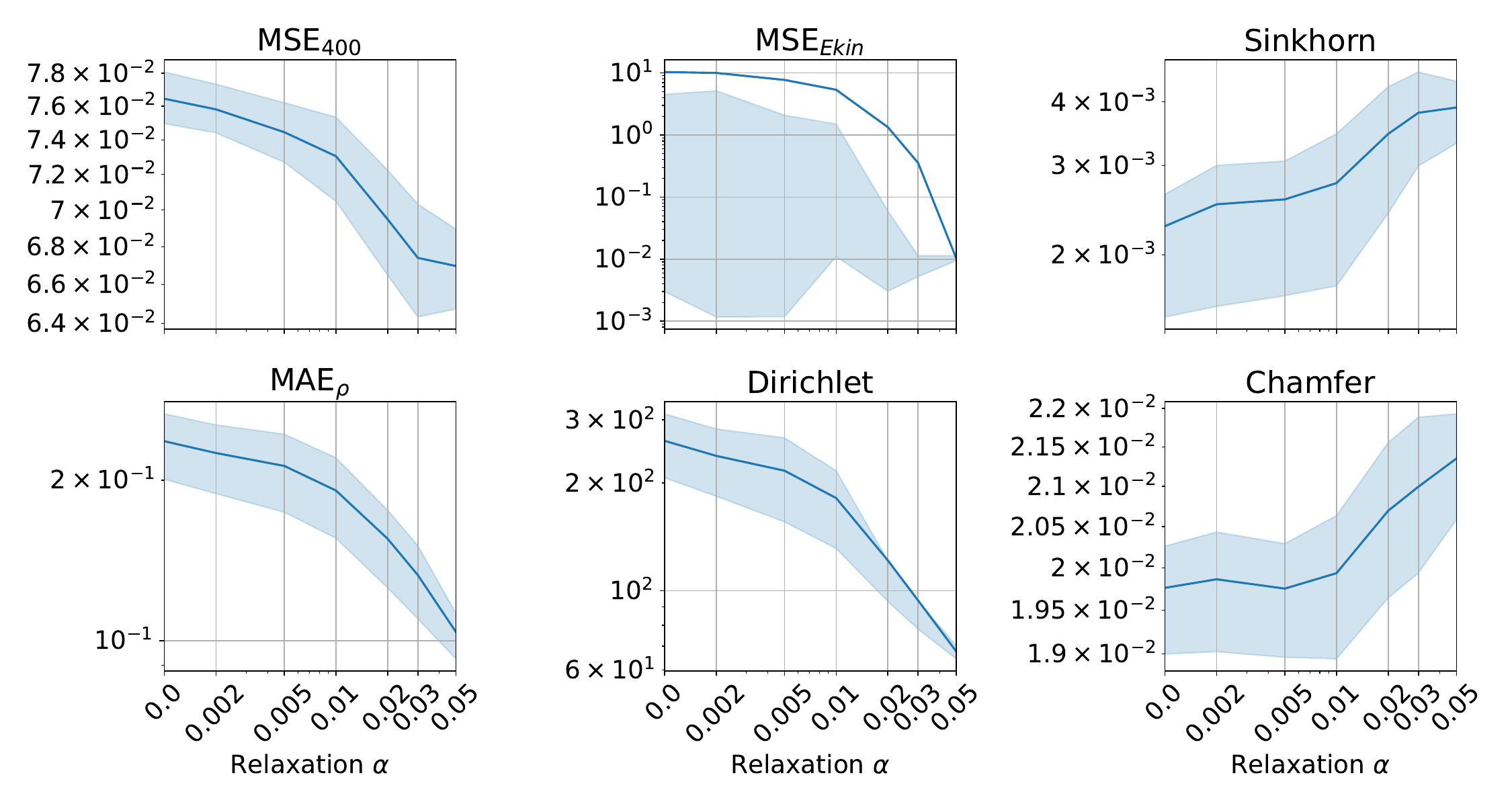}
    \caption{Ablations on LDC 2D with SEGNN-10-64 ($l=1$) regarding relaxation parameter $\alpha$. \label{fig:ldc2d_segnn_alphas}}
\end{figure}

\begin{figure}[ht]
    \centering  
    \includegraphics[width=0.48\textwidth]{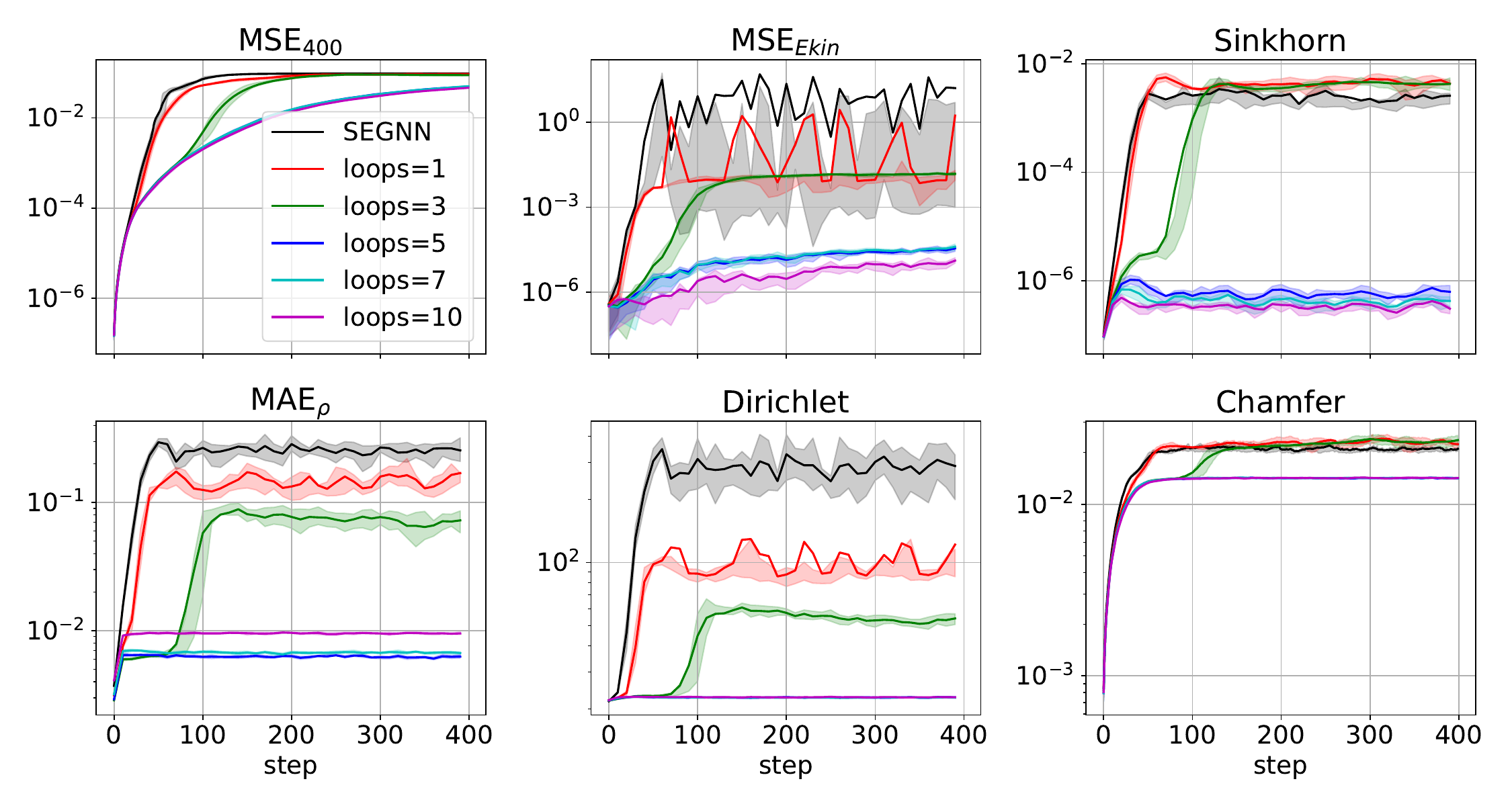}
    \hspace{10px}
    \includegraphics[width=0.48\textwidth]{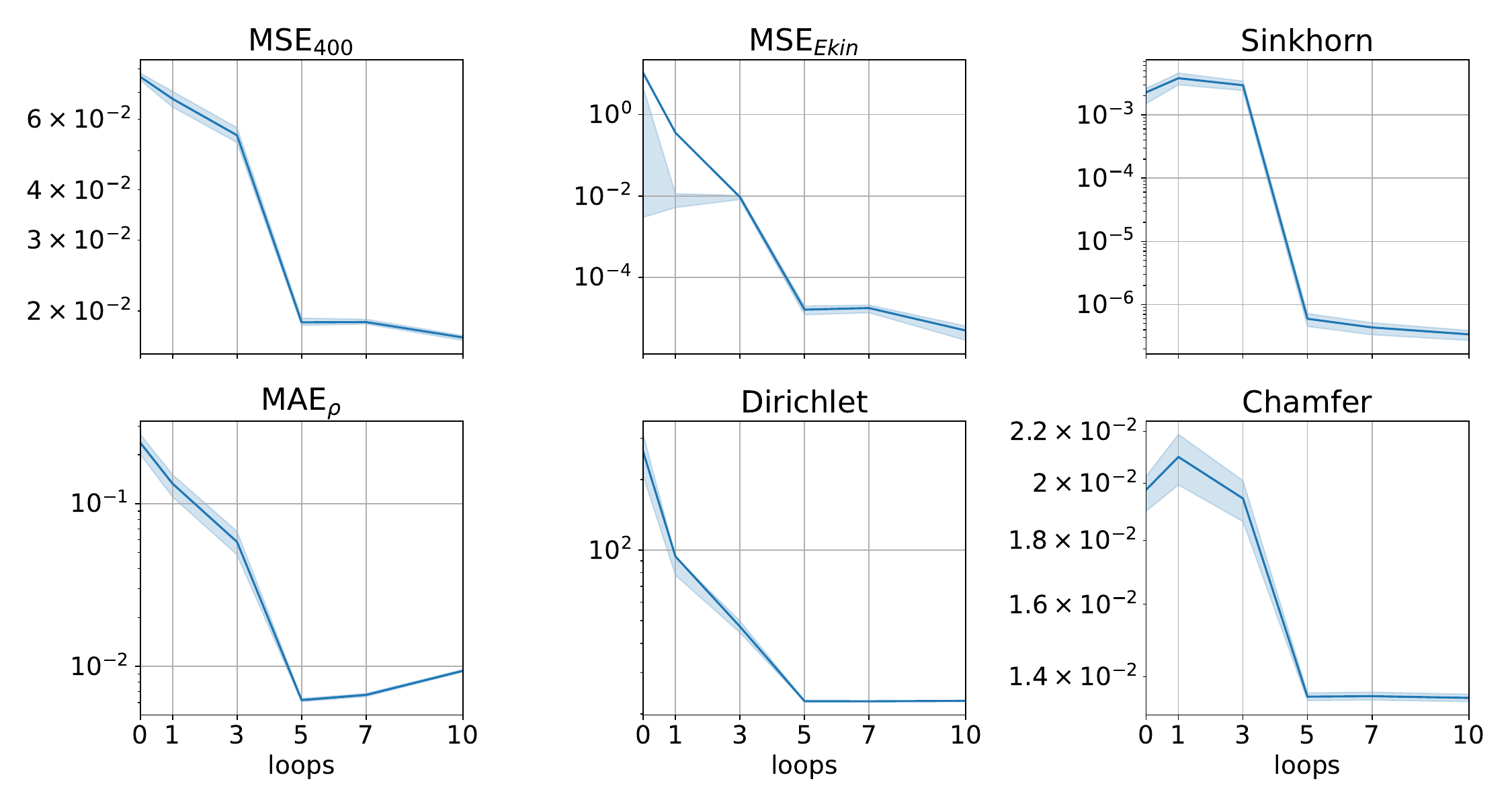}
    \caption{Ablations on LDC 2D with SEGNN-10-64 ($\alpha=0.03$) regarding the number of relaxation steps/loops. \label{fig:ldc2d_segnn_loops}}
\end{figure}

\begin{figure}[ht]
    \centering
    \includegraphics[width=0.48\textwidth]{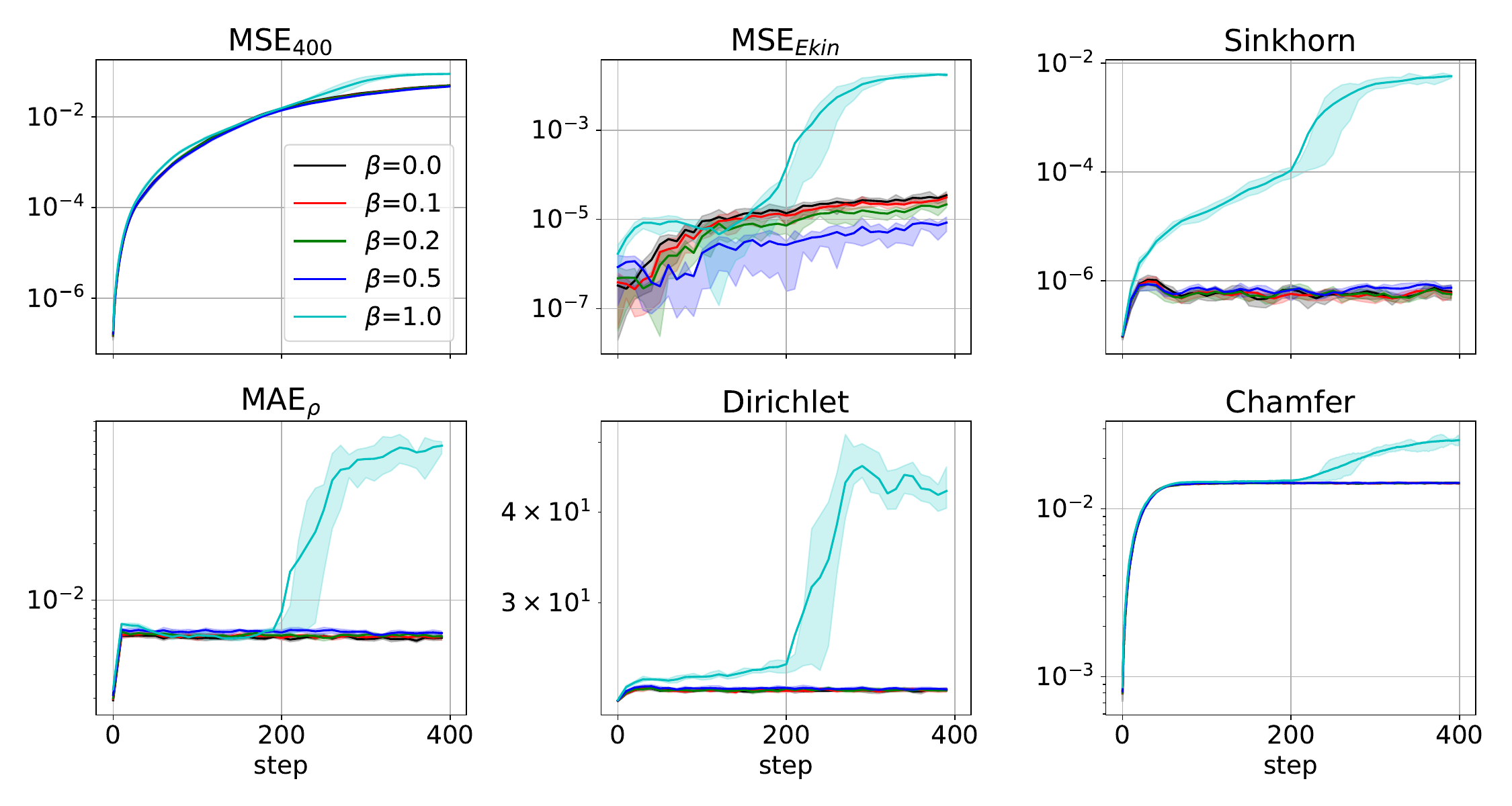}
    \hspace{10px}
    \includegraphics[width=0.48\textwidth]{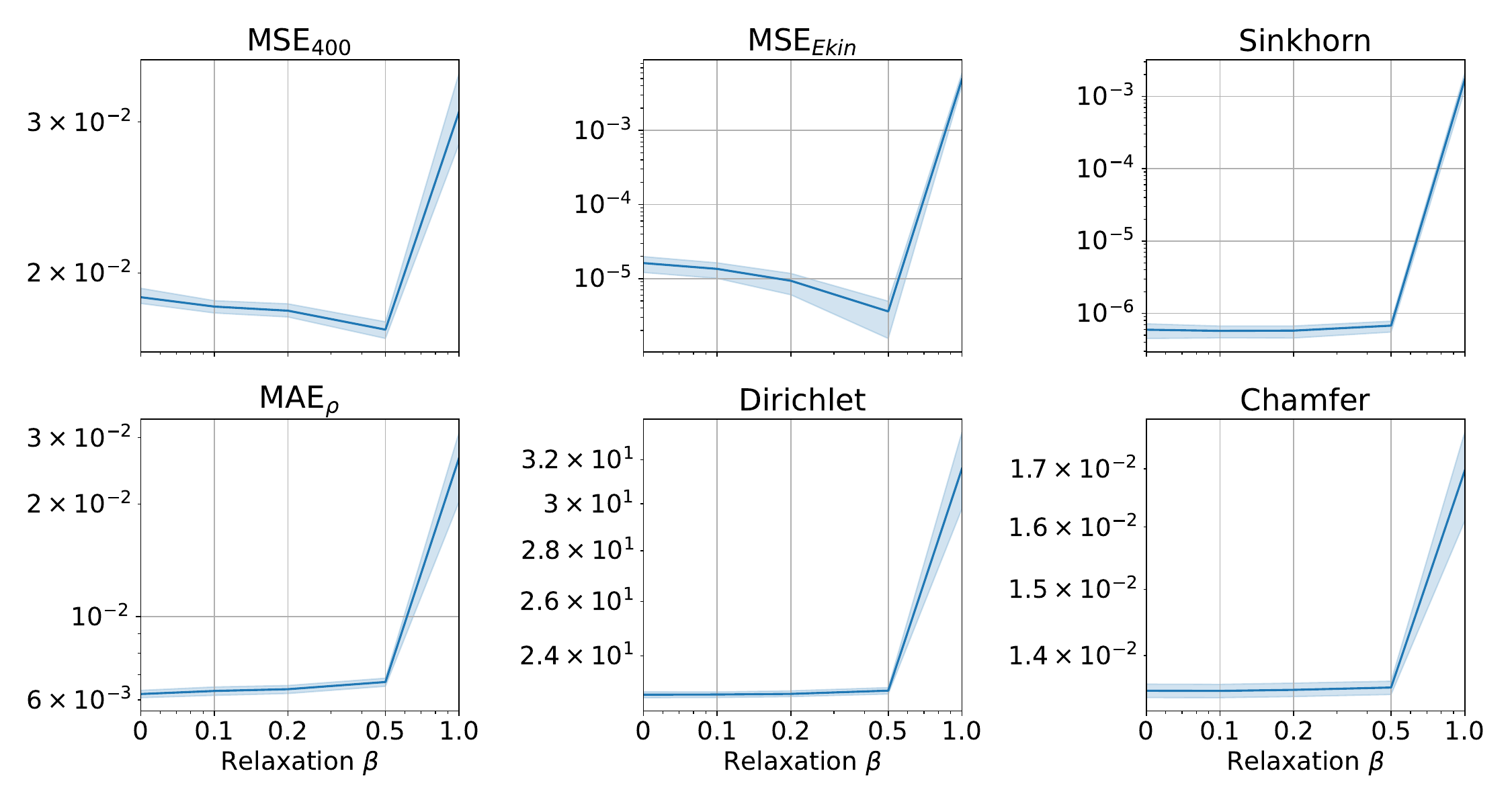}
    \caption{Ablations on LDC 2D with SEGNN-10-64 ($\alpha=0.03$, $l=5$) regarding relaxation parameter $\beta$.
    \label{fig:ldc2d_segnn_betas}}
\end{figure}

\newpage
\subsubsection{LDC 3D with GNS}
These plots agree with our choice of hyperparameters from \cref{tab:hyperparams} and show the sensitivity with respect to the relaxation parameters.

\begin{figure}[ht]
    \centering
    \includegraphics[width=0.48\textwidth]{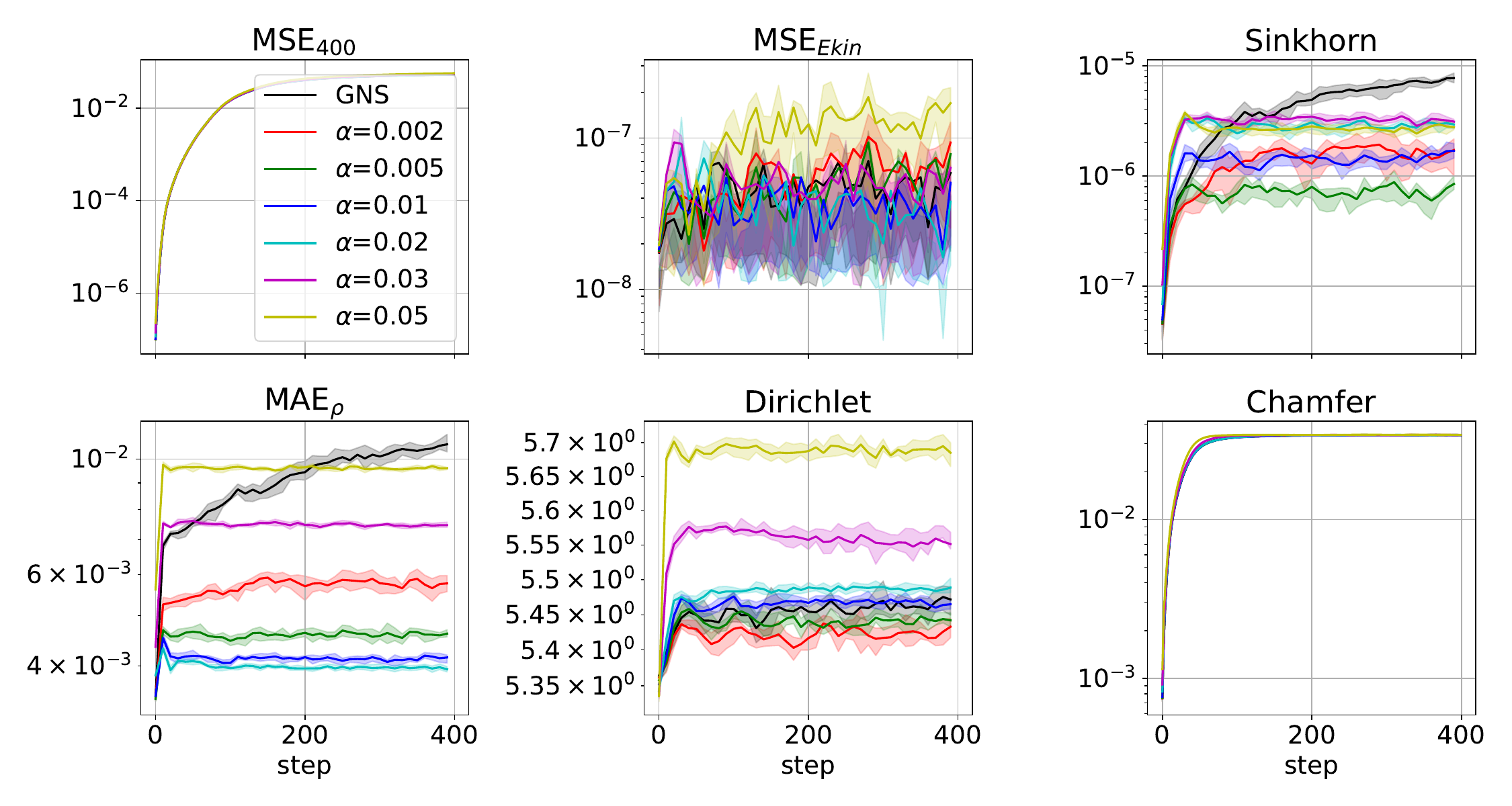}
    \hspace{10px}
    \includegraphics[width=0.48\textwidth]{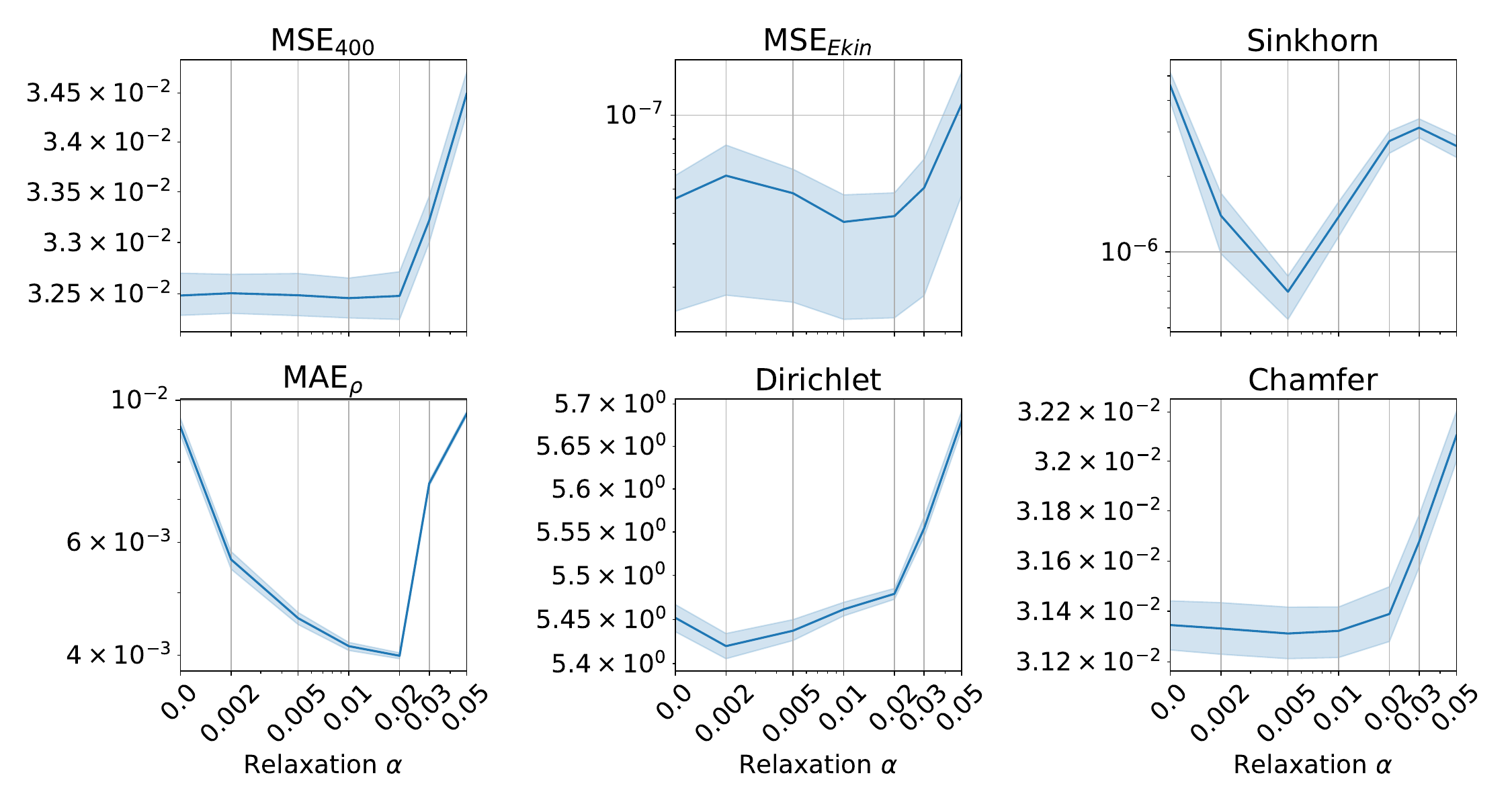}
    \caption{Ablations on LDC 3D with GNS-10-128 ($l=1$) regarding relaxation parameter $\alpha$. \label{fig:ldc3d_gns_alphas}}
\end{figure}

\begin{figure}[ht]
    \centering  
    \includegraphics[width=0.48\textwidth]{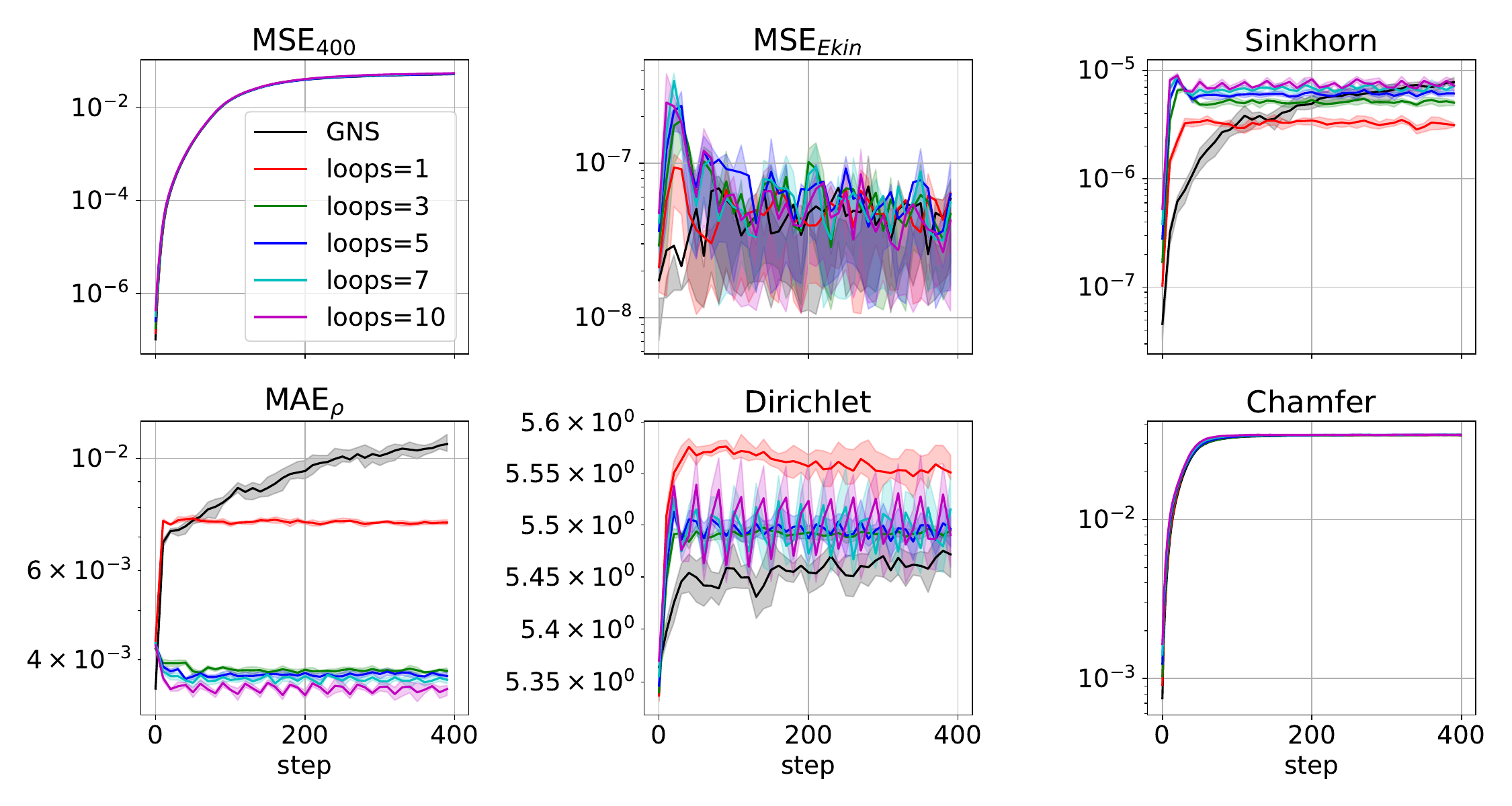}
    \hspace{10px}
    \includegraphics[width=0.48\textwidth]{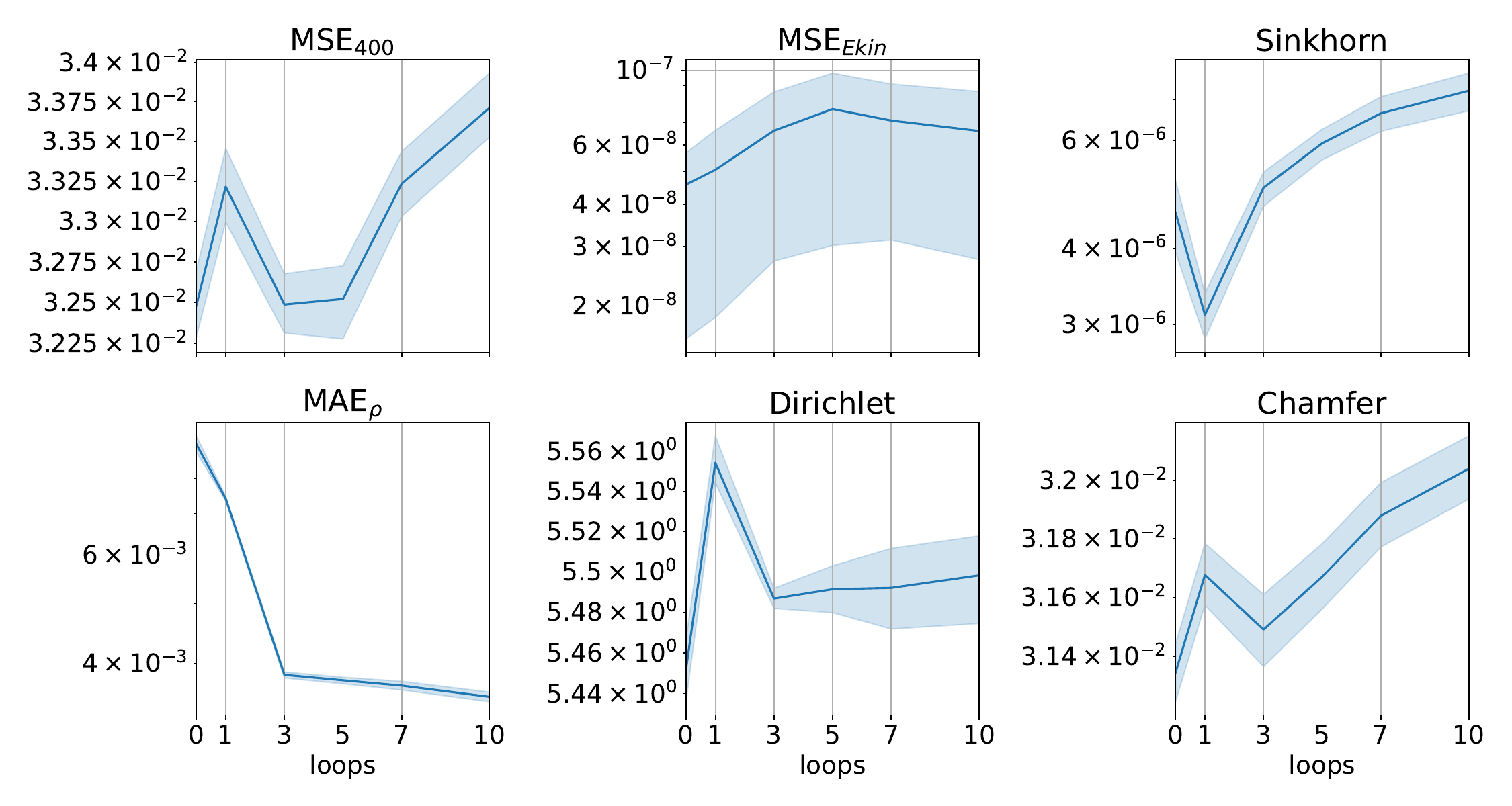}
    \caption{Ablations on LDC 3D with GNS-10-128 ($\alpha=0.02$) regarding the number of relaxation steps/loops. \label{fig:ldc3d_gns_loops}}
\end{figure}

\begin{figure}[ht]
    \centering
    \includegraphics[width=0.48\textwidth]{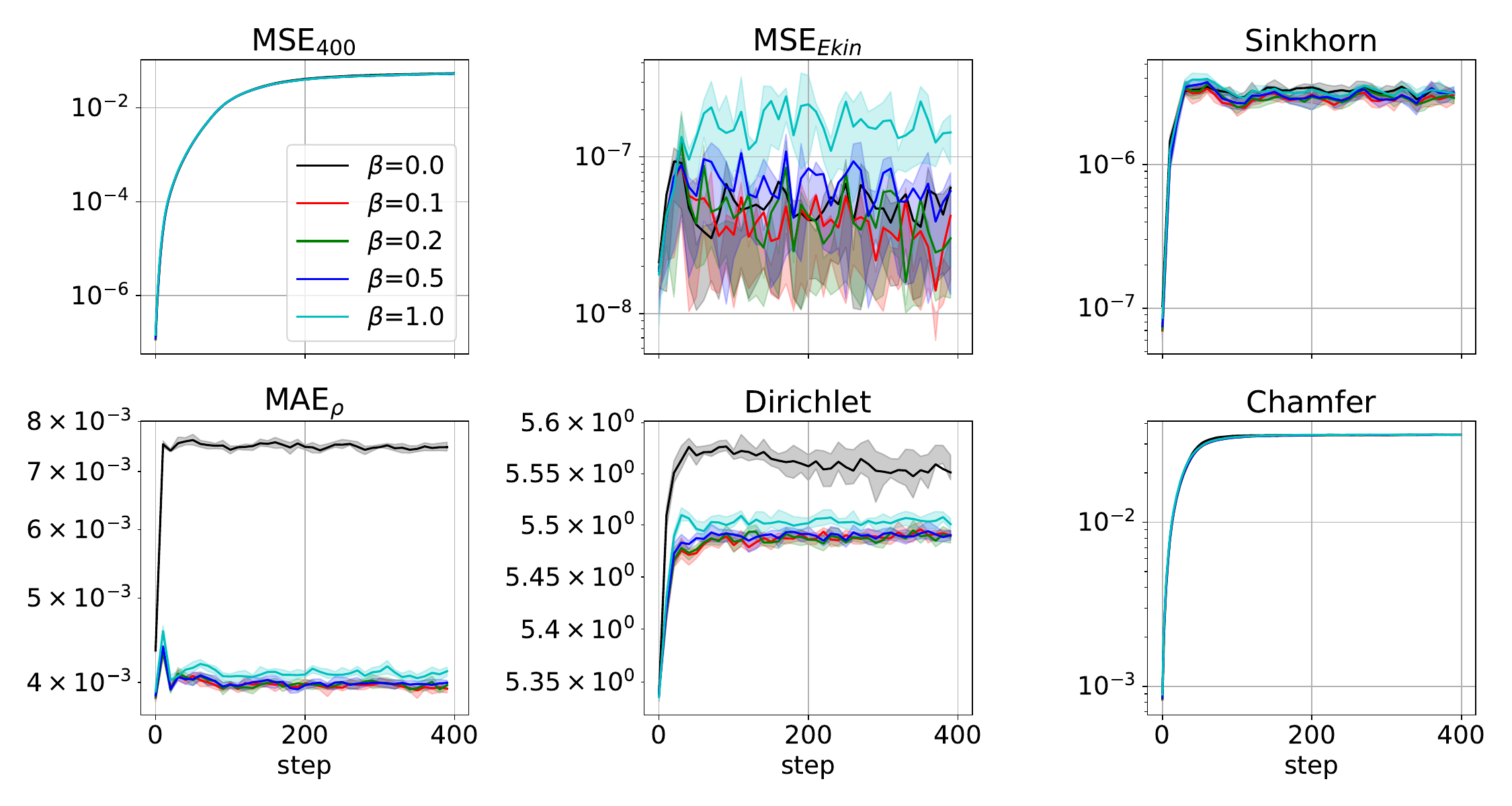}
    \hspace{10px}
    \includegraphics[width=0.48\textwidth]{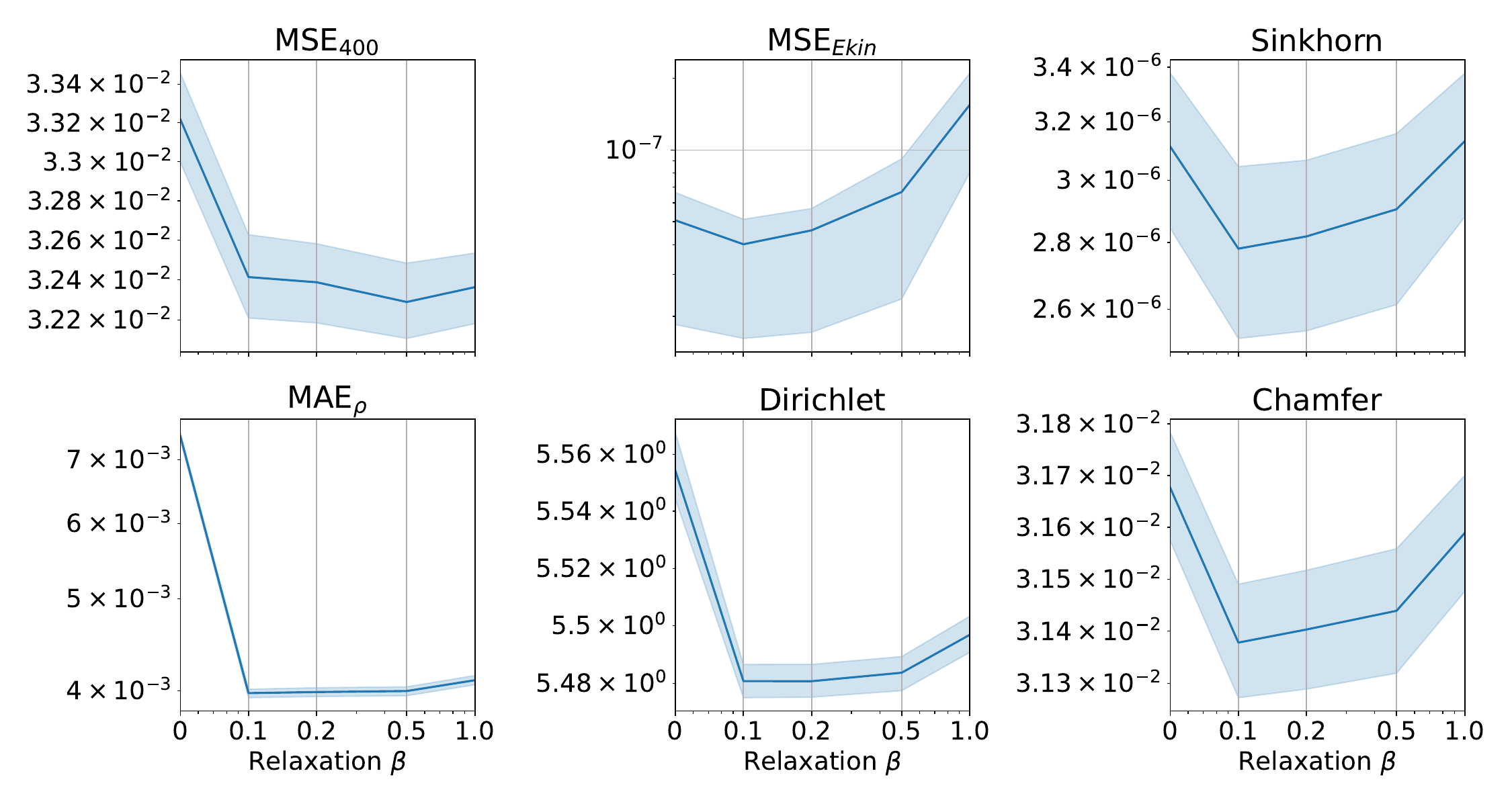}
    \caption{Ablations on LDC 3D with GNS-10-128 ($\alpha=0.02$, $l=1$) regarding relaxation parameter $\beta$.
    \label{fig:ldc3d_gns_betas}}
\end{figure}

\newpage
\subsubsection{LDC 3D with SEGNN}

\begin{figure}[ht]
    \centering
    \includegraphics[width=0.48\textwidth]{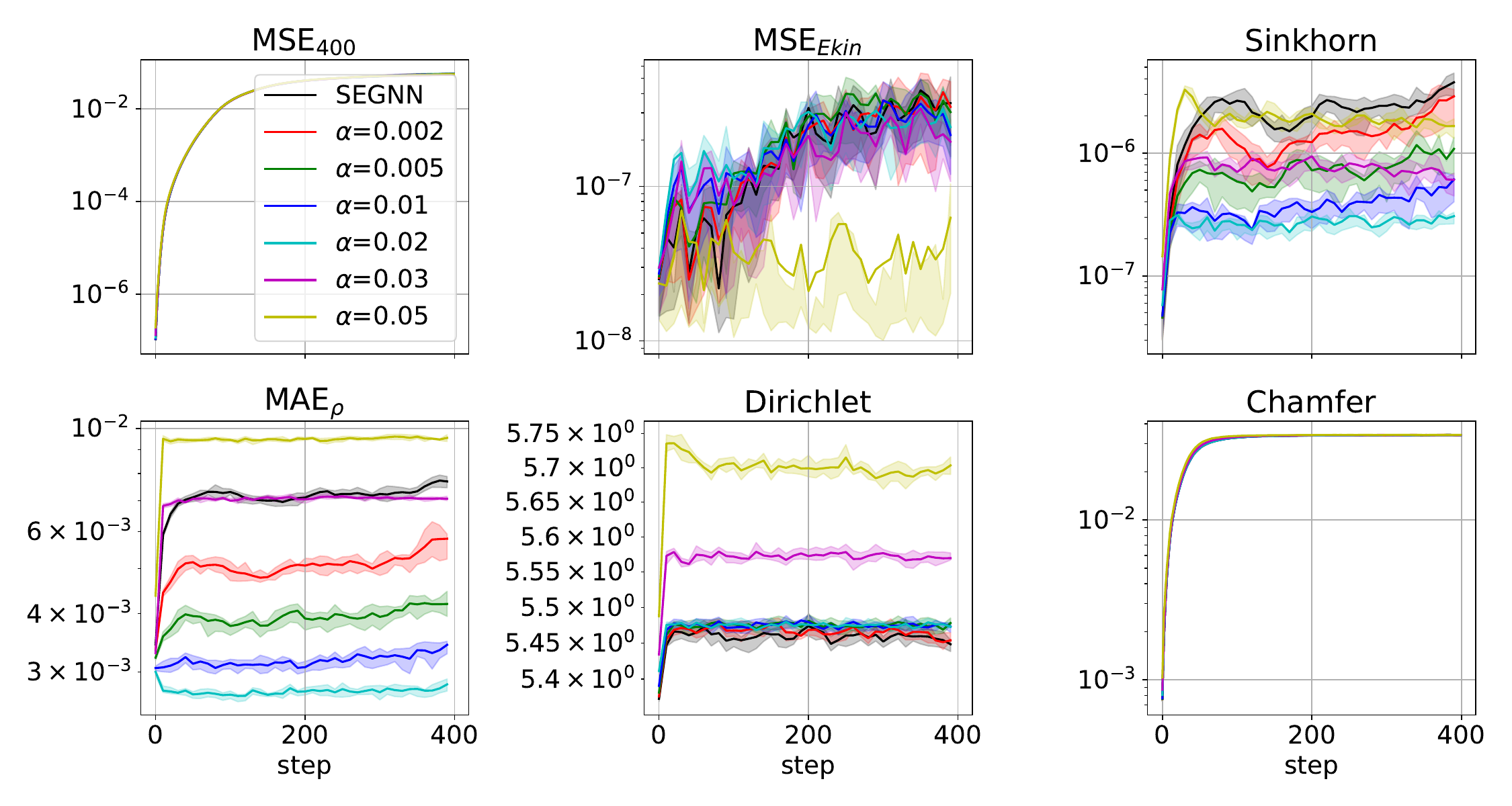}
    \hspace{10px}
    \includegraphics[width=0.48\textwidth]{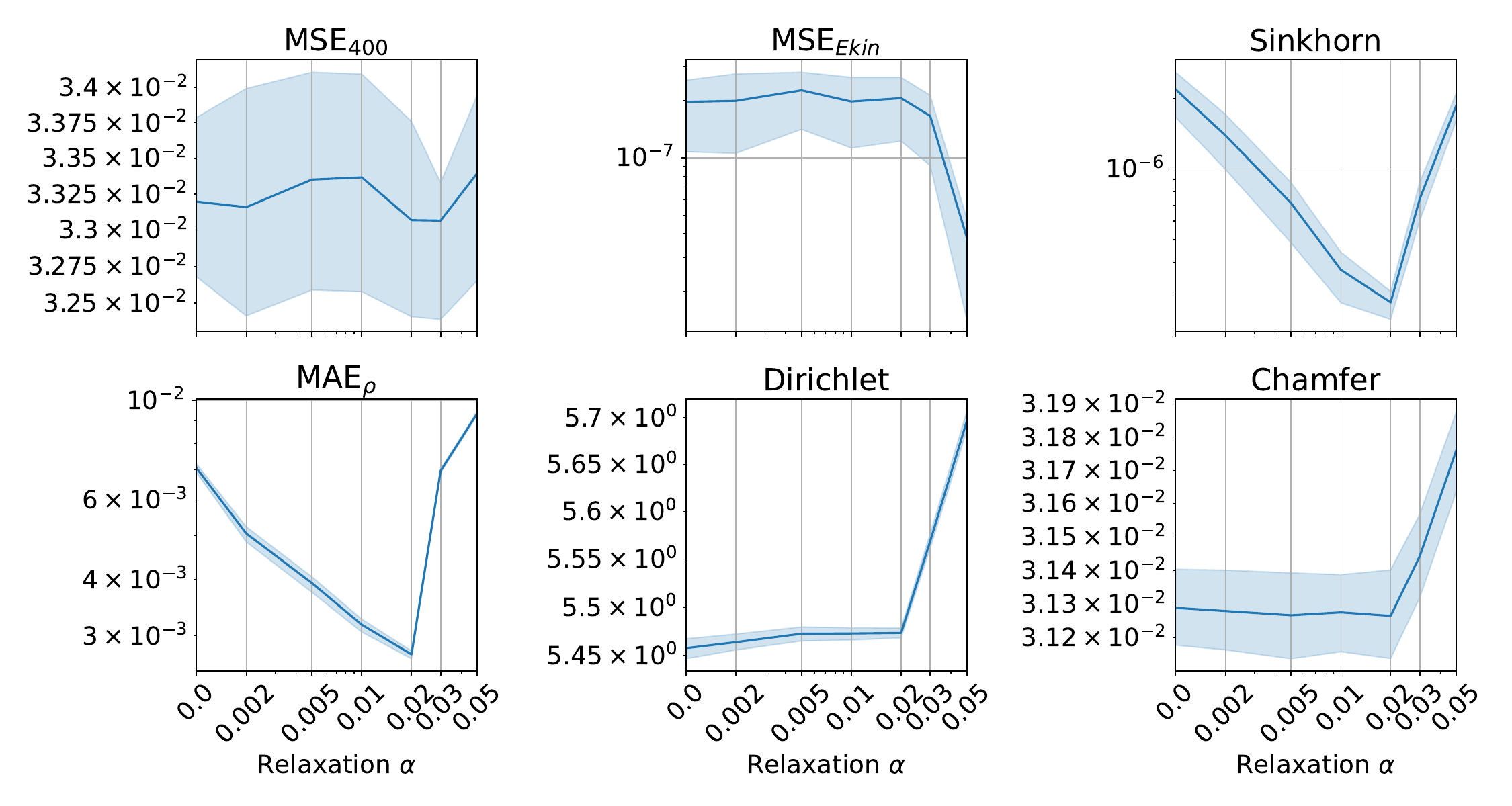}
    \caption{Ablations on LDC 3D with SEGNN-10-64 ($l=1$) regarding relaxation parameter $\alpha$. \label{fig:ldc3d_segnn_alphas}}
\end{figure}

\begin{figure}[ht]
    \centering  
    \includegraphics[width=0.48\textwidth]{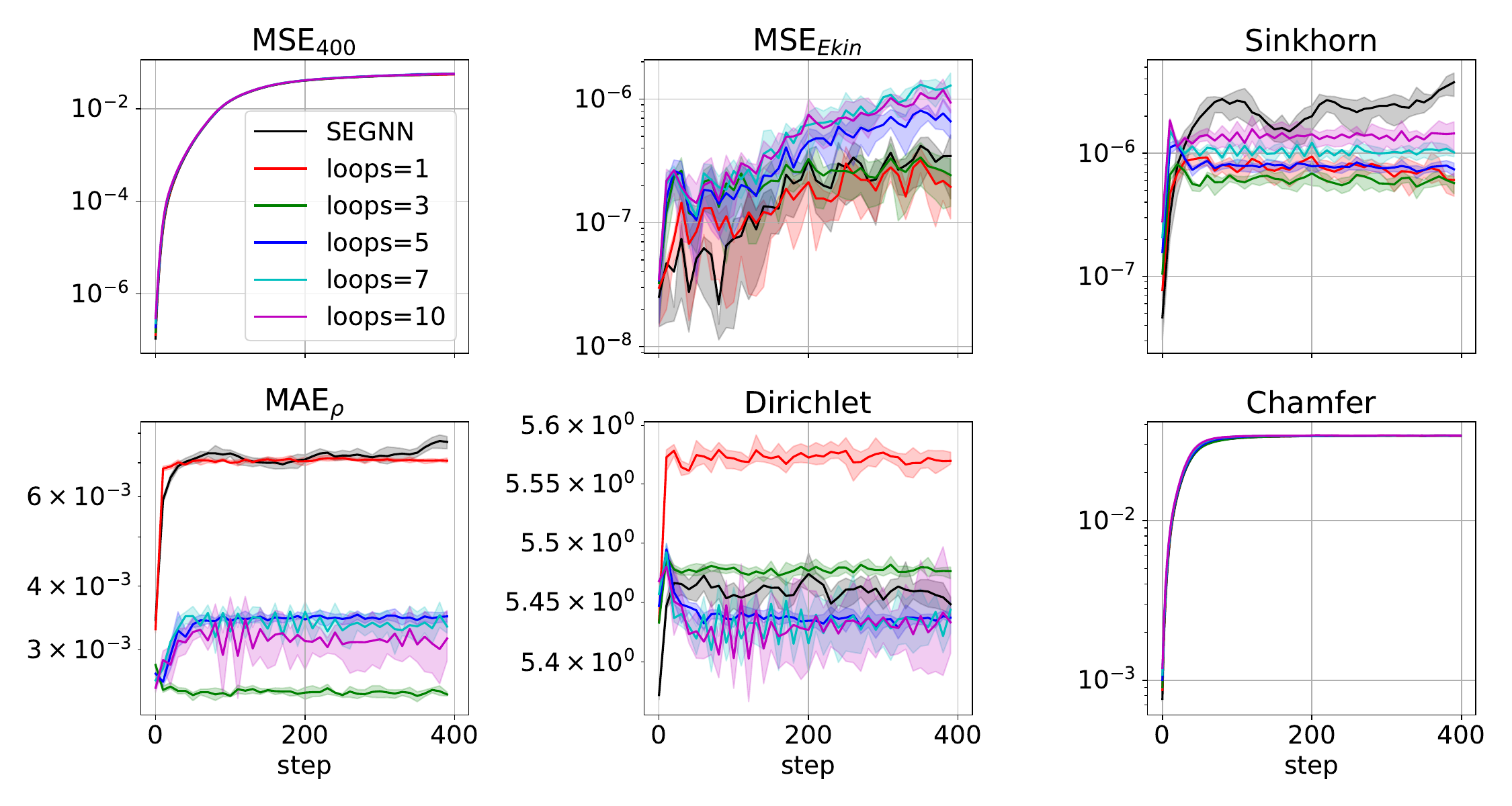}
    \hspace{10px}
    \includegraphics[width=0.48\textwidth]{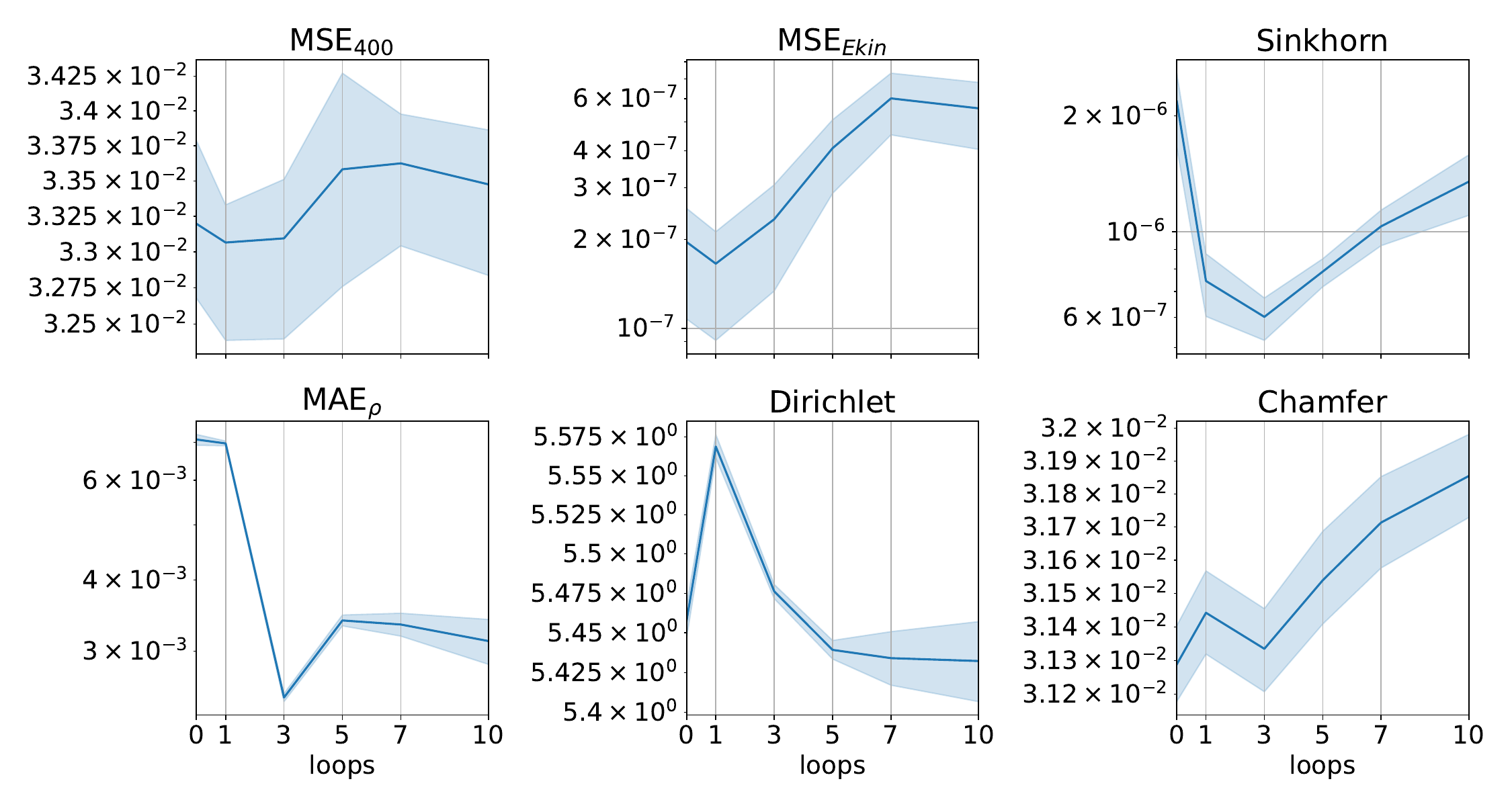}
    \caption{Ablations on LDC 3D with SEGNN-10-64 ($\alpha=0.02$) regarding the number of relaxation steps/loops. \label{fig:ldc3d_segnn_loops}}
\end{figure}

\begin{figure}[ht]
    \centering
    \includegraphics[width=0.48\textwidth]{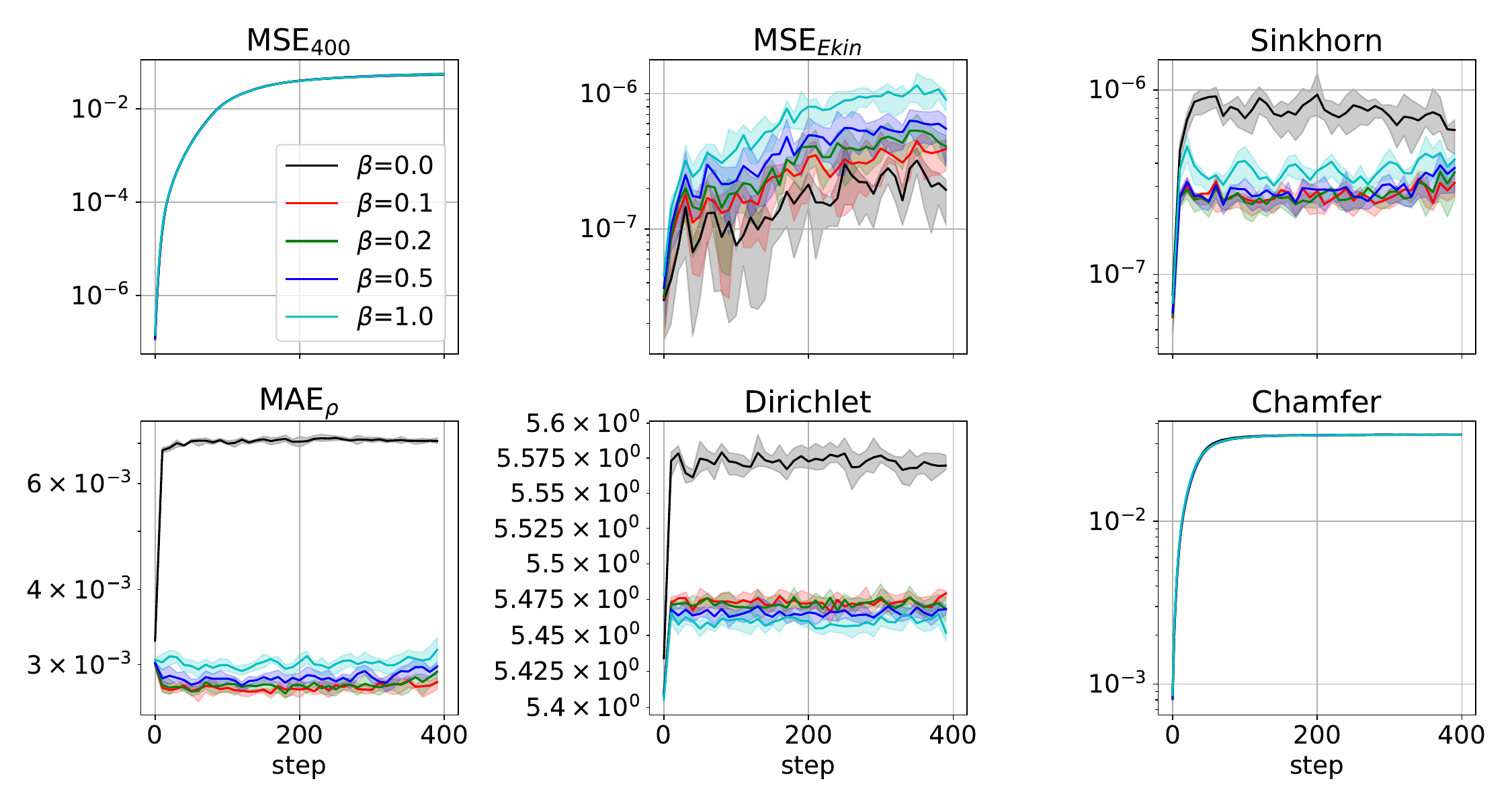}
    \hspace{10px}
    \includegraphics[width=0.48\textwidth]{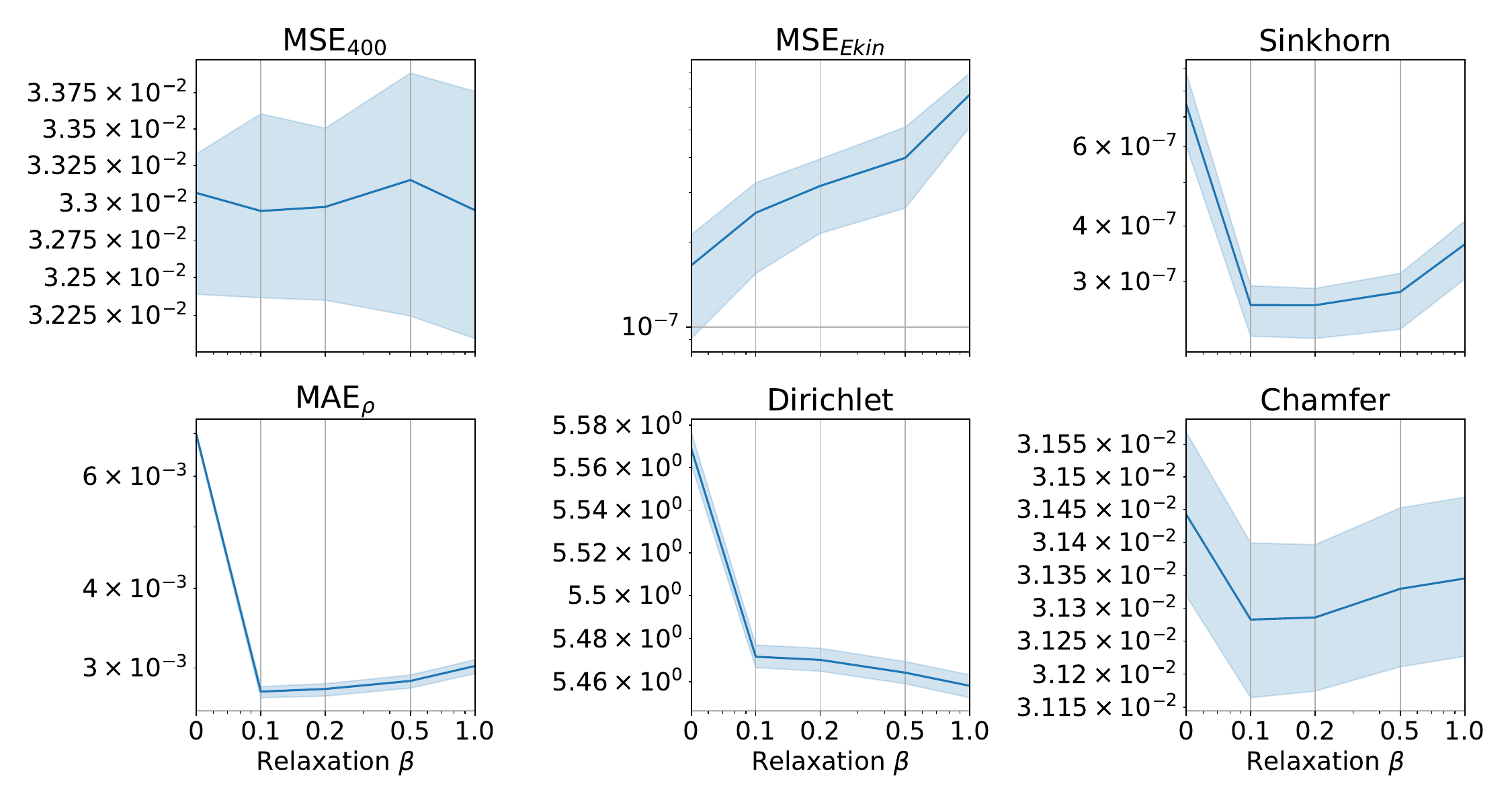}
    \caption{Ablations on LDC 3D with SEGNN-10-64 ($\alpha=0.02$, $l=1$) regarding relaxation parameter $\beta$.
    \label{fig:ldc3d_segnn_betas}}
\end{figure}

\subsection{Reverse Poiseuille Flow} \label{app:ablations_rpf}
\begin{figure}[ht]
    \centering
    \includegraphics[width=0.48\textwidth]{figs/rpf2d_gns_ext.pdf}
    \hspace{10px}
    \includegraphics[width=0.48\textwidth]{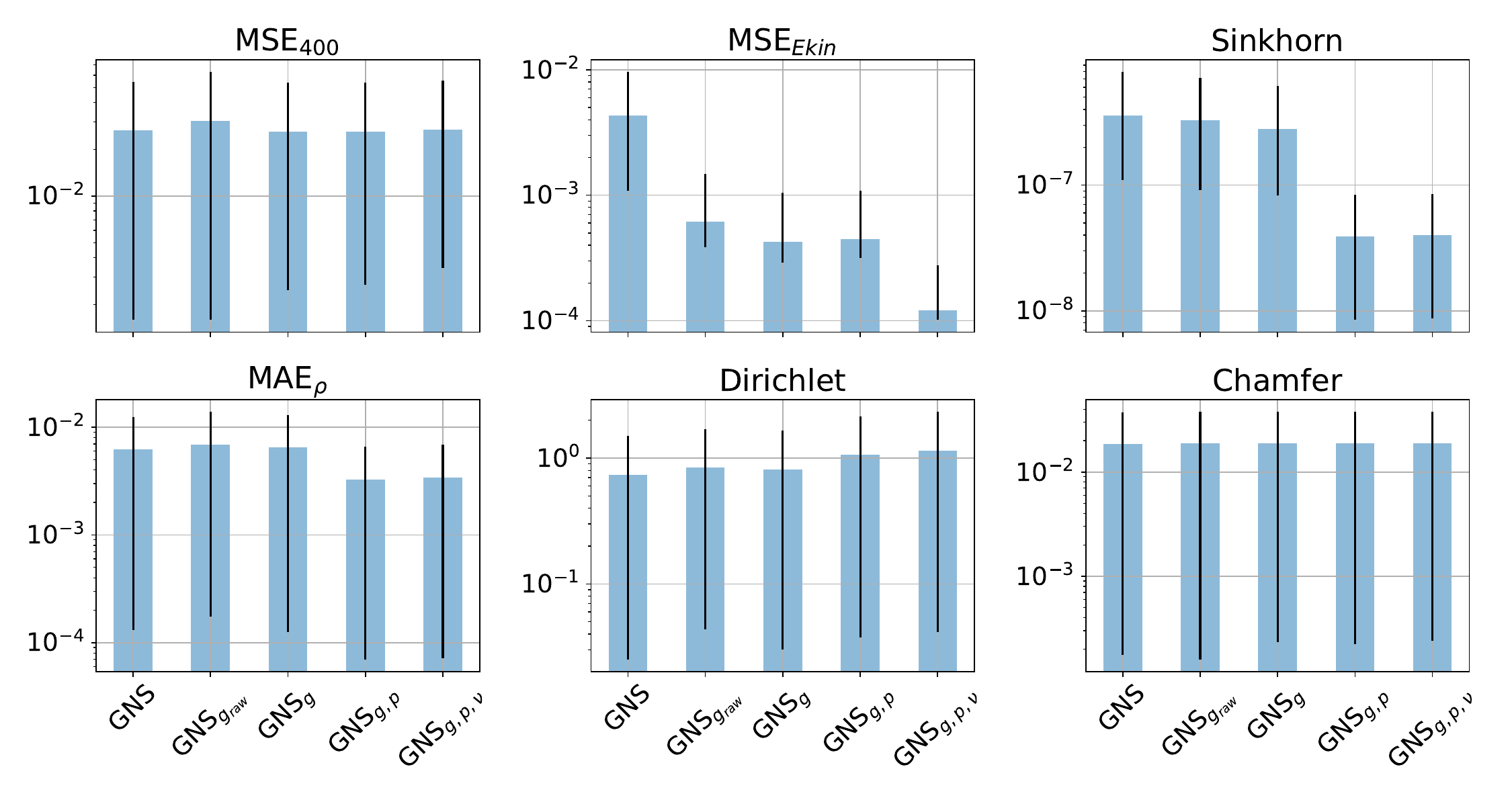}
    \caption{Ablations on RPF 2D with GNS-10-128 over the simulation length (left) and the average thereof (right). \label{fig:rpf2d_gns_ext}}
\end{figure}

\begin{figure}[ht]
    \centering  
    \includegraphics[width=0.48\textwidth]{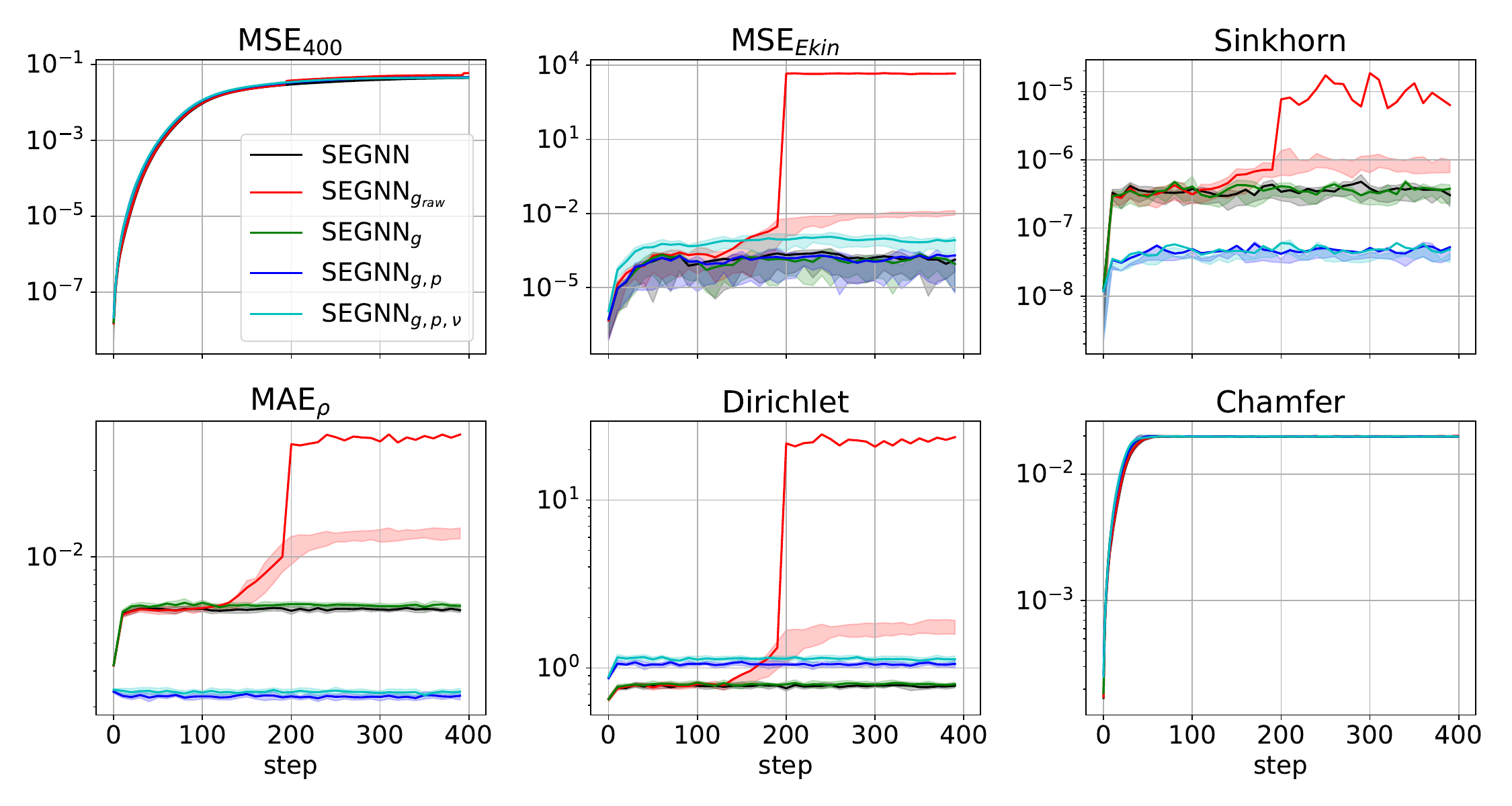}
    \hspace{10px}
    \includegraphics[width=0.48\textwidth]{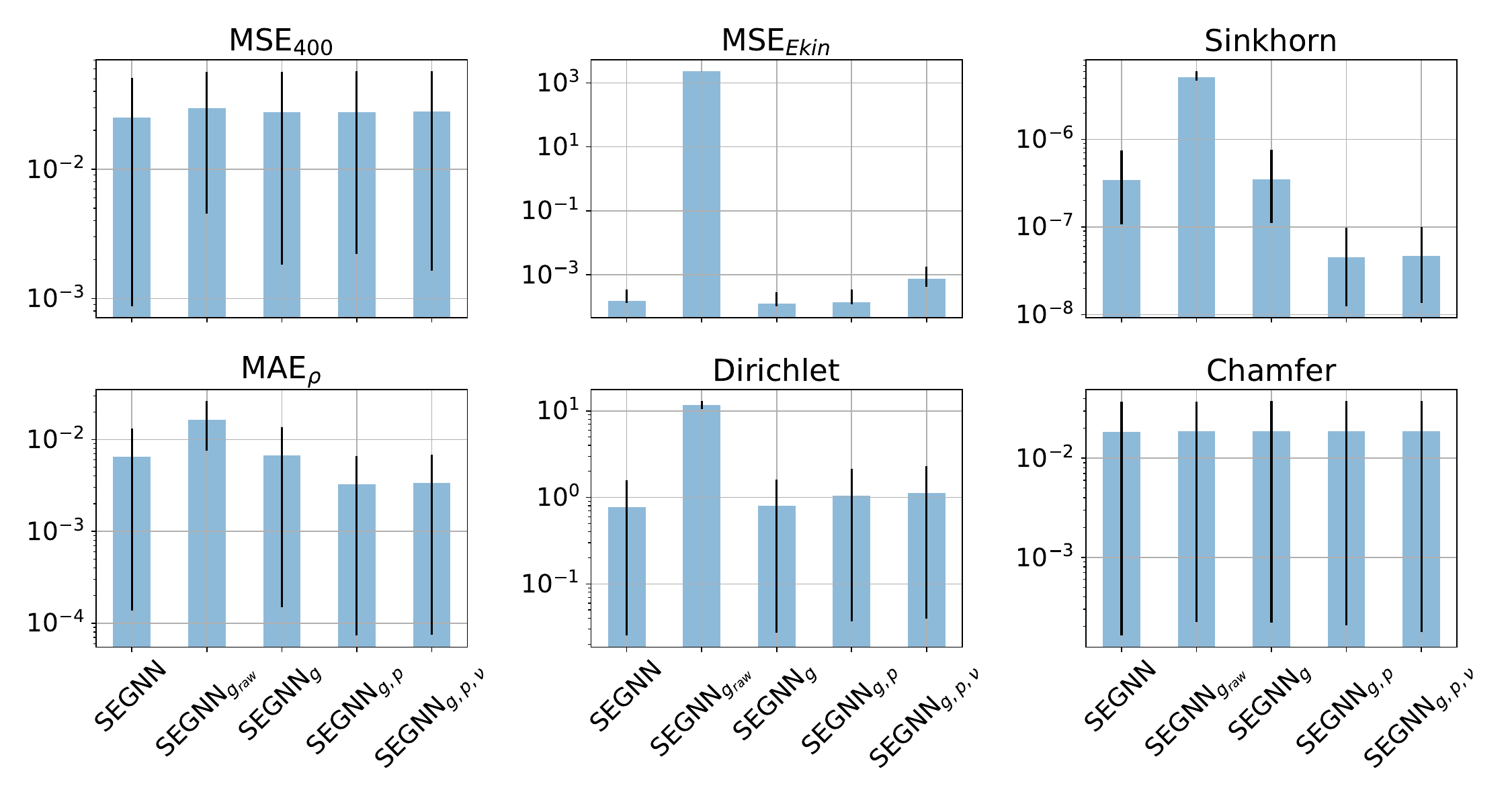}
    \caption{Ablations on RPF 2D with SEGNN-10-64 over the simulation length (left) and the average thereof (right). \label{fig:rpf2d_segnn_ext}}
\end{figure}

\begin{figure}[ht]
    \centering
    \includegraphics[width=0.48\textwidth]{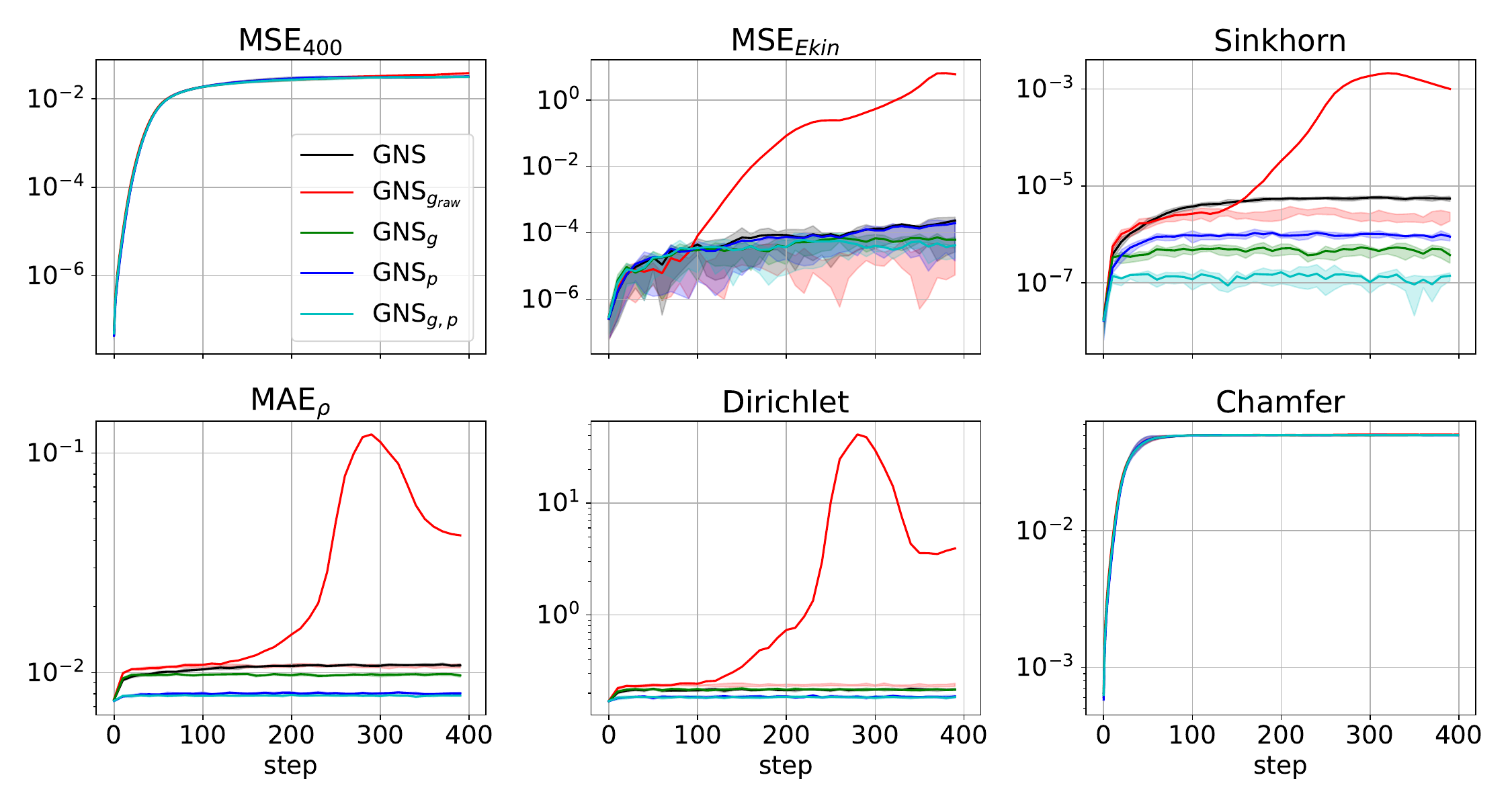}
    \hspace{10px}
    \includegraphics[width=0.48\textwidth]{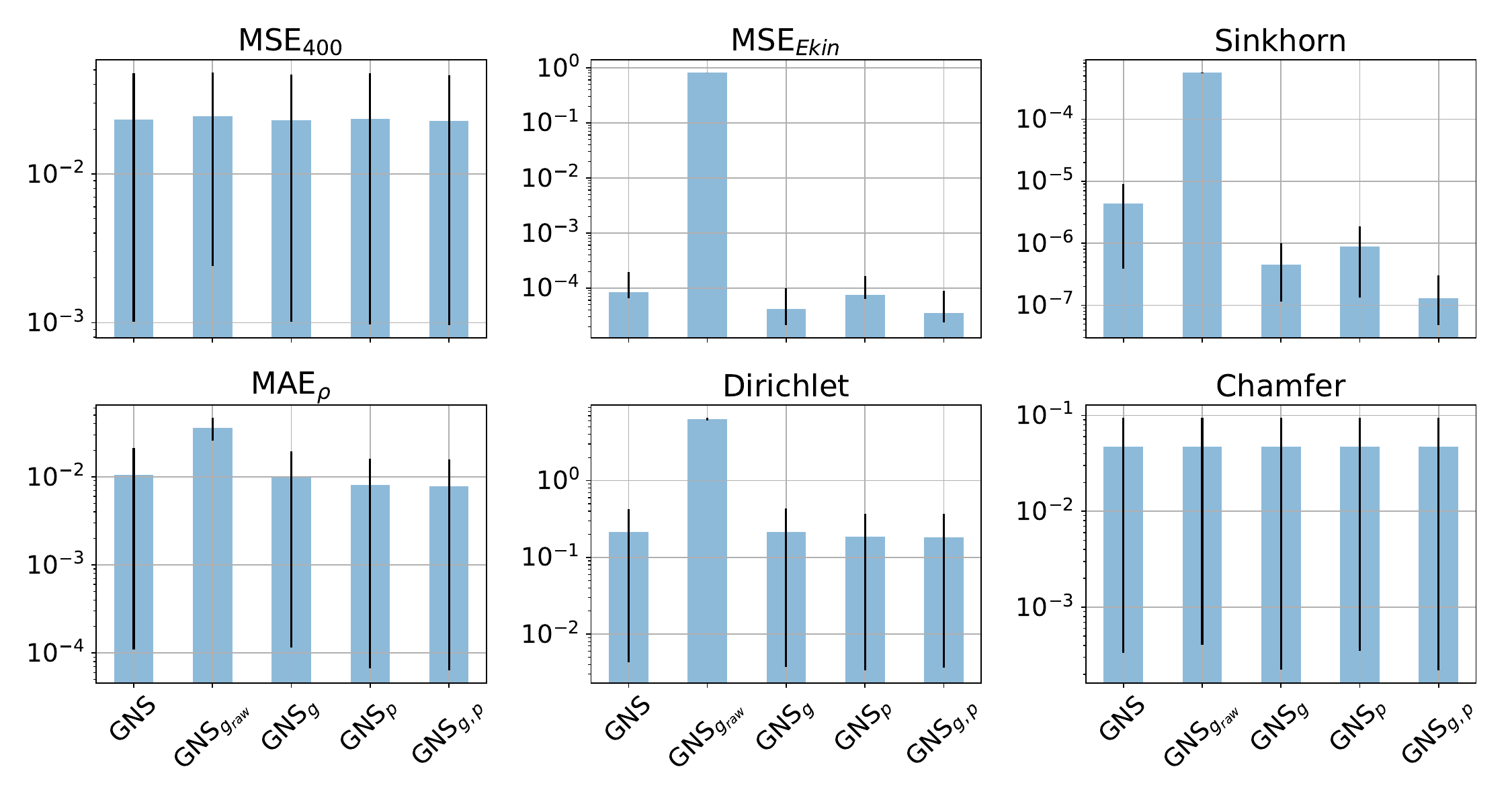}
    \caption{Ablations on RPF 3D with GNS-10-128 over the simulation length (left) and the average thereof (right). \label{fig:rpf3d_gns_ext}}
\end{figure}

\begin{figure}[ht]
    \centering  
    \includegraphics[width=0.48\textwidth]{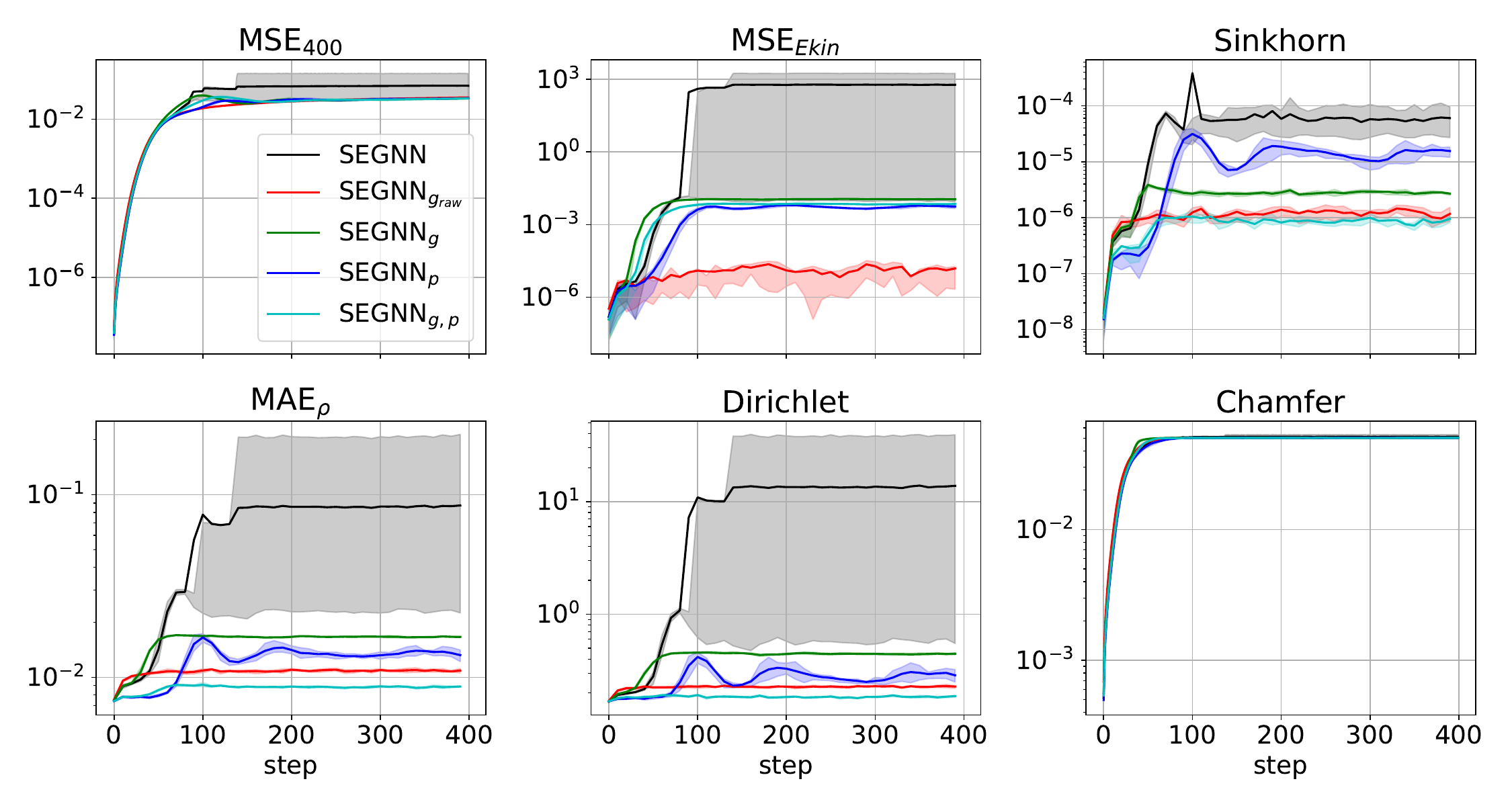}
    \hspace{10px}
    \includegraphics[width=0.48\textwidth]{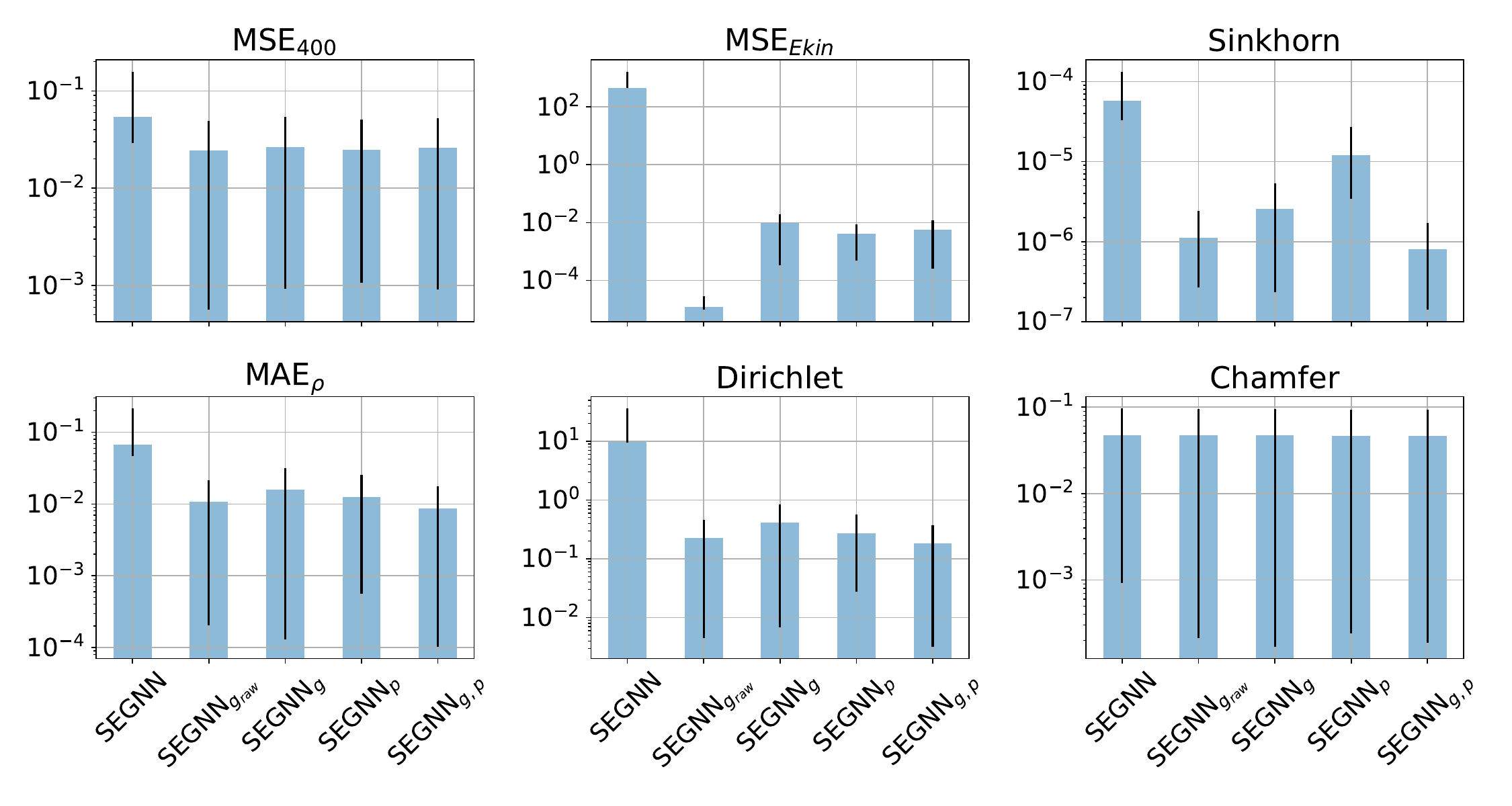}
    \caption{Ablations on RPF 3D with SEGNN-10-64 over the simulation length (left) and the average thereof (right). \label{fig:rpf3d_segnn_ext}}
\end{figure}

We compare all variants of RPF model from the main paper with the case of not smoothing the external force, denoted $\square_{g_{raw}}$. The main message with regard to excluding the external force from the training target (all methods with $\square_{g}$) is that not smoothing the force function when it has discontinuities leads to highly unstable models, see MSE$_{Ekin}$ in \cref{fig:rpf2d_segnn_ext,fig:rpf3d_gns_ext}. It is probably a matter of too few test trajectories that we do not observe such blow-ups in \cref{fig:rpf2d_gns_ext,fig:rpf3d_segnn_ext}.

\clearpage

\section{Training with Relaxations}  \label{app:relaxed_training}
We also explored to idea of incorporating the SPH relaxation during training, hoping that the learned model can be regularized toward predicting better particle distributions, which could make the SPH relaxation during inference unnecessary. We explored two degrees of freedom when training a GNS-10-128 model on the 2D LDC dataset: (a) dependence on the relaxation parameter $\alpha$, and (b) performance when trained with relaxation but evaluated with or without it. 

\paragraph{Basic setup}
We remind the reader that according to \cref{tab:hyperparams}, the optimal relaxation parameters on 2D LDC are $\alpha=0.03$ and 5 relaxation steps, but from the ablation in \cref{fig:ldc2d_gns_alphas}, we see that even one relaxation step significantly improves the dynamics. Thus, for simplicity, we use $\alpha=0.03$ with 1 relaxation step for our training with relaxation. We implemented this training scheme by adding the relaxation to every forward call of the model, i.e. when pushforward is applied, the relaxation is executed at every pushforward step.

\begin{figure}[ht]
    \centering  
    \includegraphics[width=0.48\textwidth]{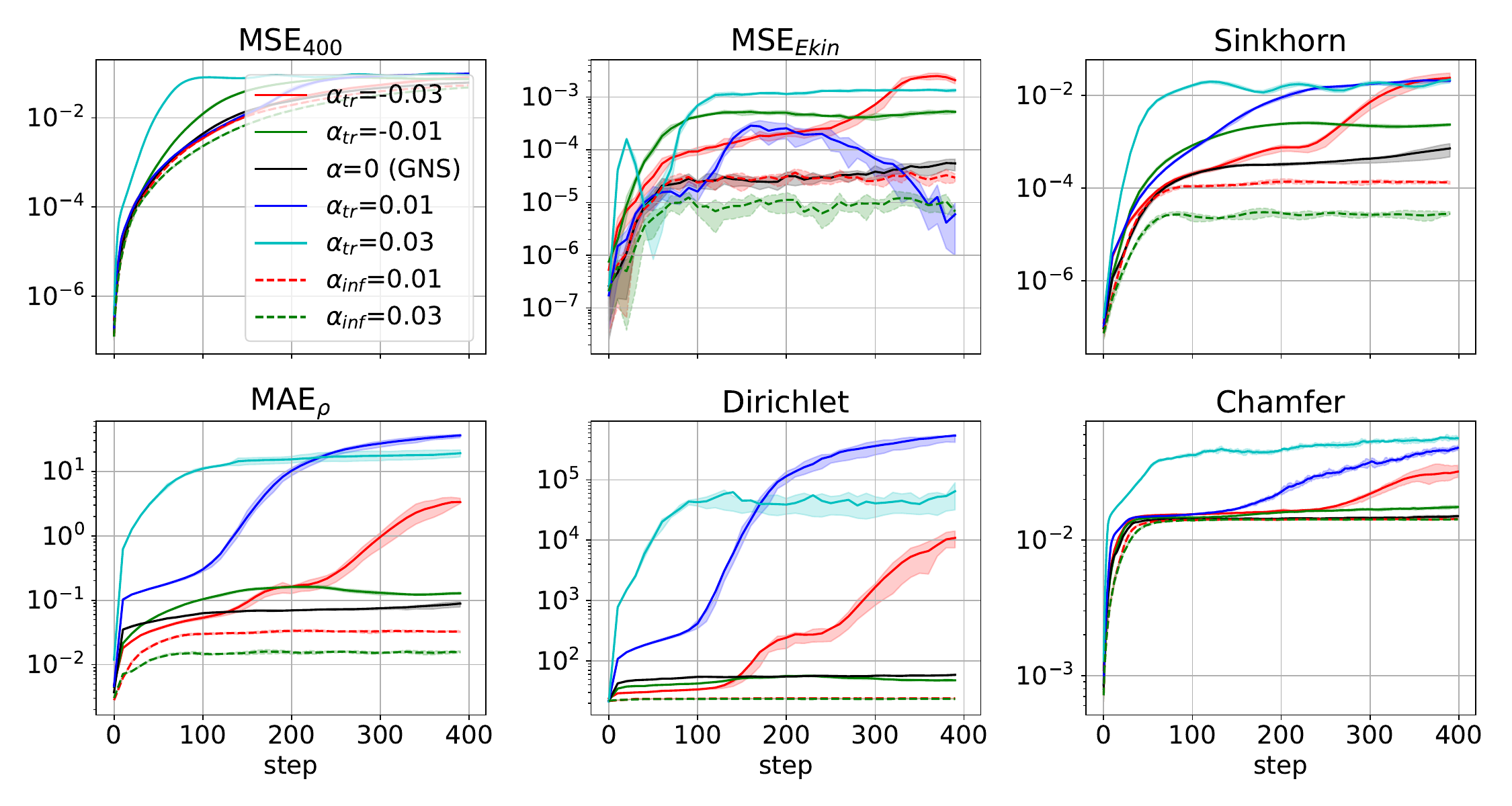}
    \hspace{10px}
    \includegraphics[width=0.48\textwidth]{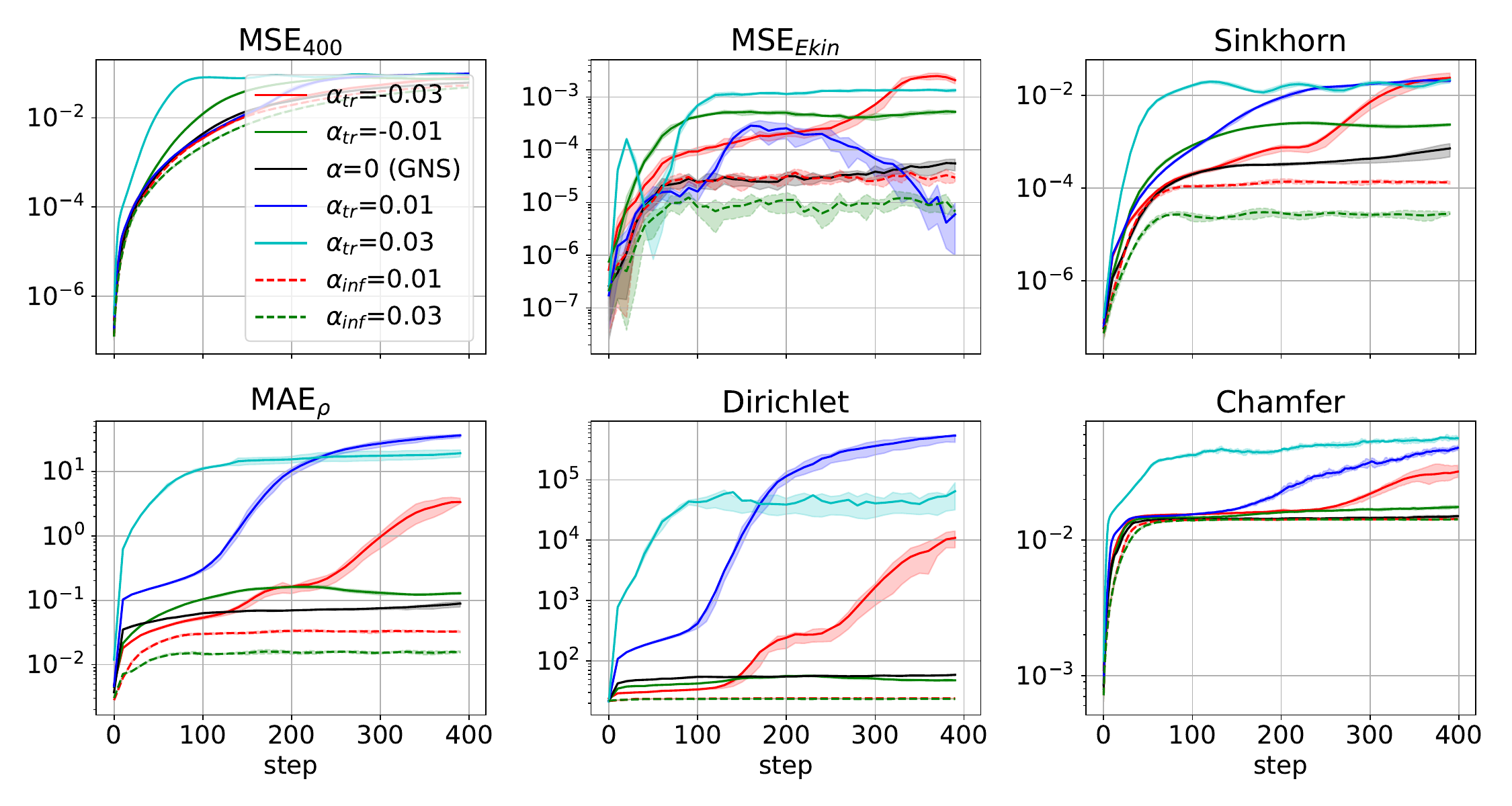}
    \caption{GNS-10-128 trained on 2D LDC with relaxation, and but evaluated \textbf{without} relaxation. We denote with $\alpha_{\text{tr}}$ that the model has experienced relaxation only during training and with $\alpha_{\text{inf}}$ only during inference. Metrics over the simulation length (left) and the average thereof (right). \label{fig:rlx_ldc2d_0}}
\end{figure}

\paragraph{Training with "negative" relaxation} 
One highly appealing idea is to train the model with what we call "negative" relaxation, i.e. flipping the sign of the relaxation term by setting $\alpha$ to a negative value, by which the model would learn to over-correct unfavorable distributions. However, the results for $\alpha<0$ in \cref{fig:rlx_ldc2d_0} are rather discouraging.

\begin{figure}[ht]
    \centering
    \includegraphics[width=0.48\textwidth]{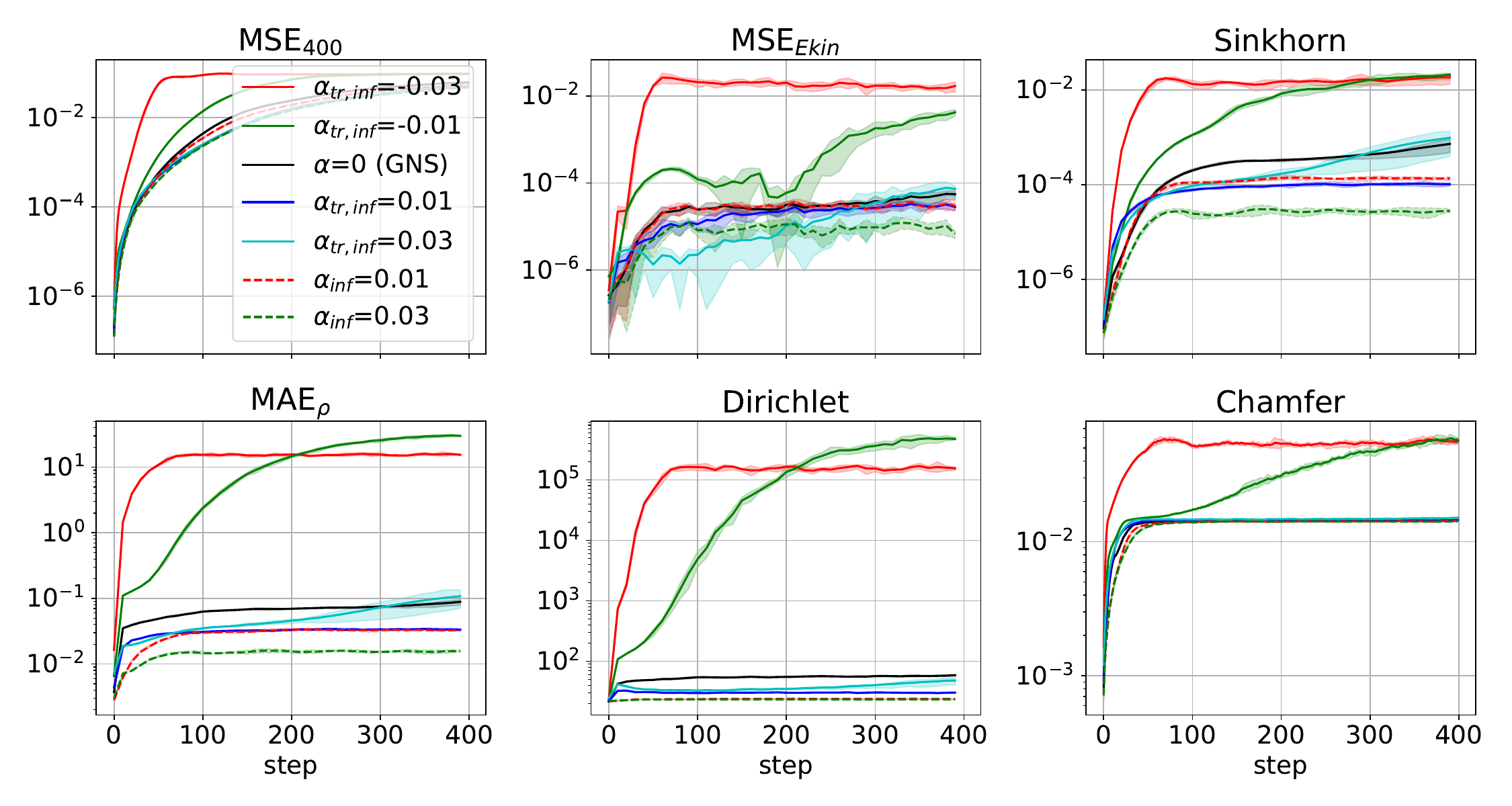}
    \hspace{10px}
    \includegraphics[width=0.48\textwidth]{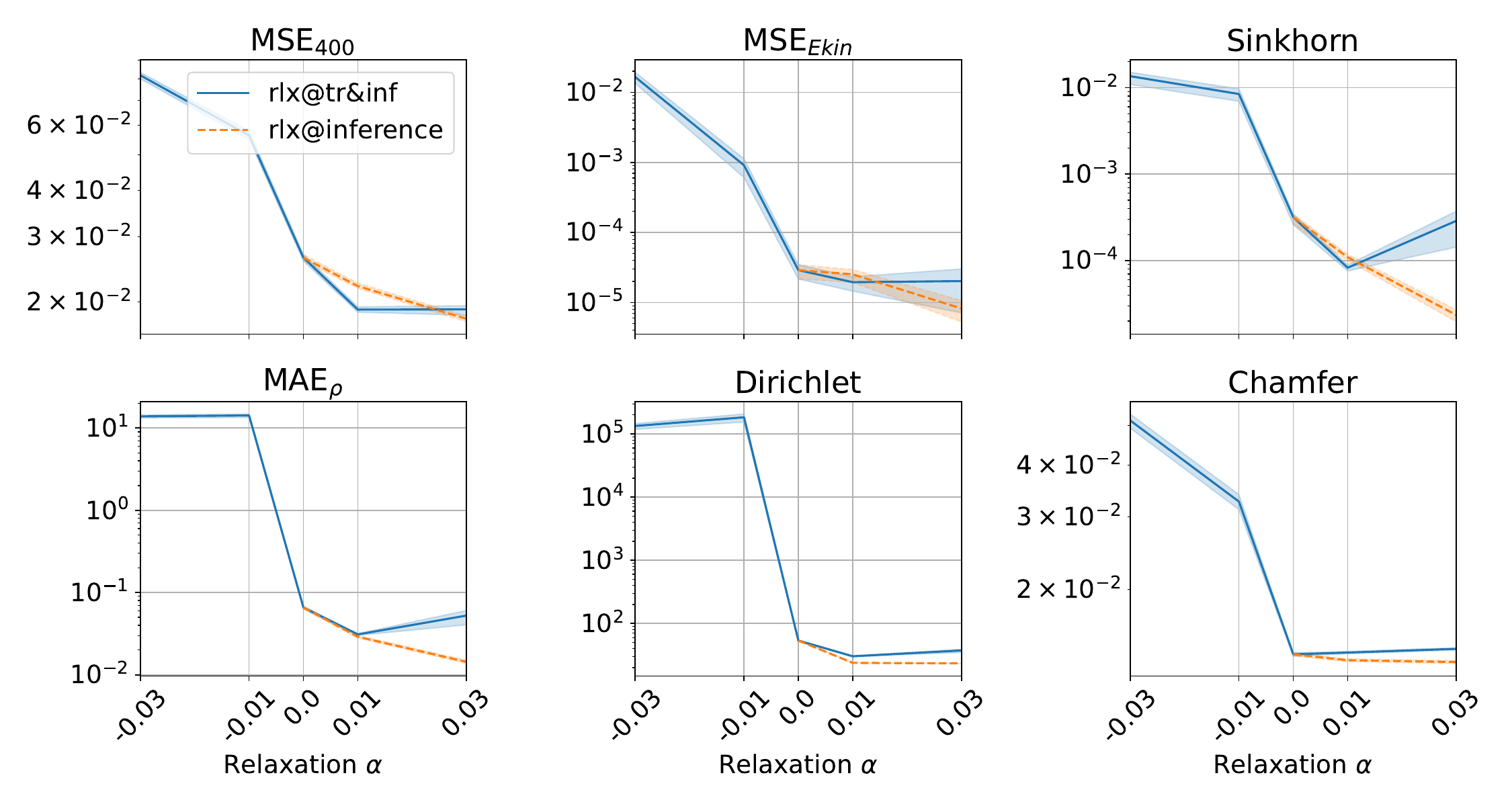}
    \caption{GNS-10-128 trained on 2D LDC with relaxation, and also evaluated \textbf{with} relaxation. We denote with $\alpha_{\text{tr,inf}}$ that the model has experienced relaxation both during training and inference and with $\alpha_{\text{inf}}$ only during inference. Metrics over the simulation length (left) and the average thereof (right). \label{fig:rlx_ldc2d_1}}
\end{figure}

\paragraph{Training and inference with relaxation}
Similar to subtracting the external force from the learning target, which we discussed in length and seems very useful, we investigated how the model would perform when it can predict an even worse particle distribution, which is then corrected through a relaxation both during training and inference, see $\alpha>0$ in \cref{fig:rlx_ldc2d_1}. But also here, we get worse results than only applying relaxation during inference. In addition, training with relaxation requires separate retraining until $\alpha$ is tuned, which is not the case with our inference time relaxation.

\end{document}